\documentclass[11pt]{article}
\usepackage{color}
\usepackage{graphicx}
\usepackage{makeidx}
\usepackage{fixmath}
\usepackage{amsfonts}
\usepackage{amsmath}
\newcommand{\beqa}{\begin{eqnarray}}
\newcommand{\eeqa}[1]{\label{#1}\end{eqnarray}}
\newcommand{\bequ}{\begin{equation}}
\newcommand{\eequ}[1]{\label{#1}\end{equation}}

\newcommand{\beq}{\begin{equation}}
\newcommand{\eeq}{\end{equation}}
\newcommand{\overliner}{\begin{eqnarray}}
\newcommand{\earr}{\end{eqnarray}}
\newcommand{\beqn}{\begin{equation*}}
\newcommand{\eeqn}{\end{equation*}}
\newcommand{\overlinern}{\begin{eqnarray*}}
\newcommand{\earrn}{\end{eqnarray*}}

\usepackage{comment}
\usepackage{float}

\begin{document}
\title{Controlling flexural waves in semi-infinite platonic crystals}
\label{firstpage}

\date{}
\author{\normalsize{{S.G. Haslinger$^1$, N.V. Movchan$^1$, A.B. Movchan$^1$, I.S. Jones$^2$ \& R.V. Craster$^3$}} \\
\footnotesize{$^1$ Department of Mathematical Sciences, University of Liverpool, Liverpool L69 7ZL, UK} \\
\footnotesize{$^2$ Mechanical Engineering and Materials Research Centre, Liverpool John Moores University,} \\ 
\footnotesize{Liverpool L3 3AF, UK }\\
\footnotesize{$^3$ Department of Mathematics, Imperial College London, London SW7 2AZ, UK} \\}

\maketitle

\begin{abstract}
\noindent
We address the problem of scattering and transmission of a plane flexural wave through a semi-infinite array of point scatterers/resonators, which take a variety of physically interesting forms. The mathematical model accounts for several classes of point defects, including mass-spring resonators attached to the top surface of the flexural plate and their limiting case of concentrated point masses. 
We also analyse the special case of resonators attached to opposite faces of the plate. 
The problem is reduced to a functional equation of the Wiener-Hopf type, whose kernel varies with the type of scatterer considered. 
A novel approach, which stems from the direct connection between the kernel function of the semi-infinite system and the quasi-periodic Green's functions for corresponding infinite systems, is used to identify special frequency regimes. We thereby demonstrate dynamically anisotropic wave effects in semi-infinite platonic crystals, with particular attention paid to designing systems to exhibit dynamic neutrality (perfect transmission) and localisation close to the structured interface.
\end{abstract}

\section{Introduction}
Since the 1980's, there has been substantial attention devoted to wave interaction with periodic structures leading to the recent surge of interest in designing metamaterials and micro-structured systems that are able to generate effects unattainable with natural media. 
These are artificially engineered super-lattice materials, designed with periodic arrays of sub-wavelength unit cells; their major concept is that their function is defined through structure. Many of the ideas and techniques originate in electromagnetism and optics but are now filtering into other systems such as the 
Kirchhoff-Love plate equations for flexural
waves. This analogue of photonic crystals, labelled as platonics by McPhedran {\it et al.} \cite{RCM_ABM_NVM}, features many of the typical anisotropic effects from photonics such as ultra-refraction, 
negative refraction and Dirac-like
cones, see  \cite{farhat10b}--\cite{mcp},
amongst others. 
Recently, structured plates have also been both modelled, and designed, to demonstrate the capability for cloaking applications \cite{farhat09}--\cite{miss2016}.

In this article, we consider a semi-infinite platonic crystal where, by patterning one half of an infinite Kirchhoff-Love plate with a semi-infinite rectangular array of point scatterers, the leading grating acts as an interface between the homogeneous and structured parts of the plate. Haslinger {\it et al.} \cite{Has2015} analysed the case of pinned points, and highlighted effects including dynamic neutrality in the vicinity of Dirac-like points on the dispersion surfaces for the corresponding infinite doubly periodic system, and interfacial localisation, by which  waves propagate along the interface. An interesting feature of the discrete Wiener-Hopf method of solution was the direct connection between the kernel function and the doubly quasi-periodic Green's function, zeros of which correspond to the aforementioned dispersion surfaces.

\begin{figure}[ht]
\begin{center}
\includegraphics[width=14cm]{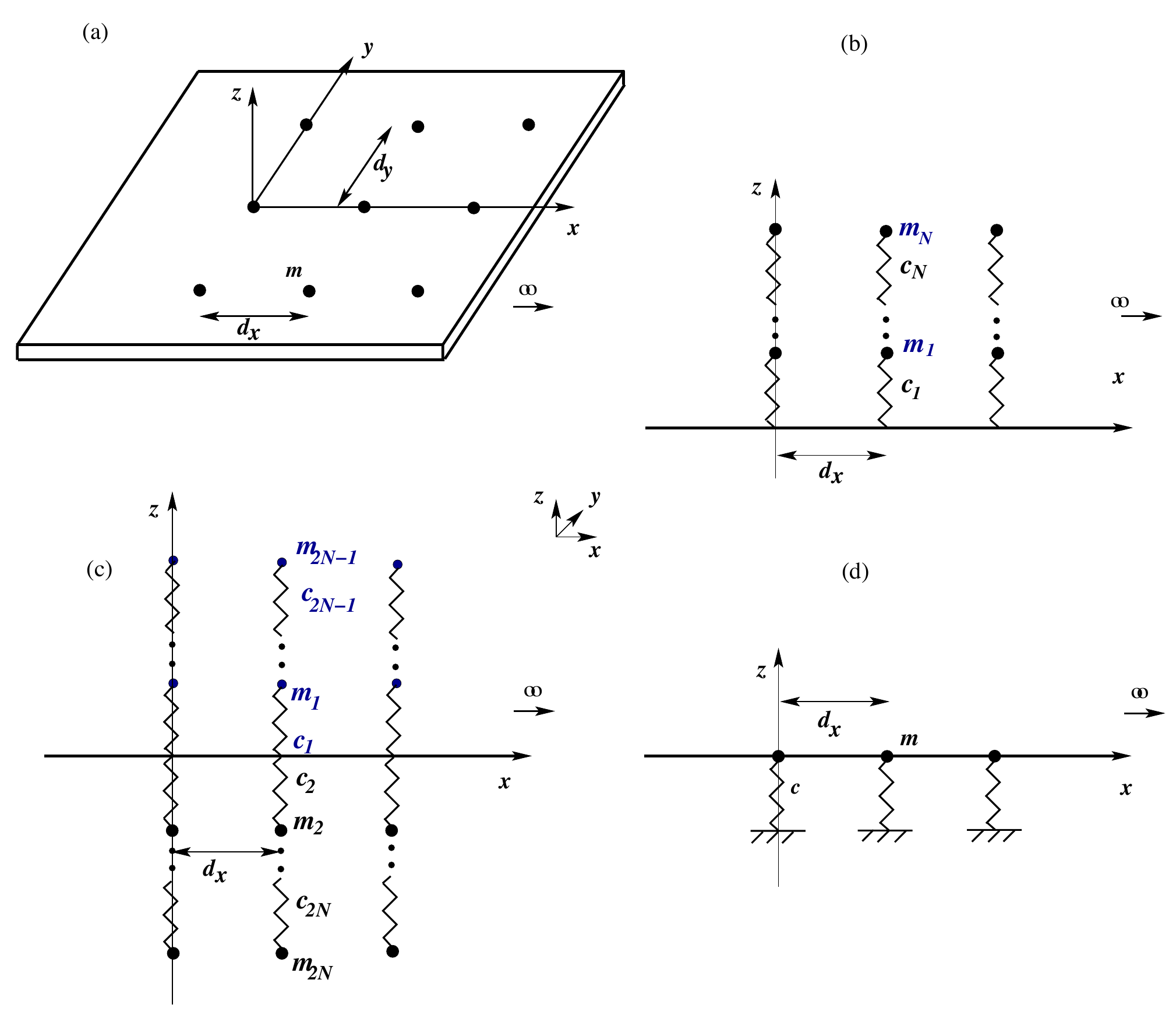}~~
\caption{\label{mass_line_res} 
Four cases for semi-infinite arrays of mass-spring resonators with periodicities $d_x, d_y$. (a) Case 1: semi-infinite array of point masses. (b) Case 2: multiple mass-spring resonators on the top surface of the plate characterised by masses $m_i$, and stiffnesses $c_i$. (c) Case 3: Multiple mass-spring resonators attached to both faces of the plate. (d) Case 4: Winkler-type foundation, the masses are embedded within the top surface of the plate.} 
\end{center}
\end{figure}

Here, we analyse four alternative physical settings for the point scatterers making up the semi-infinite periodic array, which we classify as one of two possible periodic systems; the two-dimensional ``half-plane'' with periodicity defined in both the $x$- and $y$- directions, as illustrated in figure~\ref{mass_line_res}(a), and the one-dimensional ``grating", with the periodic element confined to the $x$-axis, as illustrated in figures~\ref{mass_line_res}(b-d). All of the analysis presented in this article is for the two-dimensional periodicity, and is easily reduced to the special case of a single semi-line of scatterers for $x \ge 0$. 
\begin{itemize}
\item Case 1: point masses, characterised by mass $m$ 
\item Case 2: multiple point mass-spring resonators attached to the top surface of the plate, characterised by masses $m_i$, stiffnesses $c_i$; $i \in \mathbb{Z}^+$
\item Case 3: multiple mass-spring resonators attached to both faces of the plate 
\item Case 4: point masses with Winkler foundation (see Biot \cite{biot}), characterised by mass $m$, stiffness $c$.
\end{itemize}
It will be shown that, for certain frequency regimes, some of the cases are equivalent to one another. 

The replacement of the rigid pins with more physically interesting scatterers brings several new attributes to the model, most notably an assortment of propagation effects at low frequencies; in contrast, the case of pinned points possesses a complete band gap for low frequency vibrations up to a finite calculable value. The important limiting case of $c_1 \to \infty$ for case 2, $N =1$ (see figure~\ref{mass_line_res}b), or equivalently, $c \to 0$ for case 4 in figure~\ref{mass_line_res}(d), retrieves the periodic array of unsprung point masses. The infinite doubly periodic system of point masses has been discussed by Poulton {\it et al} \cite{Poul}, who provided dispersion band diagrams and explicit formulae and illustrations for defect and waveguide modes. 

Evans \& Porter \cite{Evans} considered one-dimensional periodic arrays of sprung point masses (case 4 in figure~\ref{mass_line_res}(d) for $-\infty < x < \infty$), including the limiting case of unsprung point masses.  The contributions by Xiao {\it et al.} \cite{Xiao} and Torrent {\it et al.} \cite{Torrent} discussed infinite doubly periodic arrays of point mass-spring resonators, as depicted in figure~\ref{mass_line_res}(b); the former for a rectangular array, and the latter for a honeycomb, graphene-like system. The authors provided dispersion relations and diagrams for the platonic crystals, and analysed the tuning of band-gaps and the association of Dirac points with the control of the propagation of flexural waves in thin plates. Examples using finite structures were also illustrated by both \cite{Xiao}, \cite{Torrent}.

In this article, we present the first analysis of semi-infinite arrays for the variety of point scatterers illustrated in figure~\ref{mass_line_res}. We demonstrate interfacial localisation, dynamic neutrality and negative refraction for the two-dimensional platonic crystals. 
The problem is formulated for the two-dimensional semi-infinite periodic array of scatterers, from which the special case of a semi-infinite line is easily recovered by replacing a quasi-periodic grating Green's function with the single source Green's function for the biharmonic operator. A discrete Wiener-Hopf method, incorporating the $z$-transform, is employed to derive a series of Wiener-Hopf equations for the various geometries. This discrete method is less common than its continuous counterpart, but it has been used by, amongst others, \cite{hills1}--\cite{tymis2}
for related problems, mainly in the context of the Helmholtz equation.

The characteristic feature of each of the resulting functional equations is the kernel which, for all of the cases featured here in figure~\ref{mass_line_res}, includes the doubly quasi-periodic Green's function, meaning that a thorough understanding of the Bloch-Floquet analysis is required. We express the kernel in a general form, and by identifying and studying special frequency regimes, we present the conditions required to predict and observe specific wave effects. This novel approach is used to design structured systems to control the propagation of the flexural waves, without evaluating the explicit Wiener-Hopf solutions, bypassing unnecessary computational challenges. We derive expressions to connect the geometries being analysed, including a condition for dynamic neutrality (perfect transmission) that occurs at the same frequency for the two-dimensional versions of both cases 3 and 4 shown in figures~\ref{mass_line_res}(c,d).

In conjunction with the Wiener-Hopf expressions for each of the cases considered, we also derive dispersion relations, and illustrate dispersion surfaces and band diagrams. Of particular importance are stop and pass band boundaries, standing wave frequencies (flat bands/low group velocity) and the neighbourhoods of Dirac-like points, which support dynamic neutrality effects. The concept of Dirac cone dispersion originates in topological insulators and has more recently been transferred into photonics (see for example \cite{zhang}--\cite{LL_JDJ_MS}). 
It is associated with adjacent bands, for which electrons obey the Schr\"{o}dinger equation, that meet at a single point called the Dirac point. 

Typically connected with hexagonal and triangular geometries in systems governed by Maxwell's equations, and most notably associated with the electronic transport properties of graphene (see, for example, Castro Neto {\it et al.} \cite{AHCN_FG_NMRP_KSN_AKG}), analogous Dirac and Dirac-like points have recently been displayed in phononic and platonic crystals (see for example \cite{Mei}--\cite{MJAS_RCM_MHM}, \cite{mcp}). 
The presence of Dirac cones is generally associated with the symmetries of the system through its geometry. When two perfect cones meet at a point, with linear dispersion, the cones are said to touch at a Dirac point. In the vicinity of a Dirac point, electrons propagate like waves in free space, unimpeded by the microstructure of the crystal. 

In platonic crystals, the analogous points generally possess a triple degeneracy, where the two Dirac-like cones are joined by another flat surface passing through what is known as a Dirac-like point. This is analogous to the terminology adopted by Mei {\it et al.} \cite{Mei} in photonics and phononics, where the existence of linear dispersions near the point ${\bf k} = {\bf 0}$ of the reciprocal lattice for the square array is the result of ``accidental" degeneracy of a doubly degenerate mode (the Dirac point, without the additional mode) and a single mode. Sometimes known as a ``perturbed" Dirac point, the accidental degeneracy does not arise purely from the lattice symmetry, as for a Dirac point, but from a perturbation of the physical parameters; in this setting, from the fourth order biharmonic operator. We identify Dirac-like points to illustrate neutrality, and ``Dirac bridges" (Colquitt {\it et al.} \cite{colq_2016}) to predict unidirectional wave propagation. We also use dispersion surfaces and the accompanying isofrequency contour diagrams to identify frequencies supporting negative refraction.

The paper is arranged as follows:
In section~\ref{form}, we formulate the problem for the two-dimensional rectangular array, using 
the discrete Wiener-Hopf technique; the special case 
of a semi-infinite grating is also identified.
We provide governing equations, and Wiener-Hopf equations for 
all cases illustrated in figure~\ref{mass_line_res}. In section~\ref{analysis},
we analyse these equations, highlighting the conditions required for 
frequency regimes to support reflection, transmission and dynamic neutrality, which we illustrate with examples.  We also demonstrate   Rayleigh-Bloch-like waves
for the semi-infinite line of scatterers. 
In
section~\ref{results},
we present special examples of waveguide transmission, whereby the structured system is designed to specifically exhibit negative refraction and interfacial localisation effects. Concluding remarks are drawn together in
section \ref{conc}.

\section{Formulation}
\label{form}
\begin{figure}[h]
\begin{center}
\includegraphics[width=7.6cm]{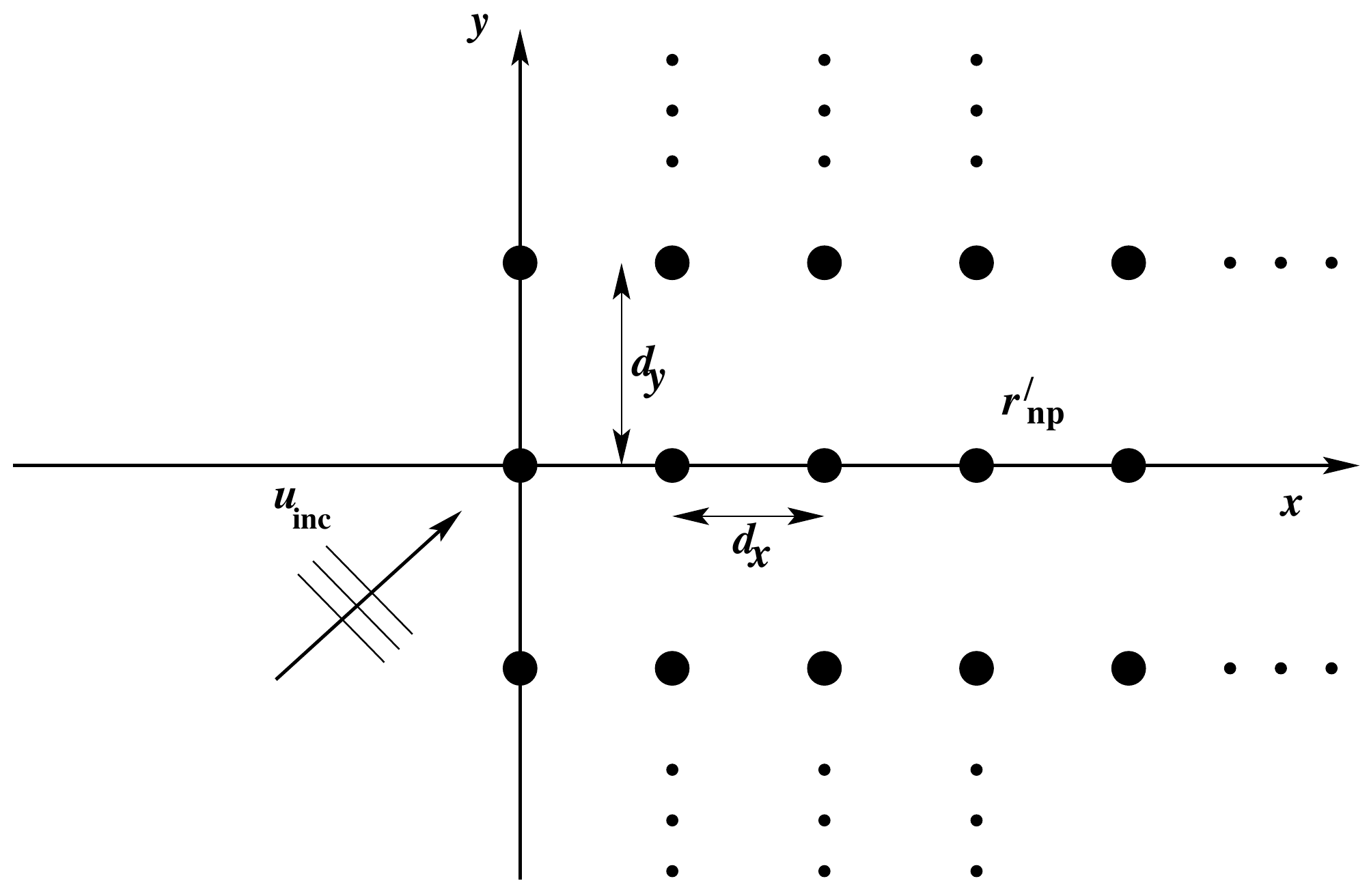}~
\includegraphics[height=3.2cm]{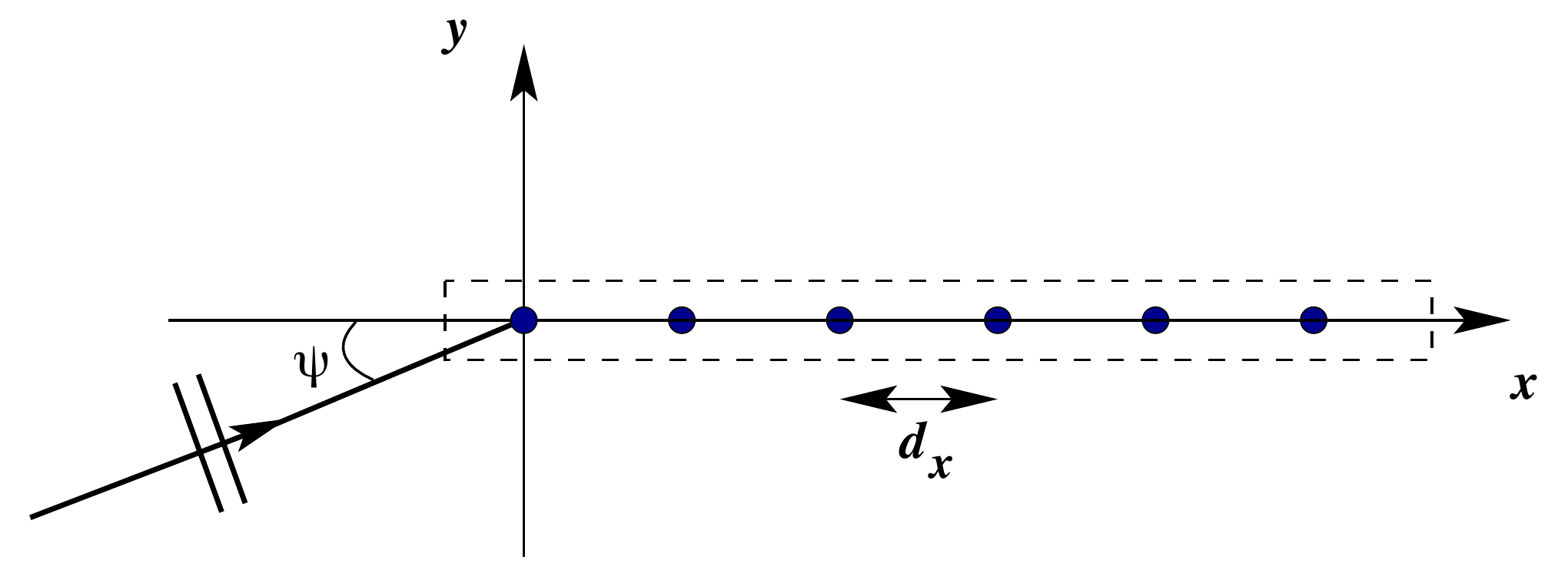}
\put(-150,50) {{\small(a)}}
\put(-45,50) {{\small(b)}}
\caption{\label{semi_masses} 
(a) A semi-infinite array of gratings of point scatterers, whose positions are denoted by ${\bf r}'_{np}$, with horizontal and vertical spacings $d_x$ and $d_y$ respectively. (b) A semi-infinite line of point scatterers for a plane wave incident at angle $\psi$.}
\end{center}
\end{figure}

A thin Kirchhoff-Love plate comprises a two-dimensional semi-infinite array of point scatterers defined by position vectors ${\bf r}'_{np} = (nd_x, pd_y)$, where $d_x, d_y$ are the spacings in the $x$- and $y$-directions respectively, and $n, p$ are integers, as illustrated in figure~\ref{semi_masses}(a). 
It is natural to consider the system as a semi-infinite array of gratings aligned parallel to the $y$-axis. By replacing each of these gratings with a single point scatterer lying on the $x$-axis, we recover the one-dimensional case of a single semi-infinite grating, as illustrated in figure~\ref{semi_masses}(b). The plate is subjected to a forcing in the form of a plane wave, incident at an angle $\psi$ to the $x$-axis.

We assume time-harmonic vibrations of the Kirchhoff-Love plate, and define equations for the amplitude of the total out-of-plane displacement field $u({\bf r})$, with ${\bf r} = (r_x, r_y)$, which can be expressed as the sum of the incident and scattered fields: 
\begin{equation}
u({\bf r}) = u_{\scriptsize{\mbox{inc}}}({\bf r}) + u_{\scriptsize{\mbox{scatt}}}({\bf r}).
\label{utot}
\end{equation} 
We express the general governing equation for $u({\bf r})$ in the form:
\begin{equation}
 \Delta^2 u({\bf r}) - \frac{\rho h  \omega^2}{D} u({\bf r}) = \Phi(\omega, m, c) \sum_{n=0}^\infty \sum_{p=-\infty}^\infty u({\bf r}'_{np}) \delta({\bf r} - {\bf r}'_{np}), \,\,\,\,\,\, {\bf r}'_{np}= (nd_x, pd_y), \,n \in \mathbb{Z}^+, \, p \in \mathbb{Z}.
 \label{utotgo}
\end{equation}
Here, $\Phi (\omega, m, c)$ is a function of radial frequency $\omega$, and the physical parameters of mass $m$ and stiffness $c$ that define the various mass-spring resonator models shown in figure~\ref{mass_line_res}. The functional forms of the various $\Phi (\omega, m, c)$ are provided later in section~\ref{funformsphi}. The characteristic physical parameters for the plate are density per unit volume $\rho$, thickness $h$ and flexural rigidity $D$ (involving Young's modulus $E$ and the Poisson ratio $\nu$), and we also adopt the use of the spectral parameter $\beta$, which has the dimension of a wavenumber:
\begin{equation}
D = \frac{Eh^3}{12(1 - \nu^2)}; \,\,\,\,\,\, \beta^2 = \omega \sqrt{\frac{\rho h}{D}}; \,\,\,\, \omega^2 =  \frac{\beta^4D}{\rho h}.
\label{physp}
\end{equation}
Note that the Kirchhoff-Love model incorporates the fourth order biharmonic operator, and gives an excellent approximation of the full linear elasticity equations for a sufficiently small value of the ratio $h/\lambda$, where $\lambda$ denotes the wavelength of the flexural
vibrations of the plate \cite{fahy}:
\begin{equation}
\lambda = \frac{2 \pi}{\beta} = 2\pi \left(\frac{\rho h \omega^2}{D}\right)^{-1/4}, \,\,\,\,\,\,\,\,\, \frac{h}{\lambda} \ll 1/6.
\end{equation}

\subsection{Governing equations and reduction to a functional equation}
Assuming isotropic scattering, we express the scattered field in the form of a sum of biharmonic Green's functions:
\begin{equation}
u_{\scriptsize{\mbox{scatt}}}({\bf r}) = \Phi(\omega, m, c) \sum_{n=0}^\infty \sum_{p=-\infty}^\infty u({\bf r}'_{np}) G(\beta |{\bf r} - {\bf r}'_{np}|),
\label{scattgo}
\end{equation}
where $G$ is the point source Green's function satisfying the equation:
\begin{equation}
\Delta^2 G(\beta, | {\bf r} - {\bf r}' |) - \beta^4 G(\beta, | {\bf r} - {\bf r}' |) = \delta ( {\bf r} - {\bf r}').
\end{equation} 
Note that for the one-dimensional case of a semi-infinite line of scatterers ${\bf r}'_{n0}$ placed on the $x$-axis, the sum over $p$ in~(\ref{scattgo}) is absent. All derivations for the two-dimensional array given below are applicable to the special case of the semi-infinite grating, with appropriate adjustments to the sums and Green's functions. 
Referring to equations~(\ref{utot})-(\ref{utotgo}), we may express the total field $u({\bf r})$ as
\begin{equation}
u({\bf r}) = u_{\scriptsize{\mbox{inc}}}({\bf r}) + \Phi(\omega, m, c) \sum_{n=0}^\infty \sum_{p=-\infty}^\infty u({\bf r}'_{np}) G(\beta |{\bf r} - {\bf r}'_{np}|).
\end{equation}
In particular, at ${\bf r} = {\bf r}'_{st}$, $s \in \mathbb{Z}^+,  t \in \mathbb{Z}$, we have the linear algebraic system
\begin{equation}
u({\bf r}'_{st}) = u_{\scriptsize{\mbox{inc}}}({\bf r}'_{st}) + \Phi(\omega, m, c) \sum_{n=0}^\infty \sum_{p=-\infty}^\infty u({\bf r}'_{np}) G(\beta |{\bf r}'_{st} - {\bf r}'_{np}|), \,\,\, s \in \mathbb{Z}^+, \,\,\, t \in \mathbb{Z}.
\label{hpalg}
\end{equation}
Recalling that we consider an incident plane wave, we define $u_{\scriptsize{\mbox{inc}}}$ in the form:
\begin{equation}
u_{\scriptsize{\mbox{inc}}} = \exp \{i (k_x r_x + k_y r_y)\}, 
\end{equation}
where ${\bf k} = (k_x, k_y) = (\beta \cos{\psi}, \beta \sin{\psi})$ is the wave vector. 
Since the scatterers are infinitely periodic in the $y$-direction, we impose Bloch-Floquet conditions for $u_{\scriptsize{\mbox{scatt}}}$ in the $y$-direction. Hence
\begin{equation}
u(r_x, r_y + q d_y) = e^{i k_y q d_y} u ({\bf r}) \,\,\, \mbox{and} \,\,\,u({\bf r}'_{np}) = u({\bf r}'_{n0})e^{i k_y p d_y}. 
\end{equation}
Thus, recalling the RHS of  equation~(\ref{hpalg}), we have
\begin{equation*}
\sum_{p=-\infty}^\infty u({\bf r}'_{np}) G(\beta |{\bf r}'_{st} - {\bf r}'_{np}|)  = u({\bf r}'_{n0}) \sum_{p=-\infty}^\infty e^{i k_y p d_y} G(\beta |{\bf r}'_{st} - {\bf r}'_{np}|) 
\end{equation*}
\begin{equation}
 = u({\bf r}'_{n0}) e^{i k_y t d_y} \sum_{p=-\infty}^\infty  e^{i k_y (p-t) d_y} \, G\left(\beta \sqrt{(s-n)^2{d_x}^2+(t-p)^2{d_y}^2}\right). 
\end{equation}
Denoting $p-t = \alpha$, this simplifies to 
\begin{equation}
\sum_{p=-\infty}^\infty u({\bf r}'_{np}) G(\beta |{\bf r}'_{st} - {\bf r}'_{np}|)  = u({\bf r}'_{n0}) e^{i k_y t d_y}\sum_{\alpha=-\infty}^\infty  e^{i k_y \alpha d_y} \, G\left(\beta \sqrt{(s-n)^2{d_x}^2+\alpha^2{d_y}^2}\right),
\end{equation}
where the sum on the right is precisely the quasi-periodic grating Green's function; we shall use the notation
\begin{equation}
G^q(\beta, |s - n|; k_y, d_x, d_y) = \sum_{\alpha=-\infty}^\infty  e^{i k_y \alpha d_y}G\left(\beta \sqrt{(s-n)^2{d_x}^2+\alpha^2{d_y}^2}\right),
\label{qpgf}
\end{equation}
when substituting back into equation~(\ref{hpalg}):
\begin{equation}
u({\bf r}'_{st}) = e^{i {\bf k} \cdot {\bf r}'_{st}} + \Phi(\omega, m, c)   \sum_{n=0}^\infty  u({\bf r}'_{n0}) e^{i k_y t d_y}\, G^q(\beta, |s - n|; k_y, d_x, d_y).
\end{equation}
Here, we use 
$$
u({\bf r}'_{st}) = e^{i k_y t d_y} u ({\bf r}'_{s0}).
$$
Thus, the algebraic system becomes
\begin{equation}
u({\bf r}'_{s0}) = e^{i k_x s d_x} + \Phi(\omega, m, c) \sum_{n=0}^\infty u({\bf r}'_{n0}) G^q(\beta, |s - n|; k_y, d_x, d_y),
\label{gov2d}
\end{equation}
where the integer $s$ denotes the $x$-position of a grating of scatterers parallel to the $y$-axis and centred on the $x$-axis, and the incident field is in the form $\exp\{i k_x s d_x\}$. Equivalently, we may write
\begin{equation}
u_s = f_s + \Phi(\omega, m, c) \sum_{n=0}^\infty u_n G^q(\beta, |s - n|; k_y, d_x, d_y),
\label{gov2d2}
\end{equation}
where we replace $u({\bf r}'_{s0})$, $u({\bf r}'_{n0})$ with $u_s, u_n$ for ease of notation, and $f_s$ represents the forcing term.

The semi-infinite sum indicates that the discrete Wiener-Hopf method is suitable, see Noble \cite{noble58a}, where the application to continuum discrete problems \cite{feld55a}, such as gratings, is presented as an exercise (4.10, p.173-4) in \cite{noble58a}. After employing the $z$-transform, we 
obtain
\begin{equation}
\sum_{s=-\infty}^{\infty} u_s z^s = 
\sum_{s=-\infty}^{\infty}\left(e^{ik_x d_x} \, z\right)^s + \Phi(\omega, m, c) \sum_{s=-\infty}^{\infty} \sum_{n=0}^{\infty} z^s  u_n G^q(\beta, |s - n|; k_y, d_x, d_y).
\end{equation}
Letting the index $s = n + \zeta$, we derive a functional equation of the Wiener-Hopf type \cite{noble58a}:
\begin{equation}
\hat{U}_- (z)  =  \hat{F} (z) + \hat{U}_+ (z) \left(\Phi(\omega, m, c) \hat{\cal G}(z) - 1 \right),
\label{whgen}
\end{equation}
where
\begin{equation}
\hat{F} (z) =  \sum_{s=-\infty}^{\infty} \left(e^{i k_x d_x} z\right)^s; \,\,\, \hat{U}_- (z)  =  \sum_{s=-\infty}^{-1} u({\bf r}'_{s0}) z^s ; \,\,\, \hat{U}_+ (z) =  \sum_{s=0}^{\infty} u({\bf r}'_{s0}) z^s,
\label{whforms_hp}
\end{equation}
$z = e^{i \theta}$ and we have introduced the notation $\hat{\cal G} (z)$ to represent  the function
\begin{equation}
\hat{\cal G} (z) = \sum_{\zeta=-\infty}^{\infty} G^q(\beta, |\zeta|; k_y, d_x, d_y) z^{\zeta},
\label{kernGr}
\end{equation}
where $G^q$ is the quasi-periodic grating Green's function given by equation~(\ref{qpgf}). 
We note that for $z = \exp{(i k_x d_x)}$, i.e. $\theta = k_x d_x$, we recover the 
doubly quasi-periodic Green's function:
\begin{equation}
\hat{\cal G} (z) = \sum_{\zeta=-\infty}^{\infty} \, \sum_{\alpha = -\infty}^{\infty} e^{i \alpha  k_y d_y} e^{i  \zeta k_x d_x} \, G\, \left(\beta \sqrt{\zeta^2{d_x}^2+\alpha^2{d_y}^2}\right).
\label{dpgreenf}
\end{equation}
We also note that for the case of a semi-infinite line of point scatterers along $x \ge 0$, we would have the same Wiener-Hopf functional equation as~(\ref{whgen}), but replace $\hat{\cal G}$ with $G^q$~(\ref{qpgf}).

\subsection{Governing equations for various point scatterers}
\label{funformsphi}
Equations~(\ref{gov2d})-(\ref{kernGr}) are the fundamental general equations for all four of the cases depicted in figure~\ref{mass_line_res}. The general kernel function
\begin{equation}
{\cal K} (z) = \Phi(\omega, m, c) \hat{\cal G} (z) - 1 
\label{genkern}
\end{equation}
differs from that of the simpler {\it pinned} semi-infinite platonic crystal analysed in \cite{Has2015} because of the change in boundary condition for the point scatterers. Whereas the rigid pins impose zero flexural displacement at ${\bf r}'_{np}$, the nonzero condition for the scatterers considered here introduce additional terms in~(\ref{genkern}). The function $\Phi_i(\omega, m, c)$, for $i = 1-4$, determines the characteristic features specific to each model, and also those common to the different cases. We now present the expressions for the various $\Phi_i(\omega, m, c)$, which we go on to explain and derive, where necessary, for each case in turn. 
\begin{itemize}
\label{items}
\item Case 1: Point masses: \,\,\,  $\Phi_1 (\omega, m, c) = m \omega^2/D$
\item Case 2: Multiple mass-spring resonators on the top surface of the plate:
\begin{itemize}
\item $N=1$: \,\,\, $\Phi_2^{N=1} \, (\omega, m, c) = \frac{cm\omega^2}{D(c - m\omega^2)}$
\item  $N=2$: \,\,\, $\Phi_2^{N=2} \, (\omega, m_i, c_i) = \frac{c_1}{D} \,\frac{c_2^2 - (c_2-m_1\omega^2)(c_2-m_2\omega^2)}{(c_1+c_2-m_1\omega^2)(c_2 - m_2\omega^2) - c_2^2}$ 
\end{itemize}
\item Case 3: Multiple mass-spring resonators on both faces of plate: 

\begin{center}
$\Phi_3^{N=1} \, (\omega, m_i, c_i) = \frac{\omega^2}{D} \left(\frac{m_1c_1}{c_1 - m_1 \omega^2} + \frac{m_2 c_2 }{c_2 - m_2 \omega^2} \right)$
\end{center}
\item Case 4: Winkler foundation point masses: \,\,\, $\Phi_4 (\omega, m, c) = (m\omega^2-c)/D$
\end{itemize}

\subsubsection{Case 1: Point masses} 
The simplest type of point scatterer is the rigid pin, defined as the limiting case of the radius of a clamped hole tending to zero. There is a large body of literature covering various problems incorporating this boundary condition. A selection of relevant papers include Movchan {\it et al.} \cite{ABM_NVM_RCM}, Evans \& Porter \cite{Evans}, \cite{Evans_2}, Antonakakis \& Craster \cite{anton1}, \cite{anton2}, Haslinger {\it et al.} \cite{SGH_NVM_ABM_RCM2}. A logical extension is to replace the pins with concentrated point masses of mass $m$, introducing an additional inertial term, and hence non-zero displacement at the point scatterer. 
We include a schematic diagram in figures~\ref{mass_line_res}(a) and~\ref{semi_masses}(b). 
The governing equation for a semi-infinite half-plane of point masses, following equation~(\ref{utotgo}), is
\begin{equation}
 \Delta^2 u({\bf r}) - \frac{\rho h  \omega^2}{D} u({\bf r}) =  \frac{m\omega^2}{D} \sum_{n=0}^\infty \sum_{p=-\infty}^\infty u({\bf r}'_{np}) \delta({\bf r} - {\bf r}'_{np}).
 \end{equation}
The total flexural displacement field $u_j = u({\bf r}'_{j0})$, of the form~(\ref{gov2d2}), is given by
\begin{equation}
u_j = f_j + \, \frac{m\omega^2}{D} \, \sum_{n=0}^{\infty} u_n G^q(\beta, |j -n|; k_y, d_x, d_y),
\label{case1_pm}
\end{equation}
and the corresponding Wiener-Hopf-type functional equation is
\begin{equation}
\hat{U}_- (z) =  \hat{F} (z) + \hat{U}_+ (z) \left( \Phi_1(\omega, m) \, \hat{\cal G}(\beta; z) - 1 \right),     \,\,\, \Phi_1(\omega, m) = \frac{m \omega^2}{D}.
\label{wh_pmass}
\end{equation}

\subsubsection{Case 2: Multiple mass-spring resonators}
\label{sec:msres}
We attach mass-spring resonators at each point ${\bf r}'_{np} = (nd_x, pd_y), n \ge 0$. Each resonator consists of $N$ point masses attached to $N$ springs. For the most general case, the finite number of masses $m_1, m_2,...,m_N$ are connected by springs of stiffness $c_1, c_2,...,c_N$ (see figure~\ref{mass_line_res}(b)).  For the sake of simplicity, we derive the governing equations, and their reduction to functional Wiener-Hopf type equations, for the cases $N = 1$ and $N=2$, but the procedure for arbitrary $N$ is a simple extension.

\subsubsection*{Single mass-spring resonator, N=1}
We assume a semi-infinite rectangular array of simple resonators consisting of point masses attached with springs to the plate at points shown in figure~\ref{semi_masses}(a), with the parameters illustrated in figure~\ref{mass_line_res}(b). We assume uniform mass $m$ and uniform stiffness $c$, and negligible effect of gravity. 
We derive the equation for $\Phi_2^{N=1} \, (\omega, m, c)$, and the accompanying governing and discrete Wiener-Hopf expressions, by applying Newton's 2nd law and Hooke's law for an arbitrary scatterer placed at ${\bf r}'_{n0} = (n d_x, 0)$. We use the fact that the quasi-periodic Green's function~(\ref{qpgf}) in equations~(\ref{gov2d})-(\ref{kernGr}) accounts for all scatterers in a grating parallel to the $y$-axis, centred at ${\bf r}'_{n0}$.

We denote the flexural displacement of the plate at ${\bf r}'_{n0}$ by $u({\bf r}'_{n0}) = u_n$. The transverse displacement of the mass $m$ is denoted by $v({\bf r}'_{n0}) = v_n$, with the forces applied to the plate by the spring, of stiffness $c$, given by $A_u({\bf r}'_{n0})$, and to the connected mass by the spring, as $A_v({\bf r}'_{n0})$.
The equation of motion for the sprung mass is written in the form: 
\begin{equation}
\label{Vn}
m \omega^2 v_n = -A_v({\bf r}'_{n0}) = c(v_n - u_n).
\end{equation} 
Transverse displacements $v_n$ and $u_n$ are evaluated with respect to ${\bf r}'_{n0}$, but for the sake of simplicity, we adopt the abbreviated notation with the subscript $n$. We write $v_n$ in terms of $u_n$ using~(\ref{Vn}), and 
referring to equations~(\ref{utotgo}),(\ref{gov2d2}) and recalling that the flexural rigidity of the plate is $D$, we have 
\begin{equation}
\label{wh_springs}
u_j =  f_j + \frac{cm\omega^2}{D(c - m\omega^2)} \sum_{n=0}^{\infty} u_n G^q(\beta, |j -n|; k_y, d_x, d_y),
\end{equation}
with
\begin{equation}
\Phi_2^{N=1} \, (\omega, m, c) = \frac{cm\omega^2}{D(c - m\omega^2)}.
\end{equation}
Note that in the limit as $c \to \infty$, equation~(\ref{wh_springs}) tends towards the equation for unsprung mass-loaded points~(\ref{case1_pm}), and $v_n = u_n$ from~(\ref{Vn}); the flexural vibrations of the plate and masses are identical for infinite stiffness, which is physically consistent with infinitely stiff springs. As $m \to \infty$, the coefficient multiplying the sum tends to $-c/D$, and this may be interpreted physically as the plate being attached to a rigid foundation with springs of stiffness $c$.

The discrete Wiener-Hopf functional equation is obtained by substituting the expression for $\Phi_2^{N=1} \, (\omega, m, c)$ into equation~(\ref{whgen}):
\begin{equation}
\hat{U}_- (z) =  \hat{F} (z) + \hat{U}_+ (z) \left( \frac{cm\omega^2}{D(c - m\omega^2)} \hat{\cal G}(\beta; z) - 1 \right).
\label{whcase1}
\end{equation} 
Recalling equation~(\ref{genkern}), the kernel ${\cal K}_2^{N=1} (z) $ for case 2, $N = 1$ is given by
\begin{equation}
{\cal K}_2^{N=1} (z) \, = \frac{cm\omega^2}{D(c - m\omega^2)} \hat{\cal G}(\beta; z) - 1 = \frac{cm\omega^2}{D(c - m\omega^2)} {\cal K}_{\mbox{\tiny pins}} (z) -1,
\label{kern1}
\end{equation}
where we introduce the notation ${\cal K}_{\mbox{\tiny pins}} (z)$ as the kernel for the case of rigid pins, see \cite{Has2015}. 

Observing that a similar expression follows for the limit case of point masses~(\ref{wh_pmass}), this connection with the pinned case enables us to employ a similar kernel factorization. First, we rewrite~(\ref{kern1}) in terms of the dimensionless parameters:
\begin{equation}
\tilde{m} = \frac{m}{\rho h {\eta}^2},  \,\,\, \tilde{c} = \frac{c {\eta}^2}{D},     \,\,\, \tilde{\beta} = \beta \eta, 
\label{dimp}
\end{equation}
where we introduce a length scale determined by the periodicity of the system $\eta = \mbox{min} \,(d_x, d_y)$, which for the sake of simplicity is taken to be $d_x$ throughout this article. 
We also introduce non-dimensional versions of Green's functions, and their arguments, which possess the dimension of $L^2$ owing to the factor $1/\beta^2$:
\begin{equation}
\tilde{\hat{{\cal G}}} (\tilde{\beta}; \tilde{z}) = \beta^2 \hat{\cal G}, \,\,\,\, \tilde{z} = z/\eta.
\end{equation}
Hence, 
\begin{equation}
\tilde{{\cal K}}_2^{N=1} (\tilde{\beta}; \tilde{z}) = \tilde{m} {\tilde{\beta}}^2 \left(\frac{1}{1 - \frac{\tilde{m}{\tilde{\beta}}^4}{\tilde{c}}} \right) \tilde{\hat{{\cal G}}} (\tilde{\beta}; \tilde{z})  -1.
\label{kcase1}
\end{equation} 
We express~(\ref{kcase1}) as 
\begin{equation}
\tilde{{\cal K}}_2^{N=1} (\tilde{\beta}; \tilde{z}) = \tilde{{\cal K}}_2^+ (\tilde{\beta}; \tilde{z}) \tilde{{\cal K}}_2^- (\tilde{\beta}; \tilde{z}),
\label{factkerngen}
\end{equation}
noting that the factorization obviously also applies to the dimensional form of the kernel.
Explicitly, we have
\begin{equation}
\tilde{{\cal K}}_2^+ (\tilde{\beta}; \tilde{z}) =  \tilde{\Phi}_2 \, (\tilde{\beta}, \tilde{m}, \tilde{c}) \,\, \tilde{\hat{{\cal G}}}_+(\tilde{\beta}; \tilde{z}); \,\,\,\,\, \tilde{{\cal K}}_2^- (\tilde{\beta}; \tilde{z}) =   \tilde{\hat{{\cal G}}}_-(\tilde{\beta}; \tilde{z}) - \frac{1}{\tilde{\Phi}_2 (\tilde{\beta}, \tilde{m}, \tilde{c}) \, \tilde{\hat{{\cal G}}}_+(\tilde{\beta}; \tilde{z})} ,
\label{factkern}
\end{equation}
where the reciprocal of the $+$ function $ \tilde{\hat{{\cal G}}}_+(\tilde{\beta}; \tilde{z})$ is a $-$ function.

\subsubsection*{Multiple mass-spring resonators, N=2}

As for the case $N = 1$, we derive the equation for $\Phi_2^{N=2} \, (\omega, m, c)$ for an arbitrary scatterer placed at ${\bf r}'_{n0} = (n d_x, 0)$, using Newton's 2nd law and Hooke's law.
In general, the transverse displacement of each mass $m_i$ is $v_n^{(i)}$, $i \in [1, N]$. For $N=2$, the equations of motion are given by 
\begin{eqnarray}
\label{Vn1Vn2}
(c_1+c_2 - m_1 \omega^2) v_n^{(1)} & =  & c_1 u_n + c_2 v_n^{(2)} \nonumber \\
(c_2 - m_2 \omega^2) v_n^{(2)} & = & c_2 v_n^{(1)}.
\end{eqnarray}
The normalised force acting on the plate at ${\bf r}'_{n0}$, in terms of the out-of-plane displacement, is given by: 
\begin{equation}
A_u({\bf r}'_{n0}) = \frac{c_1}{D} (v_n^{(1)} - u_n),
\end{equation}
where the reciprocal of the plate's flexural rigidity is the normalisation factor.
Thus, referring to equation~(\ref{wh_springs}) we may write the total flexural displacement amplitude at $u({\bf r}'_{j0}) = u_j$  as
\begin{equation}
\label{utot_res}
u_j = f_j + \frac{c_1}{D} \sum_{n=0}^{\infty} (v_n^{(1)} - u_n) G^q(\beta, |j -n|; k_y, d_x, d_y). 
\end{equation}
We eliminate $v_n^{(2)}$ from~(\ref{Vn1Vn2}) and derive the expression for $v_n^{(1)}$ in terms of $u_n$ only: 
$$
v_n^{(1)}\left(c_1 + c_2 - m_1\omega^2 - \frac{c_2^2}{c_2-m_2\omega^2} \right) = c_1u_n.
$$
Hence, we rewrite equation~(\ref{utot_res}) in the form,
\begin{equation}
u_j = f_j + \Phi_2^{N=2} \, (\omega, m_i, c_i) \sum_{n=0}^{\infty} u_n G^q(\beta, |j -n|; k_y, d_x, d_y),
\label{wh_mspm}
\end{equation}
such that
\begin{equation}
\Phi_2^{N=2} \, (\omega, m_i, c_i) = \frac{c_1}{D} \,\frac{c_2^2 - (c_2-m_1\omega^2)(c_2-m_2\omega^2)}{(c_1+c_2-m_1\omega^2)(c_2 - m_2\omega^2) - c_2^2}.
\label{phifun}
\end{equation}
As in previous cases, we employ the $z$-transform to obtain the discrete Wiener-Hopf equation:
\begin{equation}
\hat{U}_- (z) =  \hat{F} (z) + \hat{U}_+ (z) \, \left(  \Phi_2^{N=2} \, (\omega, m_i, c_i)  \hat{\cal G}(\beta; z) - 1 \right) .
\label{whcase3}
\end{equation}
This Wiener-Hopf equation resembles that of~(\ref{whcase1}) and in the limit as $c_2 \to 0$, we recover precisely that equation, and similarly~(\ref{wh_springs}) from~(\ref{phifun}).

\subsubsection{Case 3: Multiple mass-spring resonators attached to both faces of plate}
We now consider an extension of section~\ref{sec:msres} by attaching mass-spring resonators on opposite faces of the plate at the same point of the array depicted in figure~\ref{semi_masses}. This system is illustrated for the case of $2N$ mass-spring resonators in figure~\ref{mass_line_res}(c). For introducing the model, we analyse the simplest case here; two masses $m_1, m_2$ with associated spring stiffnesses $c_1, c_2$, with the index being odd for resonators attached to the top surface, and even for the bottom surface, as illustrated in figure~\ref{mass_line_res}(c). The derivations are similar to the previous sections with the flexural displacement at an arbitrary defect point ${\bf r}'_{n0} = (nd_x, 0), n \ge 0$ given by $u_n$ and the transverse displacements of the masses $m_i, i= 1,2$ given by $v_n^{(i)}$. The equation of motion of the resonator mass at a single array point is given by
\begin{equation}
\label{vqn}
m_i\omega^2v_n^{(i)} = c_i (v_n^{(i)} - u_n), \,\,\,\, i = 1, 2,
\end{equation}
where we have used Hooke's law for the right-hand side. 
Recalling the general expression for the total flexural displacement field of the plate at the point ${\bf r}'_{j0} = (jd_x, 0)$, we write
\begin{equation}
\label{ubsp}
u_j = f_j + \frac{1}{D} \sum_{n=0}^{\infty} (A_n^{(1)} +A_n^{(2)}) G^q(\beta, |j - n|; k_y, d_x, d_y),
\end{equation}
where the forces $A_n^{(i)}$ are given by
\begin{equation}
A_n^{(i)} = c_i (v_n^{(i)} - u_n), \,\,\,\,\, i = 1, 2.
\end{equation}
Similar to case 2, $N = 2$, we derive the governing equation in the form
\begin{equation}
u_j = f_j + \Phi_3(\omega, m_i, c_i) \sum_{n=0}^{\infty} u_n G^q(\beta, |j - n|; k_y, d_x, d_y),
\label{case5_gov}
\end{equation}
where
\begin{equation}
\Phi_3(\omega, m_i, c_i)  = \frac{\omega^2}{D} \left(\frac{m_1c_1}{c_1 - m_1 \omega^2} + \frac{m_2 c_2 }{c_2 - m_2 \omega^2} \right).
\end{equation}
Employing the $z$-transform in the standard way, the accompanying Wiener-Hopf representation is
\begin{equation}
\hat{U}_- (z) = \hat{F} (z) +  \hat{U}_+ (z) \left(\Phi_3(\omega, m_i, c_i) \, \hat{\cal G}(\beta; z) - 1  \right).
\label{whcase4}
\end{equation}

\subsubsection{Case 4: point masses with Winkler-type foundation}
\label{winkd}

 An alternative model for adding mass-spring resonators is shown in figure~\ref{mass_line_res}(d),  
where point masses are embedded within the plate, and additional springs, attached to a fixed foundation, are added below. Referring to equation~(\ref{Vn}) for case 2, $N = 1$, we obtain a similar equation, except that here the displacement at the point of attachment to the fixed foundation $v_n$, is zero. 
Hence,
\begin{equation}
m \omega^2 u_n = c u_n,
\end{equation}
and the solution for the total flexural amplitude at ${\bf r}'_{j0} = (jd_x, 0)$ is
\begin{equation}
\label{epsprings}
u_j = f_j + \, \frac{m\omega^2-c}{D} \, \sum_{n=0}^{\infty} u_n G^q(\beta, |j - n|; k_y, d_x, d_y).
\end{equation}
We note that with this model, taking the limit as $c \to 0$ recovers the case of concentrated point masses (case 1), in contrast to the model for case 2, $N = 1$, where $c \to \infty$ retrieved the limiting case of point masses. 
Similarly, the Wiener-Hopf equation is easily deduced from the general equation~(\ref{whgen}):
\begin{equation}
\hat{U}_- (z)  =  \hat{F} (z) + \hat{U}_+ (z) \left( \frac{m\omega^2-c}{D} \, \hat{\cal G}(\beta; z) - 1 \right),
\label{whcase2}
\end{equation}
with the kernel function ${\cal K}_4(z)$ defined by
\begin{equation}
{\cal K}_4(z) = \frac{m\omega^2-c}{D} \, \hat{\cal G}(\beta; z) - 1, \,\,\,\,\,\, \mbox{or} \,\,\,\,\,\, \tilde{{\cal K}}_4(\tilde{z})
\left(\tilde{m}\tilde{\beta}^2  - \frac{\tilde{c}}{\tilde{\beta}^2} \right)
{\tilde{\hat{\cal G}}} (\tilde{\beta}; \tilde{z}) -1.
\label{wink_dless}
\end{equation}

\section{Analysis of kernel functions: reflection, transmission and dynamic neutrality}
\label{analysis}
We identify three important frequency regimes using the kernel equation~(\ref{genkern}): reflection, transmission and dynamic neutrality. Special cases of waveguide transmission including negative refraction and interfacial localisation are illustrated in the subsequent section~\ref{results}. 
The five Wiener-Hopf expressions~(\ref{wh_pmass}), (\ref{whcase1}), (\ref{whcase3}), (\ref{whcase4}) and (\ref{whcase2}) are characterized by their respective kernels, which all take the general form
\begin{equation}
{\cal K} (z) = \Phi(\omega, m, c) \hat{\cal G}(\beta; z) - 1. 
\label{kern_gen_3}
\end{equation}
An analysis of these functions gives us insight into the behaviour of the possible solutions. 

There are three natural limiting regimes to consider for a kernel function with this structure - when it is either very large or very small, and when $\Phi \, \hat{\cal G}$ tends to 0, i.e. ${\cal K} \to -1$. Referring to the general Wiener-Hopf equation~(\ref{whgen}), the first two cases infer that $\hat{U}_+$ is respectively very small or very large; the physical interpretation of $\hat{U}_+$ is the amplitude of scattering within the platonic crystal. Thus, small $|\hat{U}_+|$ indicates reflection (blocking), and large $|\hat{U}_+|$ indicates enhanced transmission, which is of particular interest for a single line of scatterers since it manifests in the form of Rayleigh-Bloch-like modes propagating along the grating itself (see, for example, Evans \& Porter \cite{Evans} and Colquitt {\it et al.} \cite{colq_2015} for related problems). For ${\cal K} \to -1$, the general expression~(\ref{whgen}) tells us that in the limit, 
$$
\hat{U}_- + \hat{U}_+ = \hat{F}.
$$
Recalling that $\hat{F}$ represents the incident field, and $\hat{U}_- + \hat{U}_+$, the total field, we may interpret this regime as perfect transmission or dynamic neutrality; the wave propagation is unimpeded by the microstructure of the platonic crystal, a phenomenon often associated with the vicinity of Dirac or Dirac-like points \cite{RCM_ABM_NVM}. Summarising the three regimes, we have
\begin{itemize}
\item
reflection (blocking) \,\,\,\,\,\,\,\,\,\,\,\,\,\,\,\,\,\,\,\,\,\,\,\,\,\,\,\,\,\,\,\,\,\,\,\,\,\,\,\,\,\,\,\,\,\,\,\,\,\,\,\,\,\,\,\,\,\,\,\,\,\,\,\,\,\,\,\,\,\,\,  $|{\cal K}| \gg 1$
\item
waveguide transmission \,\,\,\,\,\,\,\,\,\,\,\,\,\,\,\,\,\,\,\,\,\,\,\,\,\,\,\,\,\,\,\,\,\,\,\,\,\,\,\,\,\,\,\,\,\,\,\,\,\,\,\,\,\,\,\,\,\,\,\,\,\,\,  $|{\cal K}| \ll 1$
\item
dynamic neutrality (perfect transmission)  \,\,\,\,\,\,\,\,\,\,\,\,\,\,\,  ${\cal K} \to -1$. 
\end{itemize}

By studying the Wiener-Hopf equation and its kernel function, we are able to derive conditions for observing wave effects for the different types of point scatterers. For example, it is instructive to compare the representations for point masses attached to a Winkler-type foundation~(\ref{whcase2}),(\ref{wink_dless}) with the equivalent expressions for the mass-spring resonators attached to the top of the plate given by~(\ref{whcase1}),(\ref{kern1}). Both expressions incorporate the term $m \omega^2 - c$, which defines the resonance frequency $\omega_r$ of the individual mass-spring resonators:
\begin{equation}
m \omega_r^2 - c = 0 \iff \omega_r^2   = \frac{c}{m}.
\label{resfreq}
\end{equation}
Crucially, however, the kernel functions differ in that this term is in the numerator for the Winkler case, but in the denominator for 
the mass-spring resonator case in~(\ref{whcase1}),(\ref{kern1}). This indicates that, for example, the transmission condition for the Winkler foundation 
would correspond to the regime of reflection ($|{\cal K}| \gg 1$) for the mass-spring resonator case and vice versa.

\subsection{Reflection}
Referring to the general equation~(\ref{kern_gen_3}), reflection (blocking) is predicted for frequency regimes where either $\Phi(\omega, m, c)$ or $\hat{\cal G}$, or both functions together, blow up. We recall that the kernel function is precisely $\hat{\cal G}(\beta; z)$ for the case of a semi-infinite array of rigid pins analysed by \cite{Has2015}, and that for $z = \exp\{i \theta\}$, with $\theta = k_x d_x$ and ${\bf k} = (k_x, k_y)$ the Bloch vector for a doubly periodic system, $\hat{\cal G}(\beta; z)$ is a doubly quasi-periodic Green's function. Much has been written about this Green's function in the literature; see for example, McPhedran {\it et al.} \cite{RCM_ABM_NVM}, McPhedran {\it et al.} \cite{mcp}, Poulton {\it et al.} \cite{Poul}. A very important property is that its zeros correspond to the dispersion relation for the infinite doubly periodic system of rigid pins, which possesses a complete band gap for low frequency vibrations up to a finite calculable value. Here we express $\hat{\cal G}(\beta; z)$~(\ref{dpgreenf}) in the form:
\begin{equation}
\hat{\cal G}(\beta; z) = \hat{\cal G}(\beta; {\bf k}) = \frac{i}{8 \beta^2} \left( {\cal S}_0^H (\beta; {\bf k}) + 1 + \frac{2i}{\pi} {\cal S}_0^K (\beta; {\bf k}) \right),
\end{equation}
where ${\cal S}_0^H$, ${\cal S}_0^K$ are lattice sums defined over the periodic array of point scatterers ${\bf r}'_{np}$ in the following way:
\begin{eqnarray}
{\cal S}_0^H (\beta; {\bf k}) & = & \sum_{{\bf r}'_{np} \ne \{0, 0\}} H_0^{(1)} (\beta |{\bf r}'_{np}| ) \, e^{i {\bf k} \cdot {\bf r}'_{np}}, \nonumber \\
{\cal S}_0^K (\beta; {\bf k}) & = & \sum_{{\bf r}'_{np} \ne \{0, 0\}} K_0^{(1)} (\beta |{\bf r}'_{np}| ) \, e^{i {\bf k} \cdot {\bf r}'_{np}}.
\end{eqnarray}
We also note that the lattice sum over the Hankel functions may be written in the form
\begin{equation}
{\cal S}_0^H (\beta; {\bf k}) = -1 + i {\cal S}_0^Y (\beta; {\bf k}).
\label{S0Y}
\end{equation}
The lattice sums ${\cal S}_0^H (\beta; {\bf k})$, ${\cal S}_0^Y (\beta; {\bf k})$ are only conditionally convergent, and require an appropriate method of accelerated convergence for numerical computations. We adopt the same triply integrated expressions originally used by Movchan {\it et al.} \cite{ABM_NVM_RCM}, and more recently by \cite{mcp}. The dispersion relation for the doubly periodic pinned array is then given by
\begin{equation}
\hat{\cal G}(\beta; {\bf k}) = \frac{i}{8 \beta^2} \left( {\cal S}_0^H (\beta; {\bf k}) + 1 + \frac{2i}{\pi} {\cal S}_0^K (\beta; {\bf k}) \right) = 0,
\end{equation}
and has real solutions.

The direct connection between the kernel function for the semi-infinite array of pins and the dispersion relation for the infinite doubly periodic array enables one to identify frequency regimes for reflection and transmission of the incoming plane waves. Similarly for the point scatterers featured in this article, which importantly do not impose zero displacement clamping conditions, the zeros and singularities of the kernel function give us information about, respectively, transmission and reflection, but the kernel~(\ref{kern_gen_3}) now depends on more than $\hat{\cal G}$. 

The singularities of $\hat{\cal G}$ still indicate regimes of stop-band behaviour, but there is additional reflection behaviour determined by $\Phi (\omega, m, c)$ becoming very large. 
This is evident for the simple mass-spring resonators with $N = 1$. The kernel is given by~(\ref{kern1}) where $\Phi$ blows up for the frequency corresponding to the resonance of the individual mass-spring resonators, $\omega_r^2 = c/m$. We would therefore expect to see reflection of incident waves for $\omega$ close to this resonant frequency $\omega_r$, and stop bands in the corresponding dispersion diagrams (arising for zeros of the kernel) for the mass-spring resonators. 
\begin{figure}[h]
\begin{center}
\includegraphics[width=4.8cm]{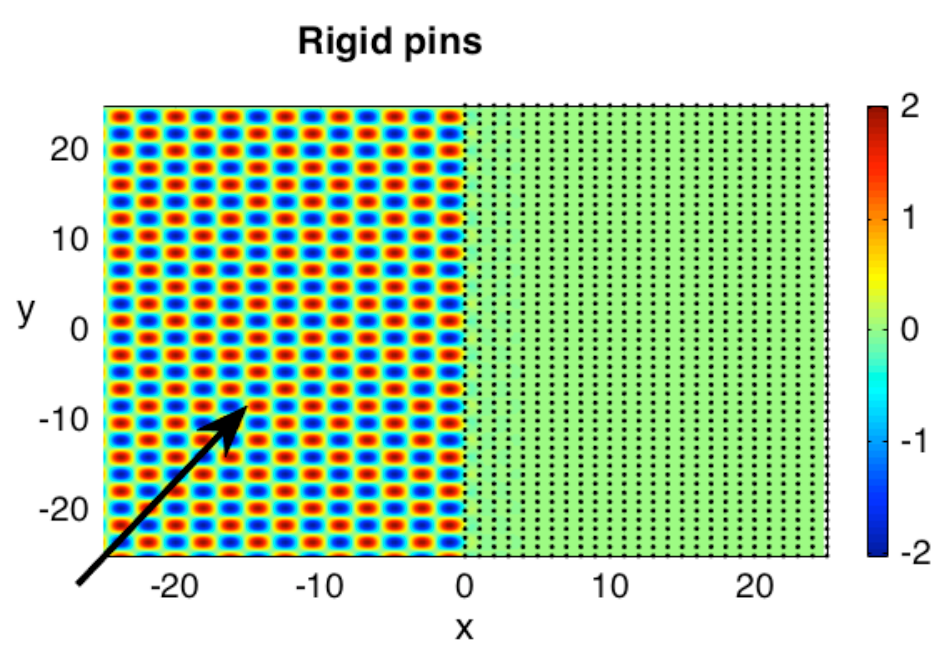}
\includegraphics[width= 4.8cm]{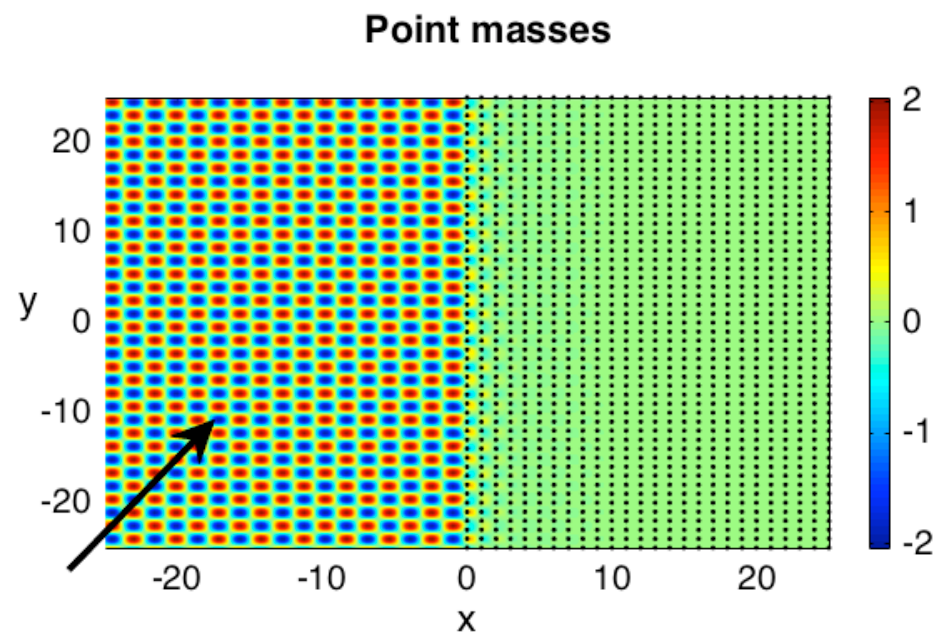}~
\includegraphics[width= 4.9cm]{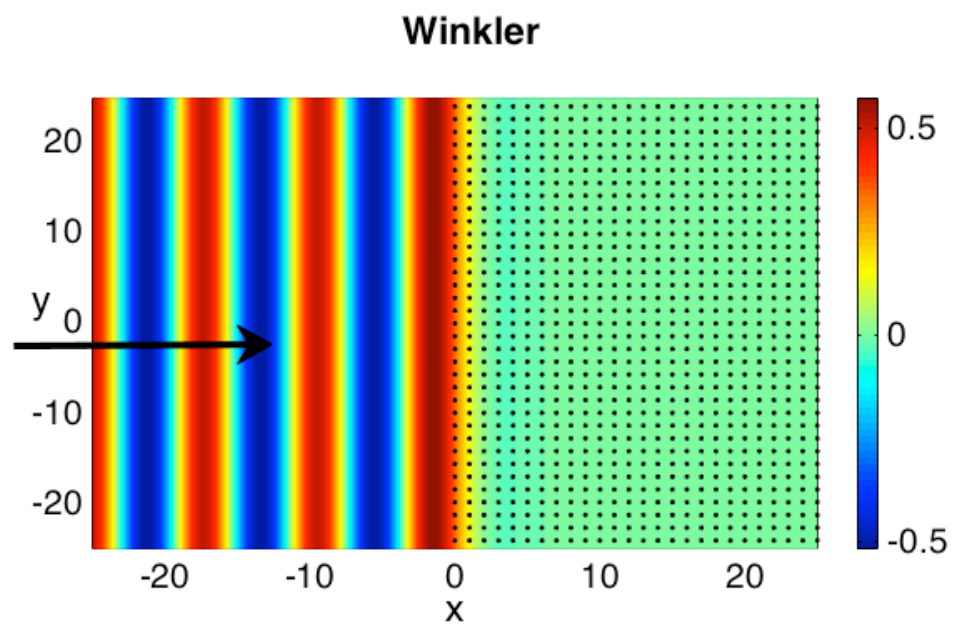}
\put(-125,35) {{\small(a)}}
\put(-77,35) {{\small(b)}}
\put(-27,35) {{\small(c)}}

\includegraphics[width= 4.6cm]{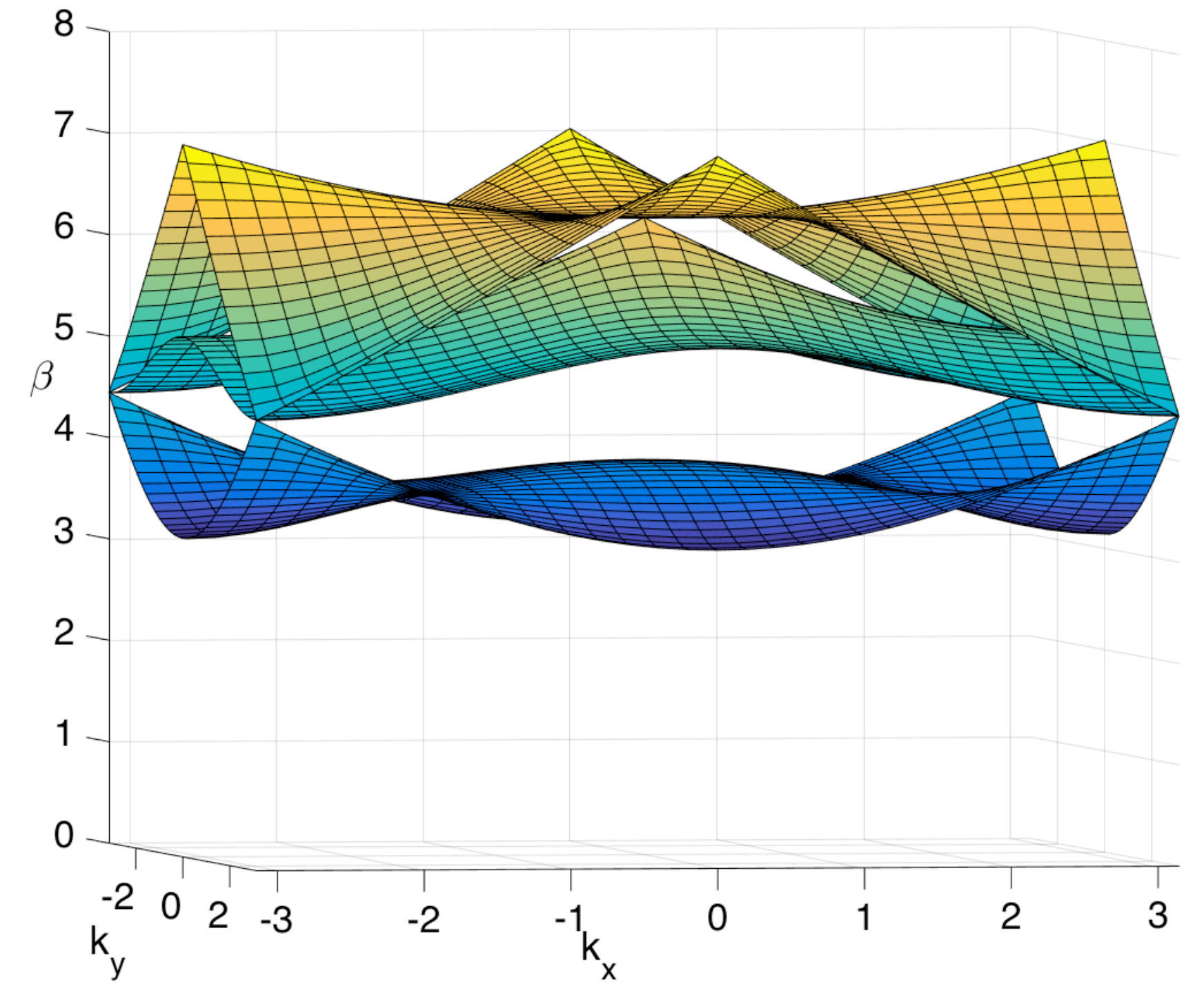}~~
\includegraphics[width= 4.6cm]{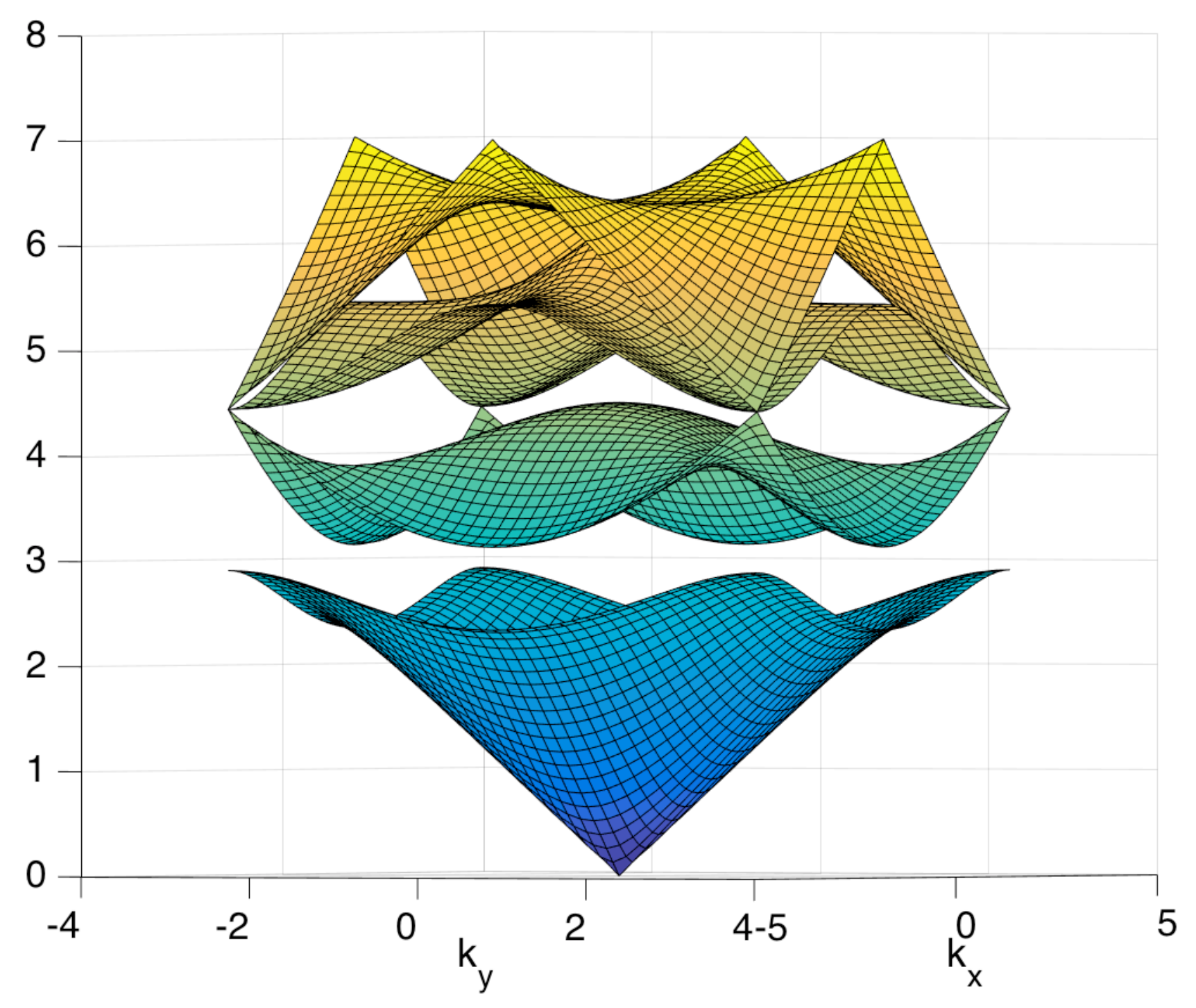}~~~
\includegraphics[width= 4.6cm]{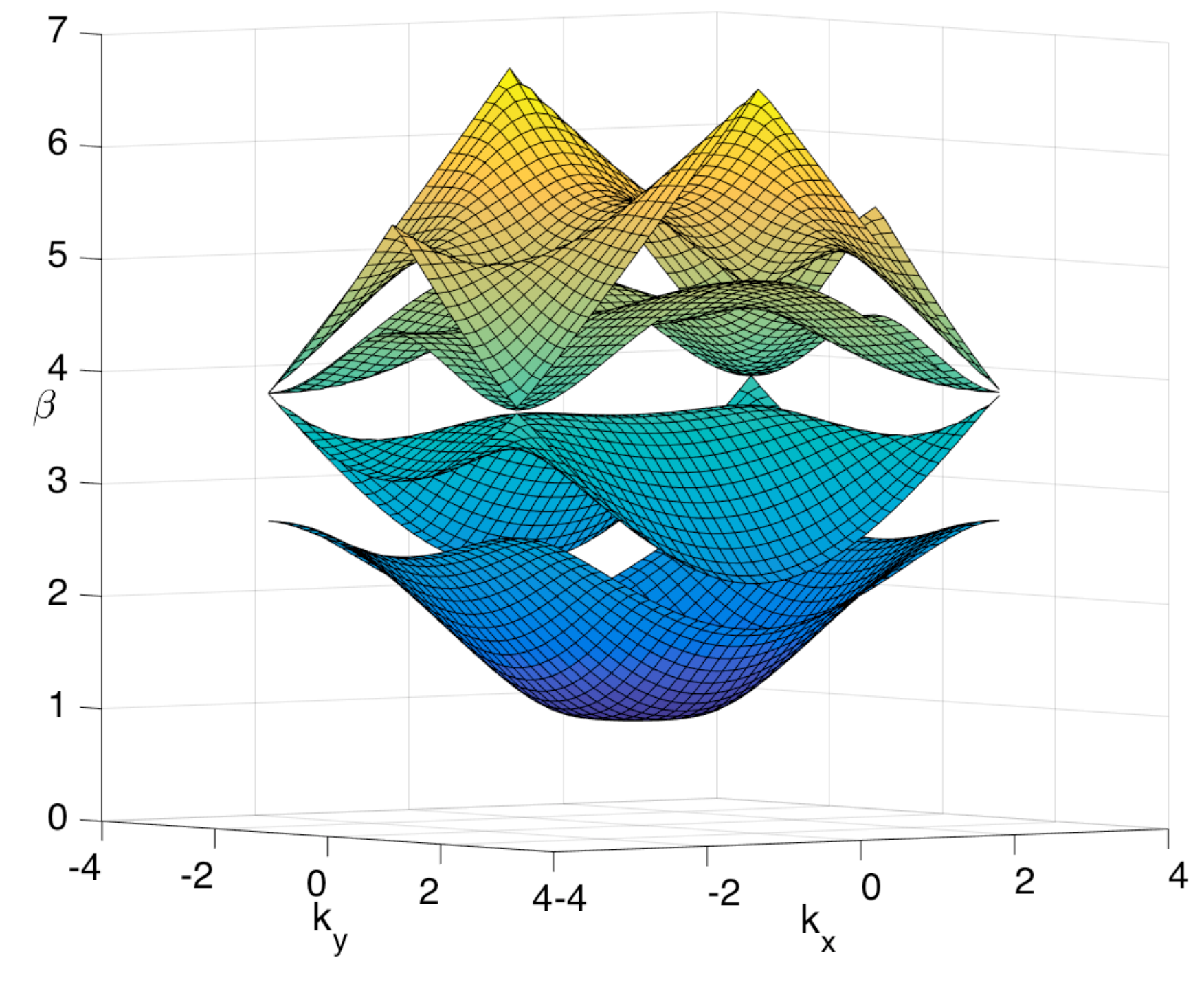}
\put(-123,-3) {{\small(d)}}
\put(-75,-3) {{\small(e)}}
\put(-25,-3) {{\small(f)}}
\caption{\label{figblock1} Band gap behaviour: Real part of total displacement field for a plane wave incident at $\psi = \pi/4$ on a semi-infinite square ($d_x = d_y =1.0$) array of (a) rigid pins for $\tilde{\beta} = 2.35$ and (b) point masses with $\tilde{m} = 1.0$ for $\tilde{\beta} = 3.02$. (c) $\psi = 0$ for a semi-infinite rectangular ($\xi = \sqrt{2}$) array for Winkler-type sprung masses with $\tilde{m} = 1.0$, $\tilde{c} = 1.9$ for $\tilde{\beta} = 0.8$. Arrows indicate direction of incident plane wave. (d-f) Dispersion surfaces for corresponding doubly periodic arrays of, respectively, parts (a-c).}
\end{center}
\end{figure}

\begin{figure}[h]
\begin{center}
\includegraphics[width= 4.9cm]{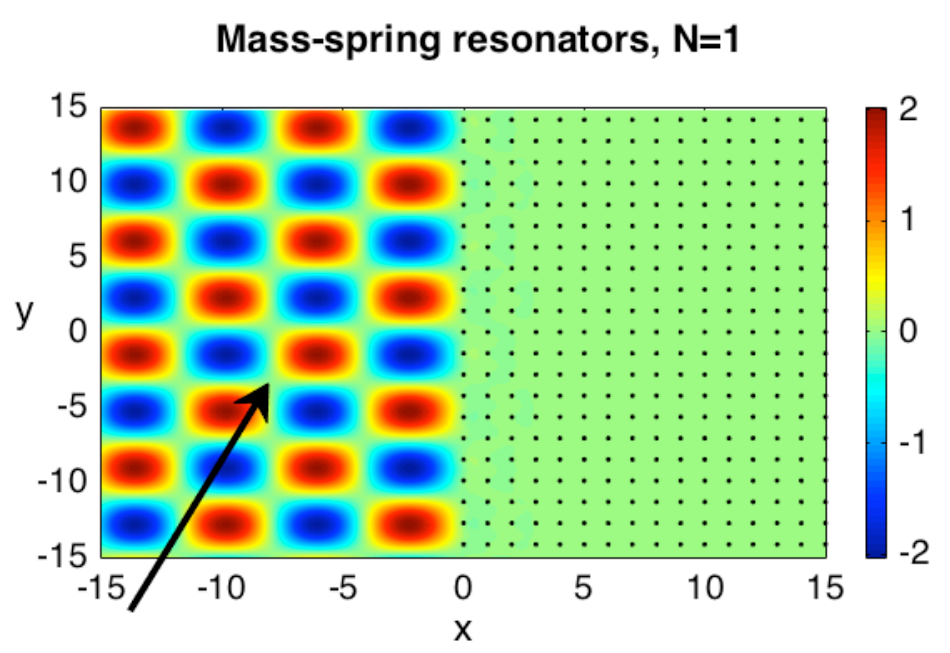}~~
\includegraphics[width= 4.95cm]{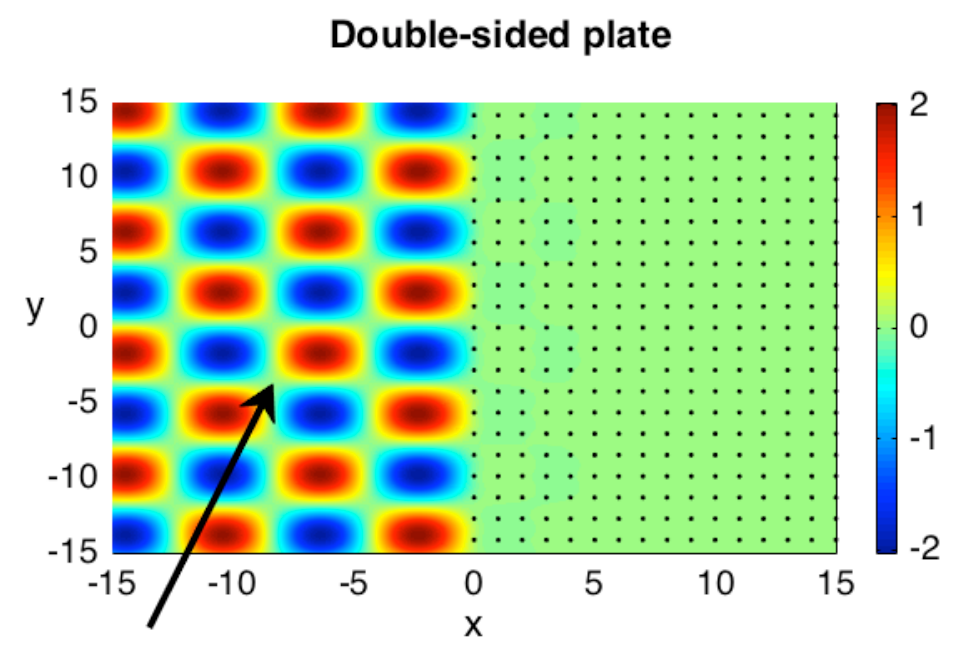}~
\includegraphics[width= 4.7cm]{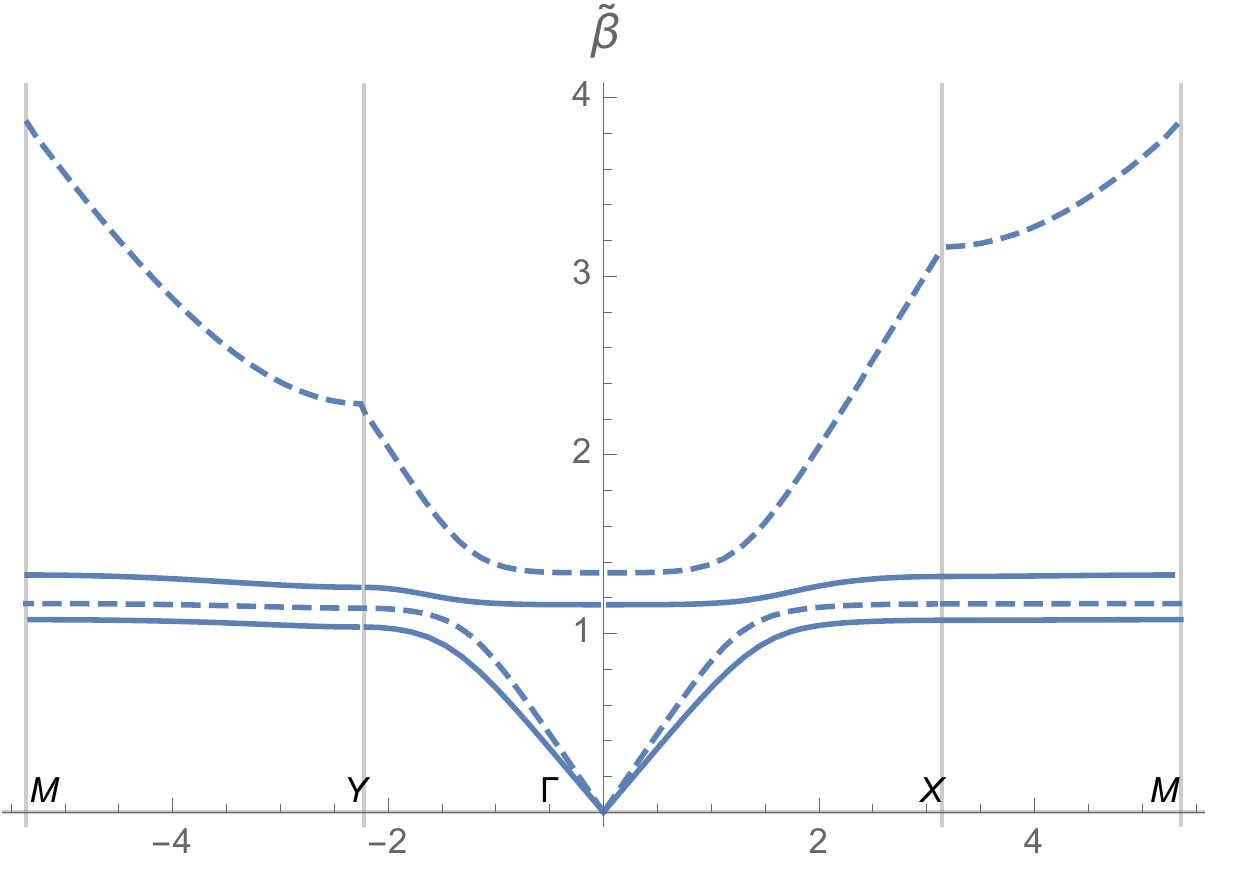}
\put(-125,37) {{\small(a)}}
\put(-77,37) {{\small(b)}}
\put(-27,37) {{\small(c)}}
\caption{\label{figblock2} Reflection: Real part of total displacement field for a plane wave incident at $\psi = \pi/4$ on a semi-infinite rectangular ($\xi = \sqrt{2}$) array of (a) mass-spring resonators for $\tilde{m} = 1.0$, $\tilde{c} = 1.910$, $\tilde{\beta} = 1.1755$ and (b) double-sided plate (DSP) for $\tilde{\beta} = 1.10$ with $m_1 = 12$, $m_2 = 108/11$, $c_1 = 400$, $c_2 = 400/3$ such that $\tilde{m}_{\scriptsize{\mbox{red}}} = 1.0$, $\tilde{c}_{\scriptsize{\mbox{red}}} = 1.910$. Arrows indicate direction of incident plane wave. (c) First two dispersion curves for resonators (dashed) and DSP (solid). The irreducible Brillouin zone is the rectangle $\Gamma X M Y$, with $\Gamma = (k_x, k_y) = (0, 0), X = (\pi, 0), M = (\pi, \pi/\sqrt{2})$, $Y = (0, \pi/\sqrt{2})$.}
\end{center}
\end{figure}

In figures~\ref{figblock1}, \ref{figblock2} we illustrate reflection (blocking) for the various  systems of figures~\ref{mass_line_res}(a-d). 
We present results for two rectangular arrays, the special case of the square array with $d_x = d_y = 1.0$, and the rectangle with aspect ratio $\xi = d_y/d_x = \sqrt{2}$. We demonstrate blocking for the square arrays of both pins and point masses in figures~\ref{figblock1}(a, b) together with their respective dispersion surfaces, and corresponding stop bands, in parts (d, e). In figure~\ref{figblock1}(c), we show the reflective behaviour of a semi-infinite rectangular array with $\xi = \sqrt{2}$ for Winkler-type sprung masses, and the corresponding dispersion surfaces are illustrated in part (f). In figure~\ref{figblock2}, we consider the same $\xi = \sqrt{2}$ for mass-spring resonators with $N=1$ in part (a) and for the plate with resonators attached to both faces (DSP) in part (b). The corresponding band diagrams are shown in figure~\ref{figblock2}(c). Here, the Brillouin zone is assumed to be the rectangle $\Gamma X M Y$, with $\Gamma = (k_x, k_y) = (0, 0), X = (\pi, 0), M = (\pi, \pi/\sqrt{2})$, $Y = (0, \pi/\sqrt{2})$. 
Note that the dispersion surfaces correspond to zeros of the kernel function, so represent the regime $|{\cal K}| \ll 1$, where the flexural waves propagate through the periodic array, which we discuss in more detail in the next section. 

The dispersion relation for the two-dimensional array of point masses follows from~(\ref{wh_pmass}): 
\begin{equation}
\left( \frac{m \omega^2}{D} \, \hat{\cal G}(\beta; z) - 1 \right) = 0,
\label{deq_masses}
\end{equation}
and in terms of dimensionless parameters, this is equivalent to 
\begin{equation}
\tilde{m}{\tilde{\beta}}^2 \tilde{\hat{{\cal G}}} (\tilde{\beta}; \tilde{z}) - 1 = 0, 
\end{equation}
which has real solutions (see figure~\ref{figblock1}(e) for the resulting surfaces). Similarly, the equation for the sprung point masses (case 2, $N = 1$) is given by 
\begin{equation}
\tilde{m} {\tilde{\beta}}^2 \left(\frac{1}{1 - \frac{\tilde{m}{\tilde{\beta}}^4}{\tilde{c}}} \right) \tilde{\hat{{\cal G}}}(\tilde{\beta}; \tilde{z}) -1 = 0,
\label{case1_wg}
\end{equation}
where the frequency dependence on the resonators is expressed in terms of dimensionless $\tilde{\beta}$ rather than $\omega$. We show the first two dispersion curves for $N = 1$ (dashed curves) in figure~\ref{figblock2}(c), where we choose dimensionless mass $\tilde{m} = 1.0$, and dimensionless stiffness $\tilde{c} = 1.910$. 
The total band-gap is centred around this resonant frequency $\omega_r$, see equation~(\ref{resfreq}), or equivalently $\tilde{\beta}_r = 1.1755$ in figure~\ref{figblock2}(c). The width of the stop band depends on the other parameter settings, which we discuss in more detail in section~\ref{dispcurvess}. 

Similar dispersion expressions for the double-sided plate and the Winkler-type sprung masses are readily obtained from their respective Wiener-Hopf/kernel equations. We show the first two dispersion curves for the DSP (solid curves) in figure~\ref{figblock2}(c), comparing the results with mass-spring resonators positioned on the top surface (dashed). For the double-sided plate, we select values of $m_1, m_2$ and $c_1, c_2$ such that the {\it reduced} mass is consistent with $\tilde{m} = 1.0$, $\tilde{c} = 1.910$. 
Reduced mass is the effective inertial mass that appears in the two-body problem of Newtonian mechanics as if it were a one-body problem. It is defined by
\begin{equation}
\frac{1}{m_{\scriptsize{\mbox{red}}}} = \frac{1}{m_1} + \frac{1}{m_2}, \,\,\, m_{\scriptsize{\mbox{red}}} \le m_1, m_2, 
\label{mred}
\end{equation}
and similarly, reduced (effective) stiffness is defined as
\begin{equation}
\frac{1}{c_{\scriptsize{\mbox{red}}}} = \frac{1}{c_1} + \frac{1}{c_2}, \,\,\, c_{\scriptsize{\mbox{red}}} \le c_1, c_2.
\label{cred}
\end{equation}
The choices for these parameter values will be explained in section~\ref{neut_eg} where it is shown that the double-sided plate supports dynamic neutrality (perfect transmission). The resulting total band gap between the first and second surfaces is clearly visible in figure~\ref{figblock2}(c) and the example of reflection for $\tilde{\beta} = 1.10$ and $\psi = \pi/4$ shown in figure~\ref{figblock2}(b) is consistent with the location of the stop band. 
The same can be said for the examples for the point masses in figure~\ref{figblock1}(b) and Winkler-sprung masses in figure~\ref{figblock1}(c). The choices of $\tilde{\beta}$ and $\psi$ illustrated coincide with the total band gaps observed for $2.94 < \tilde{\beta} < \pi$, and $0 < \tilde{\beta} < 0.94$, in figures~\ref{figblock1}(e, f) respectively.

\begin{figure}[h]
\begin{center}
\includegraphics[width=8.0cm]{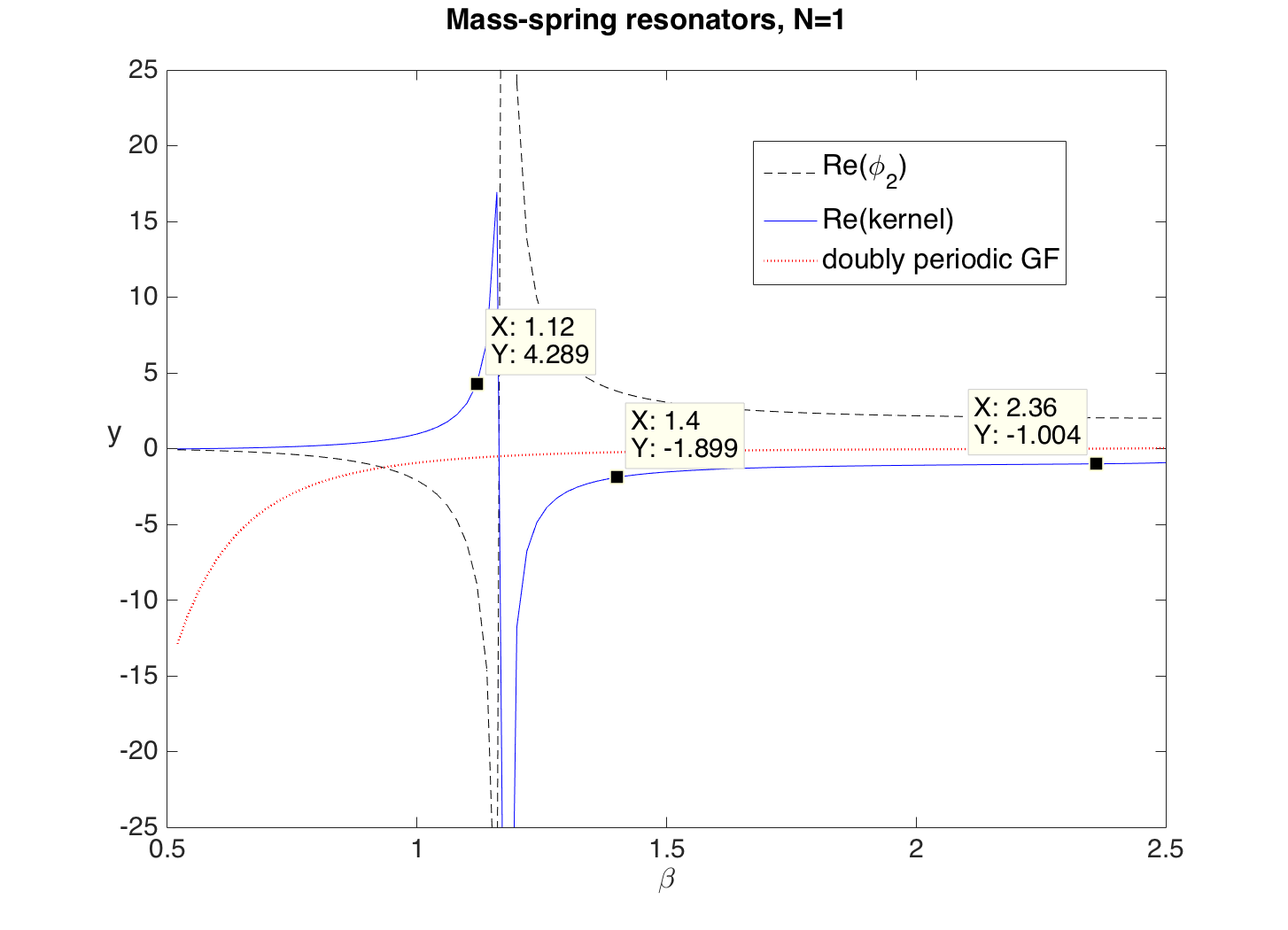}
\includegraphics[width=8.0cm]{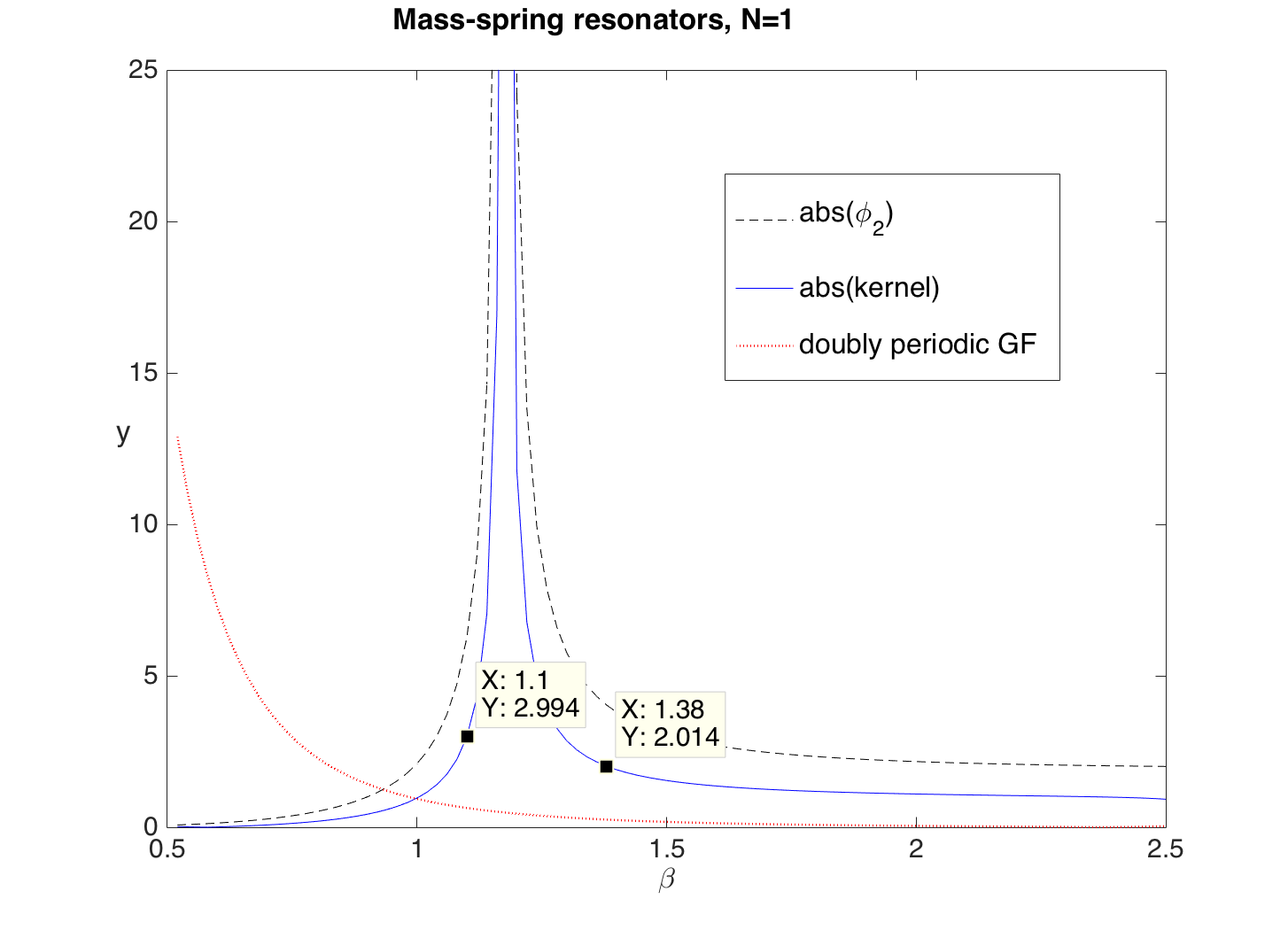}
\put(-125,62) {{\small(a)}}
\put(-47,62) {{\small(b)}}
\caption{\label{kernplot1} Transmission and reflection regimes: Plots of (a) real parts and (b) moduli of, respectively, the kernel function (solid), $\Phi (\omega, m, c)$ (dashed) and doubly quasi-periodic Green's function (dotted) for  a semi-infinite rectangular ($\xi = \sqrt{2}$) array of mass-spring resonators for $N = 1$ with $\tilde{m} = 1.0, \tilde{c} = 1.910$, $\psi = \pi/4$, $k_x = 0$ (branch $Y\Gamma$ for corresponding doubly periodic array).}
\end{center}
\end{figure}
The equations for the kernel and corresponding dispersion relations immediately give us information about the position of the stop bands. We also plot the kernel for ranges of interest to determine more information about refection and transmission regimes. For instance, for $\psi = \pi/4$, as in examples (a, b) of figure~\ref{figblock2}, we fix $k_x = 0$ (corresponding to the edge of the Brillouin zone $Y \Gamma$ for the corresponding doubly periodic system) and plot the real parts (a) and moduli (b) of the kernel (solid curves) versus $\tilde{\beta}$ in figures~\ref{kernplot1}(a, b) for the semi-infinite rectangular ($\xi = \sqrt{2}$) array of mass-spring resonators with $\tilde{m} = 1.0, \tilde{c} = 1.910$. 

The resonant frequency~(\ref{resfreq}) corresponding to maximum reflection, $\tilde{\beta}_r = 1.1755$ is clearly identified by the singularity in figures~\ref{kernplot1}(a, b). There is an approximate range of $\tilde{\beta} \in (1.12, 1.40)$ for which reflection/blocking is predicted according to $|{\cal K}| \gg 1$, which is consistent with the width of the stop band in figure~\ref{figblock2}(c). For sufficiently small $|{\cal K}|$, transmission is predicted, which coincides with solutions to the dispersion relations.  A special case of transmission is dynamic neutrality, which occurs for ${\cal K} = -1$. There is evidence in figures~\ref{kernplot1}(a, b) of an extended neutral regime for the mass-spring resonators for $\tilde{\beta} \approx 2.36$, when the solid curves in parts (a, b) tend to $\mp 1$ respectively. We discuss dynamic neutrality (which one can also think of as perfect transmission) in more detail in section~\ref{neut_eg}.

\subsection{Waveguide transmission}
For sufficiently small values of the kernel $|{\cal K}| \ll 1$, the semi-infinite crystals admit propagation of incident waves into the periodic array. Owing to the connection with corresponding infinite doubly periodic systems, this transmission property is 
predicted for the semi-infinite systems by the Bloch-Floquet analysis. 
Therefore, dispersion surfaces (as illustrated in figures~\ref{figblock1}(d-f)) give us information for predicting the propagation of waves into the inhomogeneous half of the plate. Besides standard refracted waves, there are various special cases of transmission, including dynamic neutrality (perfect transmission), interfacial localisation and negative refraction.

We demonstrate the conditions for dynamic neutrality using the expression for the kernel function, whereas previous examples of the effect given by \cite{Has2015} for the simpler pinned case, relied upon determining neighbourhoods of Dirac-like points on the accompanying dispersion surfaces. The resulting field plots yielded approximate neutrality regimes, where the direction of the propagating waves remained unchanged, but the plane wave was replaced with a total field consisting of regular regions of constructive and destructive interference of the incident and scattered fields, reminiscent of the reflected fields in figures~\ref{figblock1}(a,b) and \ref{figblock2}(a,b). The analytic prediction for perfect transmission demonstrated here is a powerful method to locate neutrality regimes, the quality of which is quantified using the condition on the scattering coefficients $|A_k| = 1$.

\subsubsection{Dynamic neutrality (perfect transmission)} 
\label{neut_eg}

The presence of Dirac-like points for the infinite system give us some insight for the corresponding semi-infinite arrays, but using the dynamic neutrality condition associated with the kernel functions enables us to predict these effects more accurately. 
Recalling the mass-spring resonators attached to both faces of the plate (DSP), we seek a condition on the kernel such that ${\cal K} \to -1$:
\begin{equation}
{\cal K}_3 = \Phi_3(\omega, m_q, c_q) \, \hat{\cal G}(\beta; z) - 1, \,\,\, \Phi_3 = \frac{\omega^2}{D} \left(\frac{m_1c_1}{c_1 - m_1 \omega^2} + \frac{m_2 c_2 }{c_2 - m_2 \omega^2} \right),
\end{equation}
which leads to the condition
\begin{equation}
\omega^2 = \frac{c_1c_2 (m_1+m_2)}{m_1m_2(c_1 + c_2)}.
\label{neut_dsp_eq}
\end{equation}
We adopt the concept of reduced mass and  stiffness, as defined by equations~(\ref{mred}), (\ref{cred}), obtaining 
\begin{equation}
\omega^2 = \frac{ \frac{1}{m_1} + \frac{1}{m_2} }{ \frac{1}{c_1} + \frac{1}{c_2} }  \,\, = \,\, \frac{c_{\scriptsize{\mbox{red}}}}{m_{\scriptsize{\mbox{red}}}} \iff m_{\scriptsize{\mbox{red}}} \omega^2 - c_{\scriptsize{\mbox{red}}} = 0.
\label{neu_cond_both}
\end{equation}
Substituting the condition for neutrality~(\ref{neu_cond_both}) into equation~(\ref{case5_gov}), we obtain
\begin{equation*}
U_p = f_p + \frac{\omega^2}{D}\, \frac{(c_{\scriptsize{\mbox{red}}}-m_{\scriptsize{\mbox{red}}}\omega^2)}{\Omega} \sum_{n=0}^{\infty} U_n G^q(\beta, |p -n|; k_y, d_x, d_y), \end{equation*}
where
\begin{equation}
\,\,\,\,\,\,\,\, \Omega = \frac{(c_1 - m_1 \omega^2)(c_2 - m_2 \omega^2)}{(c_1+c_2)(m_1+m_2)}, 
\end{equation}
an equation that is reminiscent of the simple Winkler-type foundation described by figure~\ref{mass_line_res}(d) and equations~(\ref{epsprings}),~(\ref{whcase2}) owing to the factor $(c_{\scriptsize{\mbox{red}}}-m_{\scriptsize{\mbox{red}}}\omega^2)$ in the numerator. Indeed, for specific choices of $m, c$ and $m_{\scriptsize{\mbox{red}}}, c_{\scriptsize{\mbox{red}}}$, we expect to observe neutrality in both models for the same normalised frequency $\omega$ that determines zeros of this factor.

\begin{figure}[h]
\begin{center}
\includegraphics[width=5.4cm]{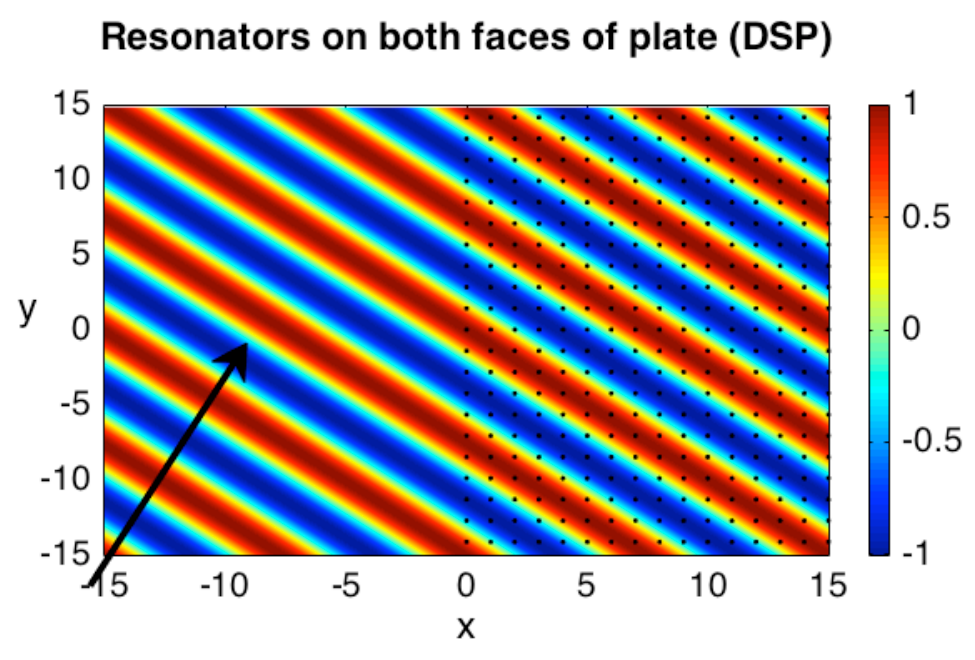}~
\includegraphics[width=5.5cm]{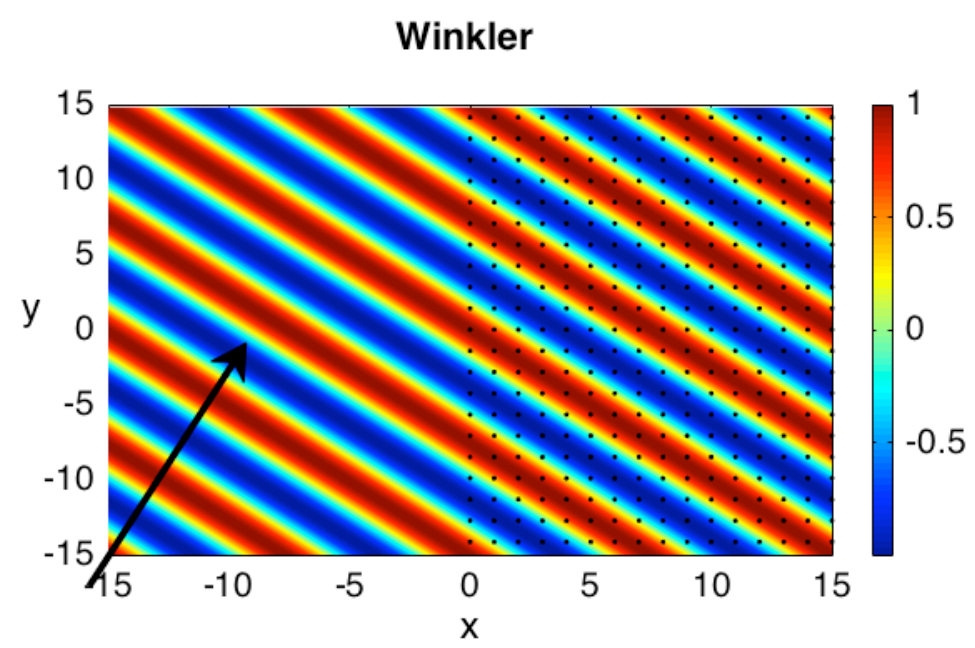}
\includegraphics[width=5.2cm]{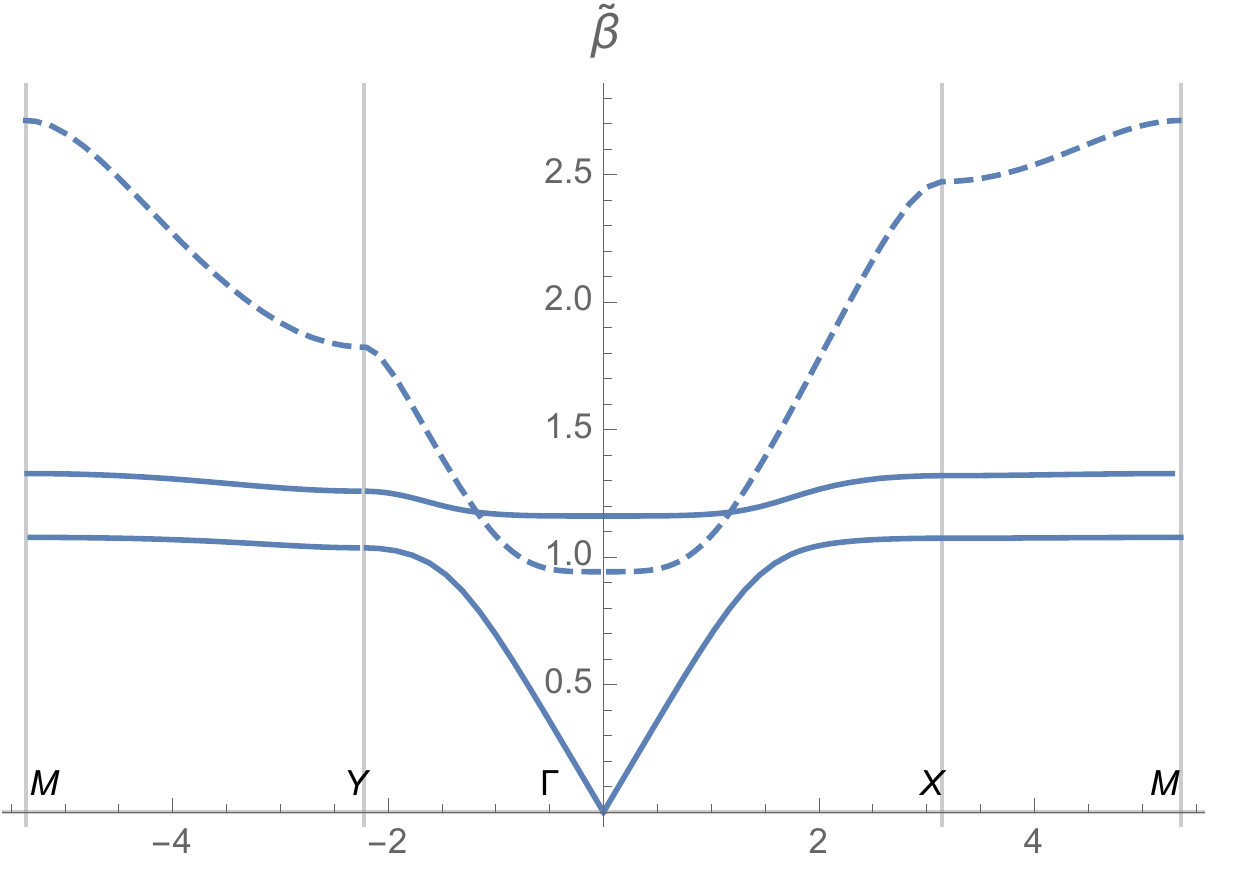}
\put(-140,39) {{\small(a)}}
\put(-85,39) {{\small(b)}}
\put(-30,39) {{\small(c)}}
\caption{\label{neueffect} Perfect transmission: Real part of the total displacement field for a semi-infinite rectangular array of point scatterers with, $d_x = 1.0, d_y = \sqrt{2}$, $\psi = \pi/4$, $\tilde{\beta} = 1.1755$ for (a) double-sided plate with $\tilde{m}_{\scriptsize{\mbox{red}}} = 1.0, \tilde{c}_{\scriptsize{\mbox{red}}} = 1.910$, (b) Winkler-type sprung point masses with $\tilde{m} = 1.0$, $\tilde{c} = 1.910$.  Arrows indicate direction of incident plane wave. (c) Band diagram for Winkler-sprung masses (dashed) and double-sided plate (solid).}
\end{center}
\end{figure}
We illustrate this surprising correspondence in figure~\ref{neueffect}.  Setting $\tilde{m} = 1.0$, $ \tilde{c} = 1.910$ for the Winkler-foundation point masses, we assign values to the parameters $m_1 = 12$, $m_2 = 108/11$, $c_1 = 400$, $c_2 = 400/3$ such that 
$m_{\scriptsize{\mbox{red}}} = 5.4, c_{\scriptsize{\mbox{red}}} = 100$, which ensures that $\tilde{m}_{\scriptsize{\mbox{red}}} =1.0$, $\tilde{c}_{\scriptsize{\mbox{red}}} =1.910$, matching the dimensionless parameter settings for the Winkler foundation case. The neutrality value of $\tilde{\beta} = 1.1755$ follows from equations~(\ref{neu_cond_both}) and~(\ref{physp}):
\begin{equation}
m_{\scriptsize{\mbox{red}}} \omega^2 - c_{\scriptsize{\mbox{red}}} = 0,\,\,\, \omega = \sqrt{c_{\scriptsize{\mbox{red}}}/m_{\scriptsize{\mbox{red}}}}; \,\,\,\, \beta^4 = \rho h \omega^2/D.
\end{equation}

We observe virtually perfect transmission (neutrality) for both physical models for $\psi = \pi/4$ with $\tilde{\beta} = 1.1755$, which coincides with the intersection of the respective dispersion curves shown in part (c) of figure~\ref{neueffect}. The first band for the Winkler case is shown by the dashed dispersion curve, whilst the first two dispersion curves for the double-sided plate are indicated using solid lines. Recall that this specific choice of $\tilde{\beta}$ is also important for the case of mass-spring resonators attached to the top surface of the plate, demonstrating stop-band behaviour in figure~\ref{figblock2}(a). The $m \omega^2 - c$ factor is common to both models, but the relationship between the two models is reciprocal; where Winkler gives perfect transmission at $\tilde{\beta}^* = 1.1755$, the mass-spring resonators attached to the top surface, lead to reflection.

\begin{figure}[h]
\begin{center}
\includegraphics[width=7.1cm]{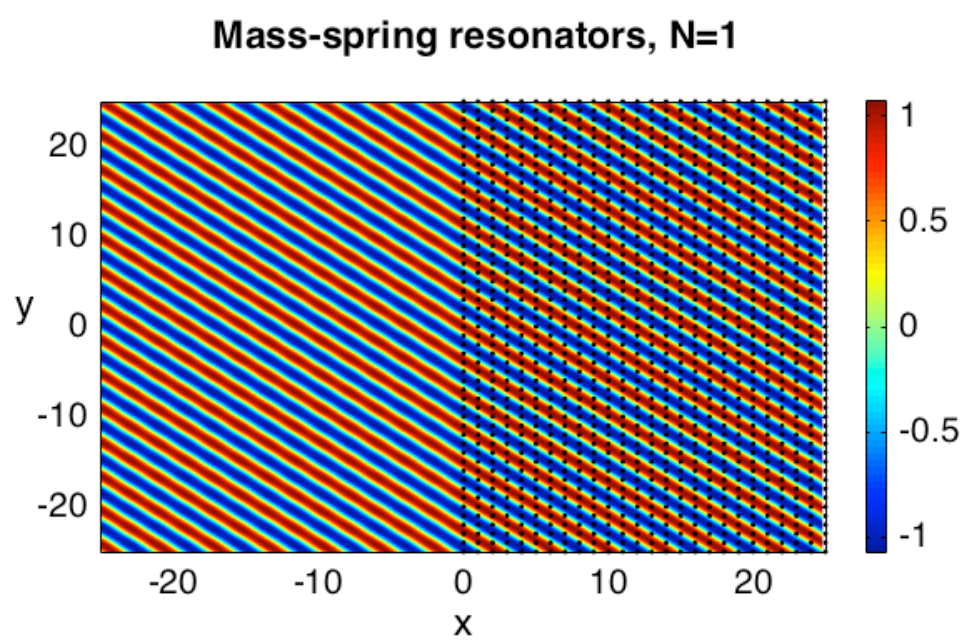}~
\includegraphics[width=7.1cm]{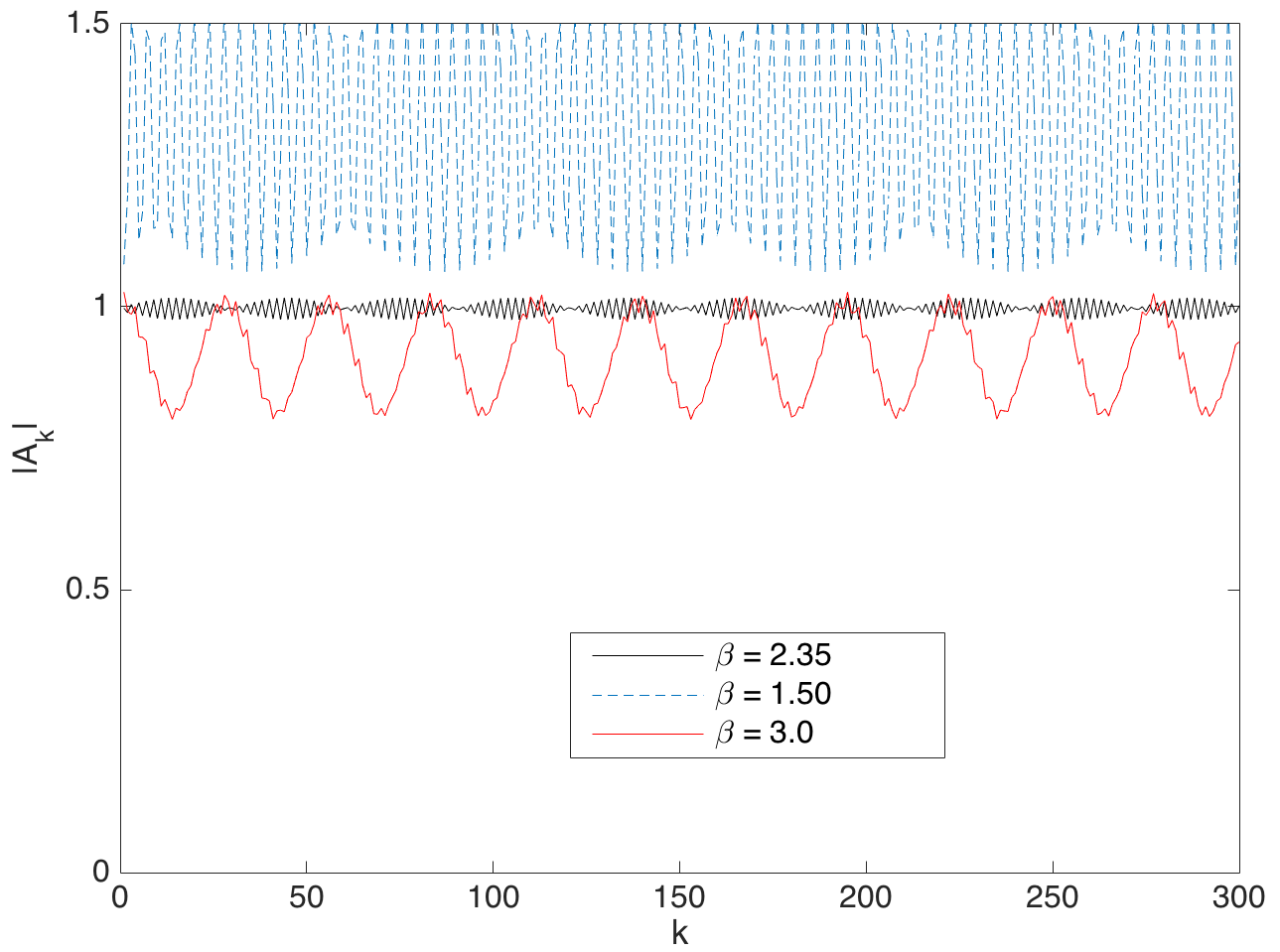}
\put(-122,58) {{\small(a)}}
\put(-43,58) {{\small(b)}}
\caption{\label{neu_ms1} Perfect transmission: (a) Real part of the total displacement field for a semi-infinite rectangular array ($\xi = \sqrt{2}$) of mass-spring resonators for $N = 1$, $\psi = \pi/4$ with $\tilde{m} = 1.0, \tilde{c} = 1.910$, $\tilde{\beta} = 2.35$. (b) Moduli of scattering coefficients $|A_k|$ for $\tilde{\beta} = 2.35$ (solid), $1.50$ (dashed), $3.0$ (lower solid).}
\end{center}
\end{figure}

Referring to figure~\ref{kernplot1}(a), the plot of the real part of kernel function (solid curve) shows a potential dynamic neutrality regime for mass-spring resonators  with $\tilde{\beta} \approx 2.36$, for which the kernel function ${\cal K} \approx -1$. For $\tilde{\beta} = 2.35, \psi = \pi/4, \xi = \sqrt{2}$, we plot the total displacement field in figure~\ref{neu_ms1}(a). 
We observe an excellent example of dynamic neutrality where the incident plane wave appears undisturbed by the interaction with the point scatterers, retaining both its direction and amplitudes. The moduli of the scattering coefficients $|A_k|$ are plotted for three values of $\tilde{\beta}$ in figure~\ref{neu_ms1}(b). For the perfect transmission frequency $\tilde{\beta} = 2.35$, $|A_k| \approx 1$ (solid) satisfying the necessary condition. For both $\tilde{\beta} = 1.50$ (dashed), and $\tilde{\beta} = 3.0$ (solid lower curve), the coefficients indicate transmission but not neutrality.

\subsubsection{Waveguide transmission for semi-infinite line of scatterers}
\label{dispcurvess}
According to Wilcox \cite{wilcox}, localised waves travelling along a grating, in the absence of an incident wave, are called Rayleigh-Bloch waves. Evans \& Porter \cite{Evans} presented conditions for the existence of Rayleigh-Bloch waves along a one-dimensional periodic array of point masses or Winkler-sprung masses in a Kirchhoff-Love plate.  
Identifying Rayleigh-Bloch regimes is also interesting for the semi-infinite array problems presented here, since evidence of the characteristic localisation will be apparent for normally incident plane waves for corresponding choices of $\beta$. 

\subsubsection*{Point masses}
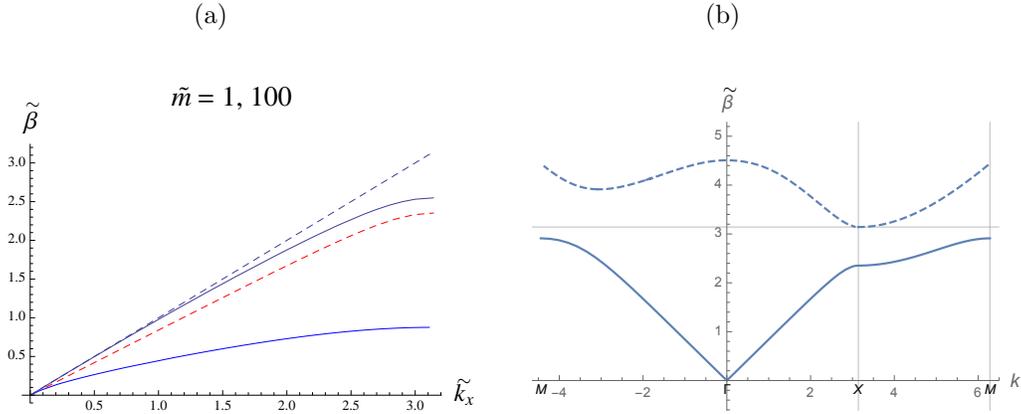
\begin{figure}[h]
\begin{center}
\includegraphics[width=6.2cm]{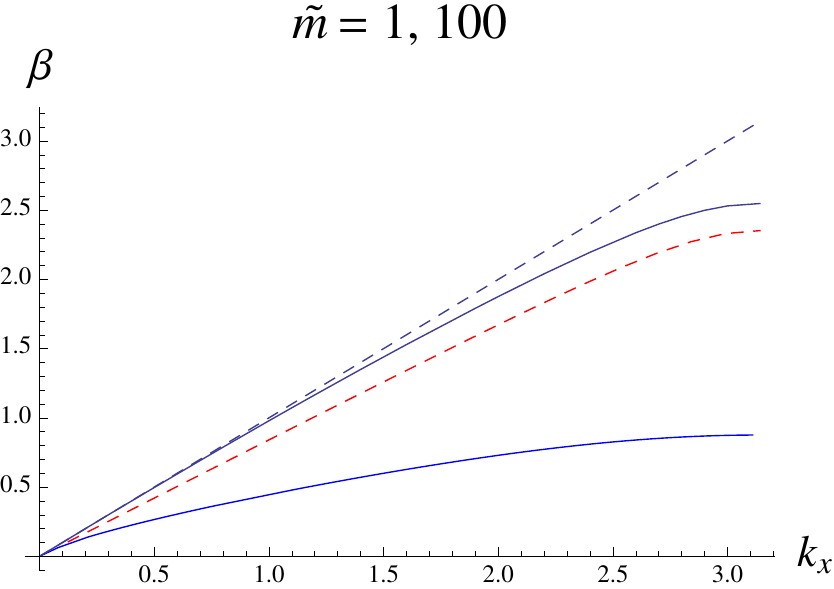}~~~~~
\includegraphics[width=6.5cm]{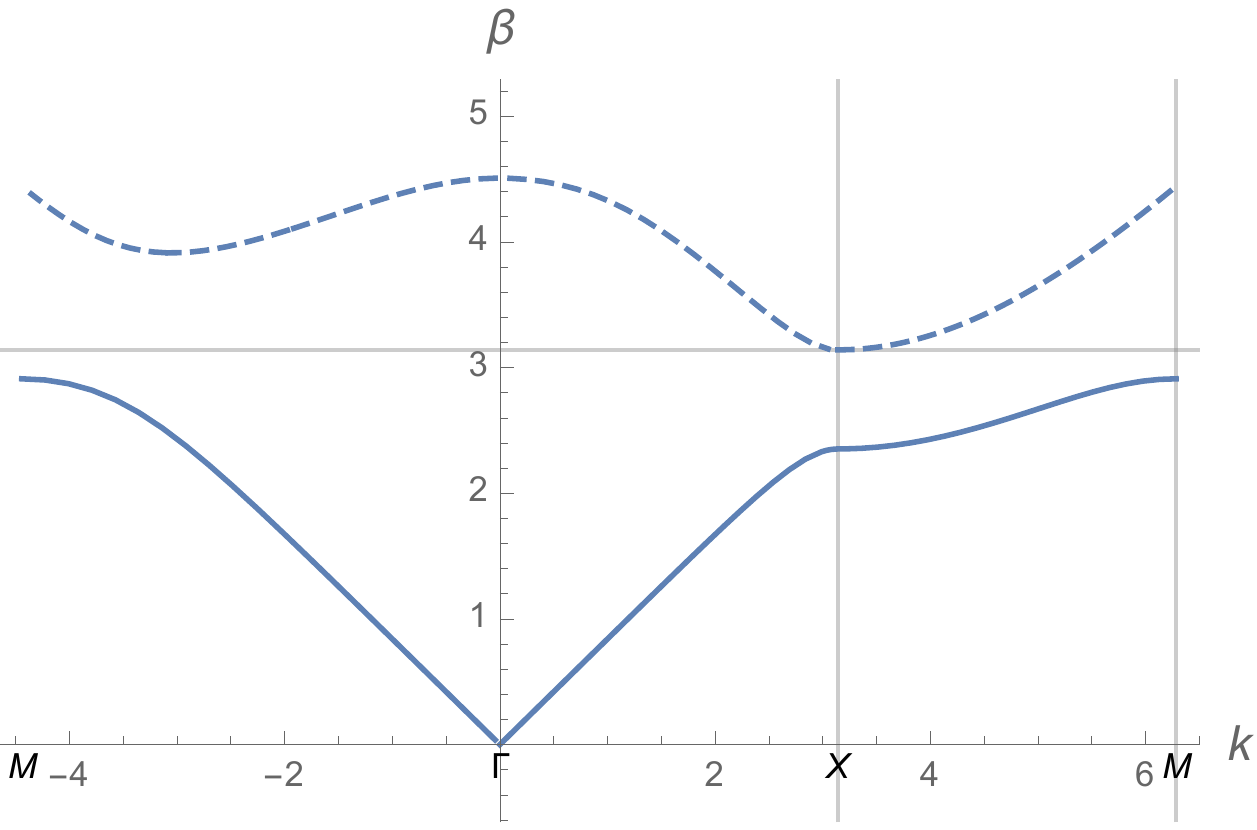}
\put(-110,52){{\small(a)}}
\put(-42,52){{\small(b)}}
\put(-132.4,40.5){{\tiny $\sim$}}
\put(-75.1,3.8){{\tiny $\sim$}}
\put(-39.7,42.5){{\tiny $\sim$}}
\caption{\label{dc1}
(a) Dispersion curves for $\tilde{m} = 1.0$ (solid upper) and $\tilde{m} = 100$ (solid lower) for $0 \le \tilde{k_{x}} \le \pi$. The dashed straight line represents both the solutions for the homogeneous plate, and the singularities of the dispersion relation for the array of masses. The dashed curve represents the dispersion curve for $\Gamma X$ for the doubly periodic square array with $\tilde{m} = 1$, $d_x = d_y = 1.0$, which is shown in (b), with irreducible Brillouin zone $\Gamma XM$, $\Gamma = (k_x, k_y) = (0, 0), X = (\pi, 0), M = (\pi, \pi)$.}
\end{center}
\end{figure}

For case of rigid pins, no Rayleigh-Bloch modes exist for real $\beta > 0$ but for point masses (case 1), there is always a solution for a Rayleigh-Bloch wave for positive mass $m$. This is evident from the approximate dispersion curves for an infinite grating shown in figure~\ref{dc1}(a), obtained for the case $|{\cal K}_1| \ll1$:
\begin{equation*}
\frac{m\omega^2}{D} \, G^q(\beta; z) - 1 = 0,
\end{equation*}
and in dimensionless parameters
\begin{equation}
\tilde{m}{\tilde{\beta}}^2 \tilde{G}^q(\tilde{\beta}; \tilde{k_x}) - 1 = 0,
\end{equation}
where $\tilde{k_x} = k_x d_x$.

Two curves for $\tilde{m} = 1.0$ and $\tilde{m} = 100$ are shown in figure~\ref{dc1}(a), along with the first two bands for the infinite square array of point masses with $\tilde{m} = 1.0$ and $d_x = d_y = 1.0$ in figure~\ref{dc1}(b), recalling that the dispersion equation  and surfaces for this case were given in equation~(\ref{deq_masses}) and figure~\ref{figblock1}(e). The similarity between the curves in  figure~\ref{dc1}(a) for the line of masses, and the $\Gamma X$ branch (i.e. with $k_y = 0$) for the two-dimensional system is striking; in both cases, we see linear-like dispersion for $\tilde{\beta}$ close to the origin, with the group velocity approaching zero as $\tilde{k_x} \to \pi$. For direct comparison, we include precisely the $\Gamma X$ branch of the first band from figure~\ref{dc1}(b) as the dashed curve in figure~\ref{dc1}(a).

The presence of the acoustic mode at low frequencies contrasts with that of rigid pins. As Poulton {\it et al.} \cite{Poul} commented,  as the dimensionless mass $\tilde{m} \to \infty$, this acoustic band becomes flatter and flatter (compare $\tilde{m} = 1.0$ with $\tilde{m} = 100$), finally collapsing into the axis $\tilde{\beta} = 0$ in the limit, thereby recovering the case of rigid pins. One other interesting feature of figure~\ref{dc1}(b), and figure~\ref{figblock1}(e), is that the $XM$ branch of the second band coincides with the dispersion curve for the homogeneous plate regardless of the value of $\tilde{m}$. This indicates that the propagation of the flexural waves in the mass-loaded plate is unaffected by the loading in this direction, which is consistent with the dynamic neutrality regime we observe in the vicinity of the Dirac-like point at $M$ in figure~\ref{figblock1}(e). This was first pointed out by McPhedran {\it et al.} \cite{RCM_ABM_NVM} who observed that the second band is ``sandwiched" between two planes of the dispersion surfaces for the homogeneous plate, where the lattice sum $S_0^Y$~(\ref{S0Y}) diverges.

\begin{figure}[h]
\begin{center}
\includegraphics[width=4.6cm]{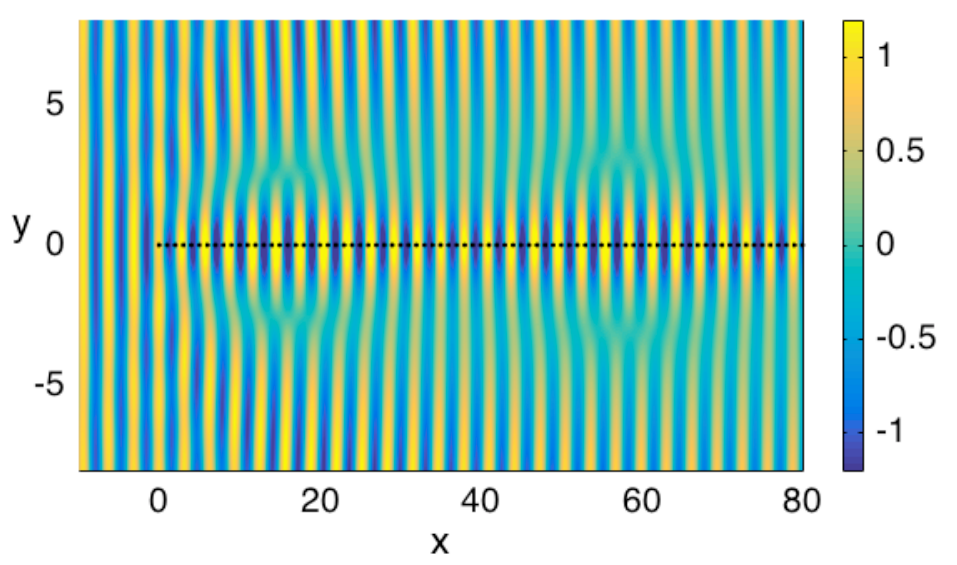}
\includegraphics[width=4.6cm]{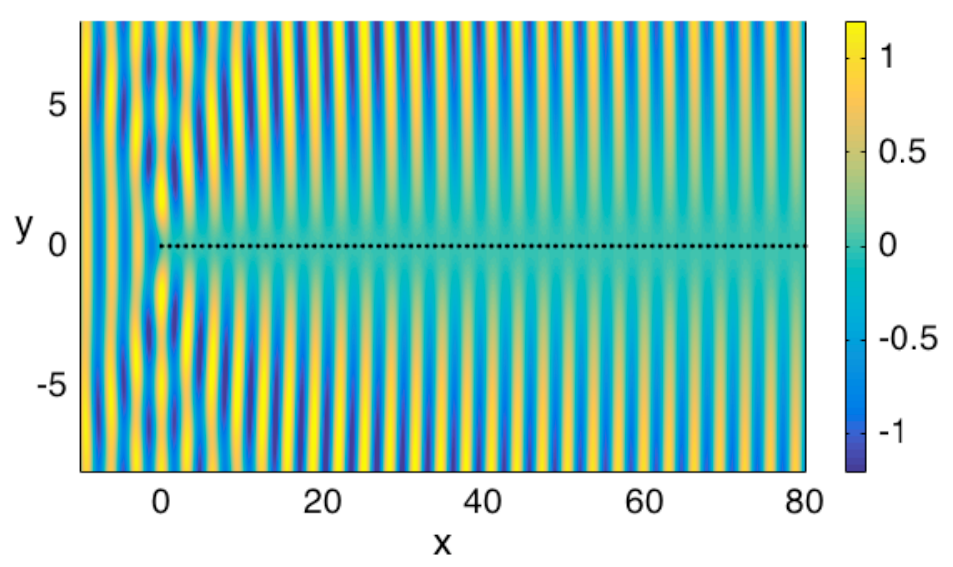}
\includegraphics[width=4.6cm]{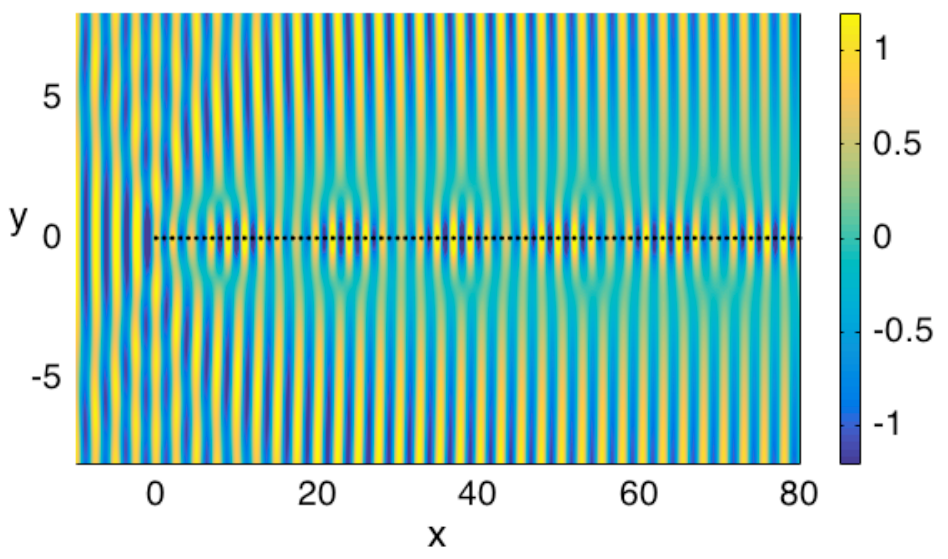}
\put(-125,32) {{\small(a)}}
\put(-75,32) {{\small(b)}}
\put(-25,32) {{\small(c)}}
\caption{\label{fig2} 
A plane wave is normally incident on an array of 1000 point scatterers with spacing $d_x = 1.0$. Real part of the total displacement field for $\tilde{\beta} = 2.0$ for (a) point masses with $\tilde{m} = 1.0$, (b) rigid pins, and (c) for $\tilde{\beta} = 2.5$ and point masses with $\tilde{m} = 1.0$.}
\end{center}
\end{figure}

The waveguide transmission regime predicted by zeros of the kernel is demonstrated for the semi-infinite line of scatterers in the form of Rayleigh-Bloch-like standing waves. This is illustrated in figures~\ref{fig2}(a, b), where we plot the real part of the total displacement field for a plane wave normally incident on a truncated semi-infinite grating of 1000 point scatterers with $\tilde{\beta} = 2.0$, comparing (a) point masses of $\tilde{m} = 1.0$ with (b) rigid pins. 
As we expect from the dispersion information, the masses support a Rayleigh-Bloch-like wave, whilst the pins exhibit blockage. To reduce the leakage of this mode away from the masses in a perpendicular direction, one should select a higher frequency, as illustrated for $\tilde{\beta} = 2.5$ in figure~\ref{fig2}(c), where the dispersion curve in figure~\ref{dc1}(a) is flatter.

\subsubsection*{Mass-spring resonators with $N = 1$}
\label{mass-spring_N=1}
\begin{figure}[h]
\begin{center}
\includegraphics[width=6.5cm]{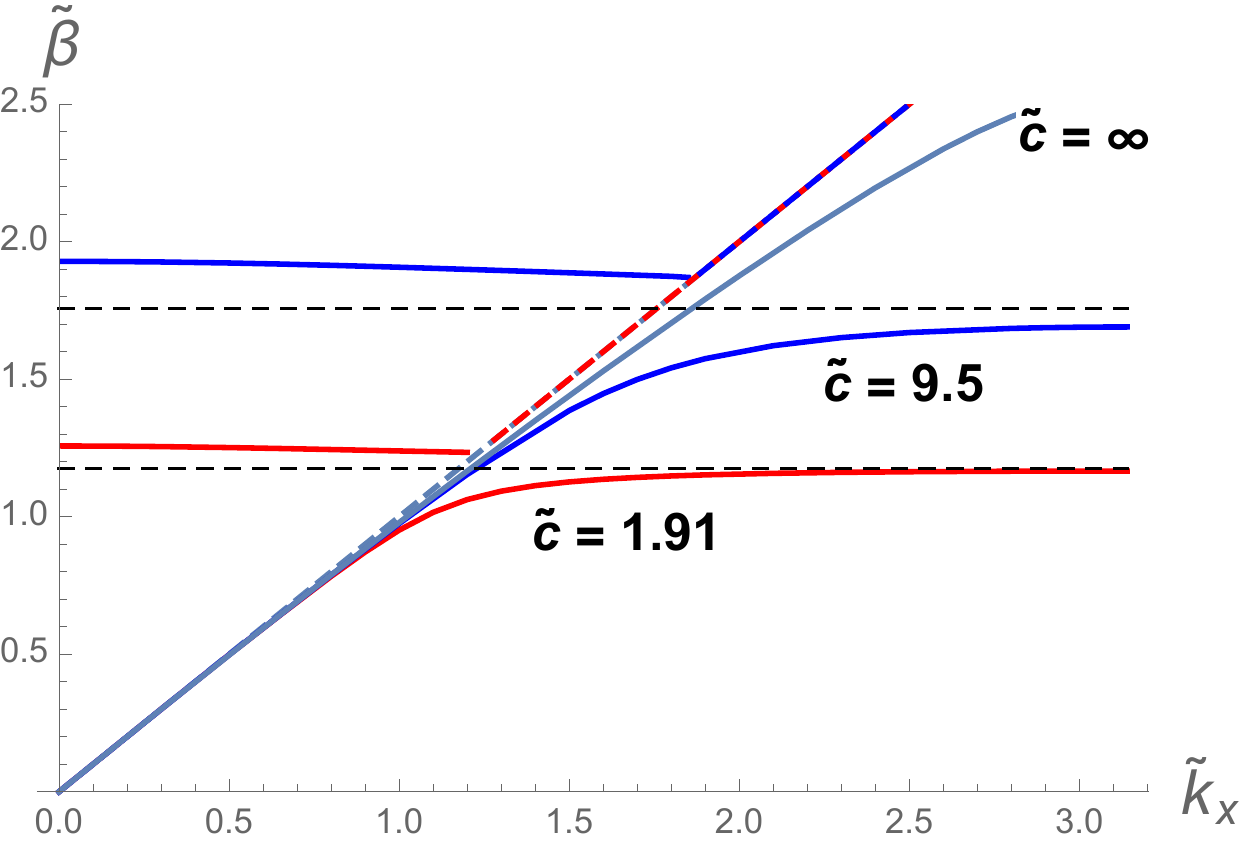}~~
\includegraphics[width=6.6cm]{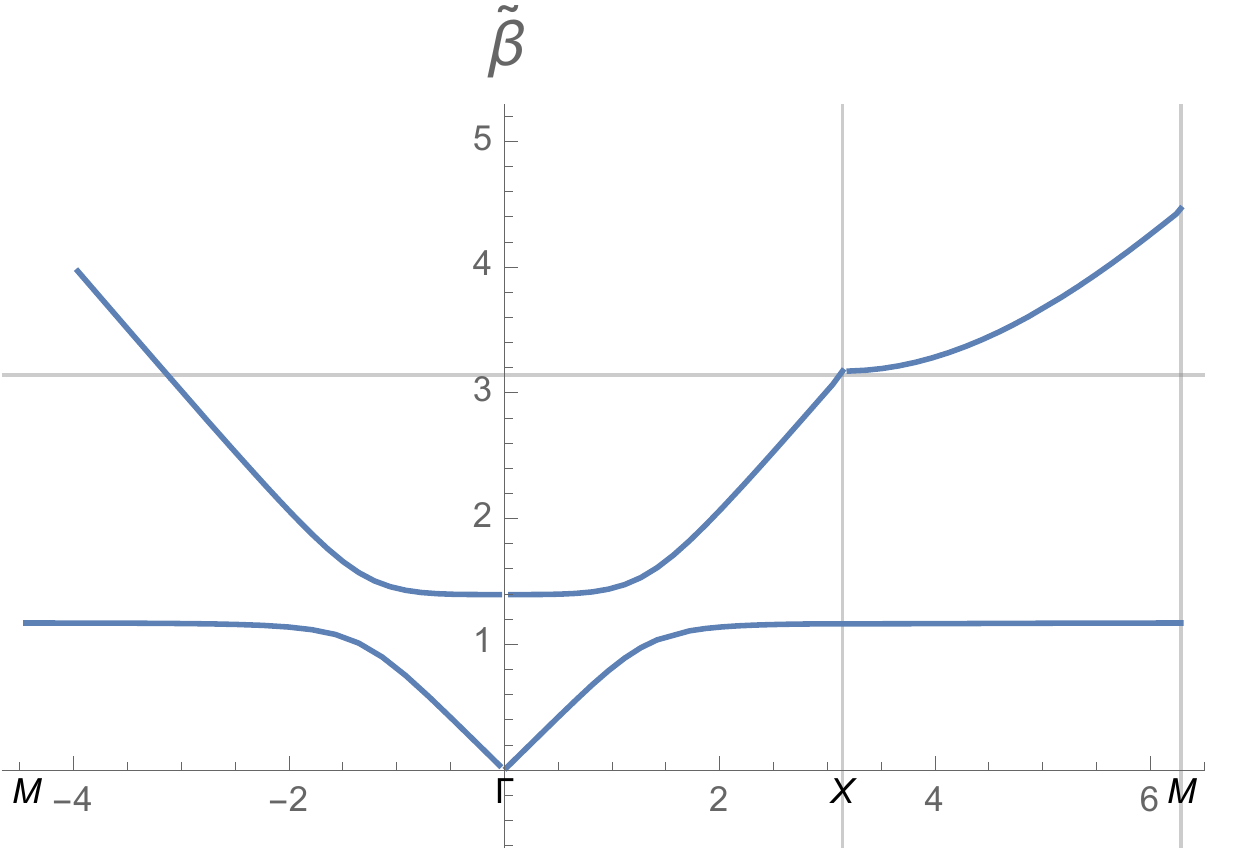}
\put(-110,50) {{\small(a)}}
\put(-42,50) {{\small(b)}}
\caption{\label{dc_springs} 
(a) Dispersion curves for an infinite line of mass-spring resonators with $\tilde{m}=1.0$ for two values of dimensionless stiffness $\tilde{c} =1.910$, 9.5 and the limit case $\tilde{c} = \infty$. Asymptotes $\tilde{\beta} = \tilde{\beta}^*$ and $\tilde{\beta} =\tilde{k_x}$ are dashed straight lines. (b) Band diagram for square array of sprung point masses with $\tilde{m} = 1.0$, $\tilde{c} = 1.910$ with irreducible Brillouin zone a triangle $\Gamma X M$, $d_x = d_y = 1.0$.}
\end{center}
\end{figure}

The dispersion relation for the semi-infinite line of mass-spring resonators, as for the analogous half-plane, is $|{\cal K}| \ll1$:
\begin{equation}
\tilde{m} {\tilde{\beta}}^2 \left(\frac{1}{1 - \frac{\tilde{m}{\tilde{\beta}}^4}{\tilde{c}}} \right) \tilde{G^q}(\tilde{\beta}; \tilde{k_x}) -1 = 0.
\label{case_wg1d_1}
\end{equation}
As expected, the introduction of springs brings new features to the dispersion picture for the waveguide transmission regime. The dispersion diagrams for, respectively, a line and doubly periodic square array are shown in figures~\ref{dc_springs}(a, b). 
The crucial difference is the term in the denominator in~(\ref{case_wg1d_1}), which contributes singularities for $\tilde{m}\tilde{\beta}^4/\tilde{c} = 1$. Physically, these solutions $\tilde{\beta}^*$ coincide with resonances arising for each mass-spring resonator, and result in branching of the dispersion curves, for both the 1d-case in figure~\ref{dc_springs}(a), and the square array with $d_x = d_y = 1.0$ in figures~\ref{dc_springs}(b), \ref{figblock2}(c). 

For a fixed dimensionless mass $\tilde{m}=1.0$, we consider dimensionless stiffness $\tilde{c}= 1.910$
and 9.5, labelled in figure~\ref{dc_springs}(a), which correspond to, respectively, stiffnesses $c = 100, 500$. We also show the limiting case, as $c \to \infty$, of point masses for the same $\tilde{m} = 1.0$. 
To the right of the straight (dashed) line $\tilde{\beta} = \tilde{k_x}$, the dispersion curves for the mass-spring resonators resemble the analogous curve for the unsprung point masses. However, these curves veer away from the asymptote $\tilde{\beta} = \tilde{k_x}$ for comparatively lower values of $\tilde{\beta}$, tending towards the horizontal asymptote $\tilde{\beta} =  {\tilde{\beta}}^*$,  
which separates the two branches of the dispersion curve for a fixed $\tilde{c}/\tilde{m}$. 

The contribution from the grating Green's function $\tilde{G^q}(\tilde{\beta}; \tilde{k_x})$ also brings singularities associated with the ``light line" $\tilde{\beta} = \tilde{k_x}$, meaning that two intersecting asymptotes are associated with each dispersion curve. 
The notion of ``light surfaces" and ``light lines" is well known in the modelling of Bloch-Floquet waves. Originating in electromagnetism, light lines identify frequencies for which light propagates in the surrounding homogeneous medium (usually air), and are now well used in problems of acoustics and elasticity for the unstructured parts of the systems in those physical settings. 
The branch to the left of $\tilde{\beta} = \tilde{k_x}$ contributes a second set of solutions for small $\tilde{k_x}$, which also appear to tend very slowly towards the asymptote $\tilde{\beta} =  {\tilde{\beta}}^*$ from above, before hitting, and then following the ``light line" $\tilde{\beta} = \tilde{k_x}$. This behaviour for the line of scatterers is consistent with the $\Gamma X$ branches of the first two bands of the doubly periodic system, illustrated in figure~\ref{dc_springs}(b). This dispersive property of the mass-spring resonator systems suggests that the semi-infinite array of sprung masses supports a neutrality effect for normally incident ($k_y = 0$) plane waves for $\tilde{\beta} > {\tilde{\beta}}^*$.

\begin{figure}[h]
\begin{center}
\includegraphics[width=13.8cm]{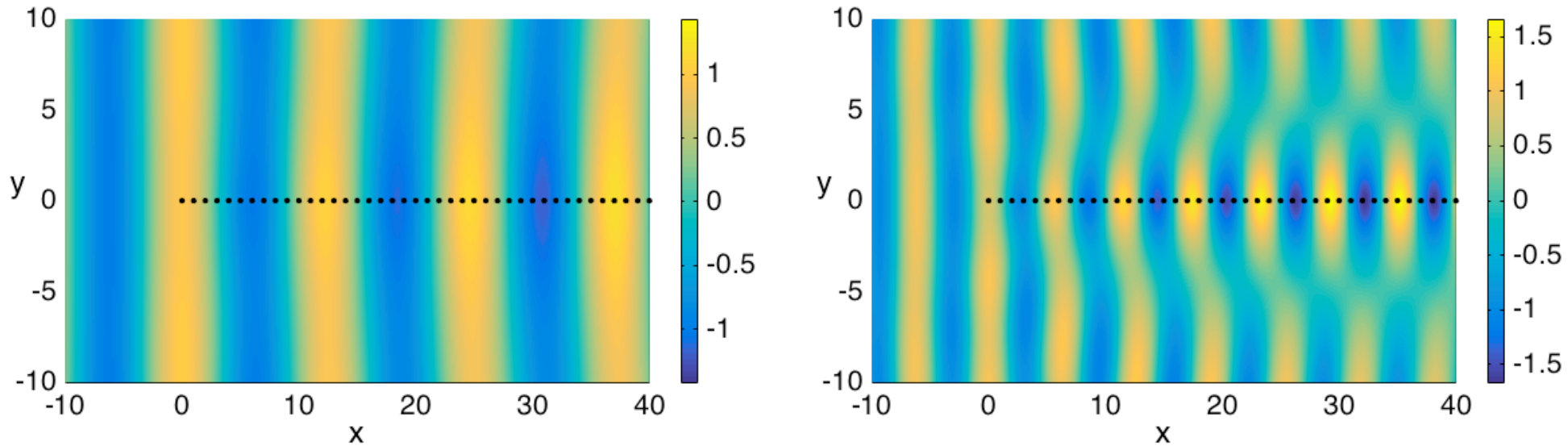}
\put(-140,40){{\small(a)}}
\put(-70,40) {{\small(b)}}

\includegraphics[width=13.8cm]{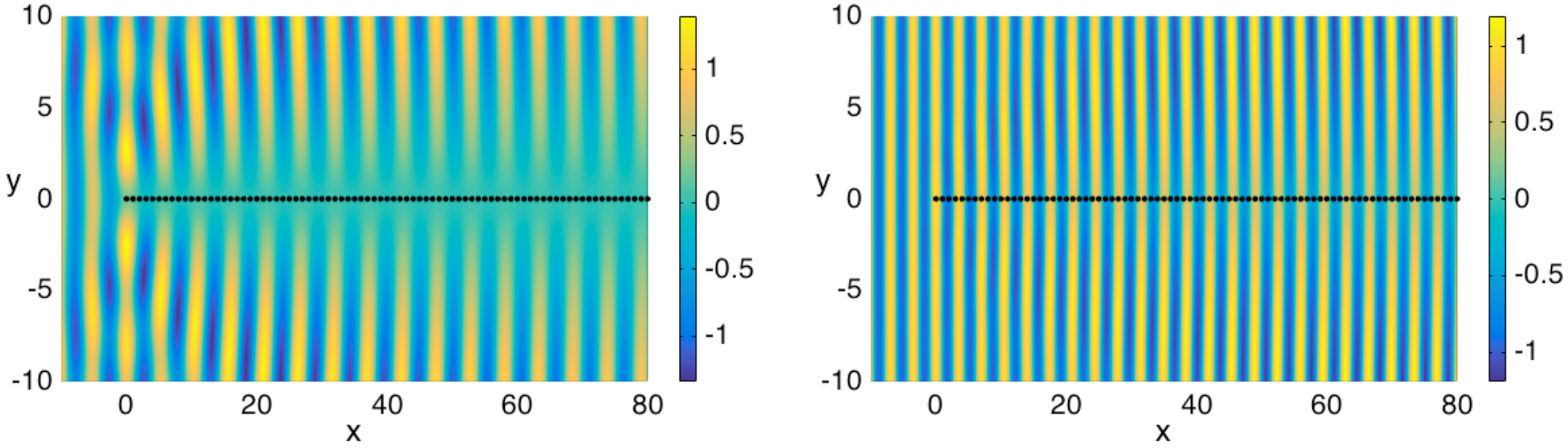}
\put(-140,40){{\small(c)}}
\put(-70,40) {{\small(d)}}

\caption{\label{nicomps} 
Real part of total displacement field for a plane wave normally incident on a line array of 1000 point sprung masses with $\tilde{m} = 1.0, \tilde{c} = 1.910$, and spacing $d_x = 1.0$ for (a) $\tilde{\beta} = 0.5$, (b) $\tilde{\beta} = 1.0$, (c) $\tilde{\beta} = 1.20$. (d) $\tilde{\beta} = 1.80$. }
\end{center}
\end{figure}

In figure~\ref{nicomps}, we consider the evolution of the total displacement fields for a semi-infinite line as the frequency parameter $\tilde{\beta}$ is increased, whilst keeping $\psi = 0$, $\tilde{m}=1.0$, $\tilde{c} = 1.910$ constant. For $\tilde{\beta} = 0.5$, we observe a long wavelength and a slight phase delay near the location of the point scatterers; compare the edge and centre of the wavefronts in figure~\ref{nicomps}(a). This difference becomes more pronounced in figure~\ref{nicomps}(b) for $\tilde{\beta} = 1.0$. Referring to the relevant dispersion curve in figure~\ref{dc_springs}(a), the sprung masses' curve is slightly further away from the ``light line" for $\tilde{\beta} = 1.0$ than for $\tilde{\beta} = 0.5$. 

The increase of $\tilde{\beta}$ to 1.20 takes us into the stop band clearly identified in the dispersion diagram. This is illustrated by the displacement field for the resonators in figure~\ref{nicomps}(c), where the point scatterers appear to block the propagation of the normally incident waves. As $\tilde{\beta}$ is increased to 1.35, we observe strong localization along the grating for normal incidence, and the corresponding dispersion curve now appears to travel along $\tilde{\beta} = \tilde{k_x}$ in figure~\ref{dc_springs}(a); we observe an increase in the moduli of the scattering coefficients, and the phase difference between the centre and edge of the wavefronts is the opposite way round to the case of frequencies below the stop band. 

For larger values of $\tilde{\beta}$, we see transmission consistent with coincidence of the dispersion curve and the straight line $\tilde{\beta} = \tilde{k_x}$. This is indicated in figure~\ref{nicomps}(d) for $\tilde{\beta} = 1.80$, where the phase difference has switched, such that the centre of the wavefront is slightly ahead of the edge; for $\tilde{\beta} = 2.4$ the phase difference  disappears entirely as perfect transmission of the plane wave is attained, similar to the example for $\tilde{\beta} = 2.35$, $\psi = \pi/4$ for the half-plane of mass-spring resonators shown in figure~\ref{neu_ms1}(a). 

\begin{figure}[h]
\begin{center}
\includegraphics[width=5cm]{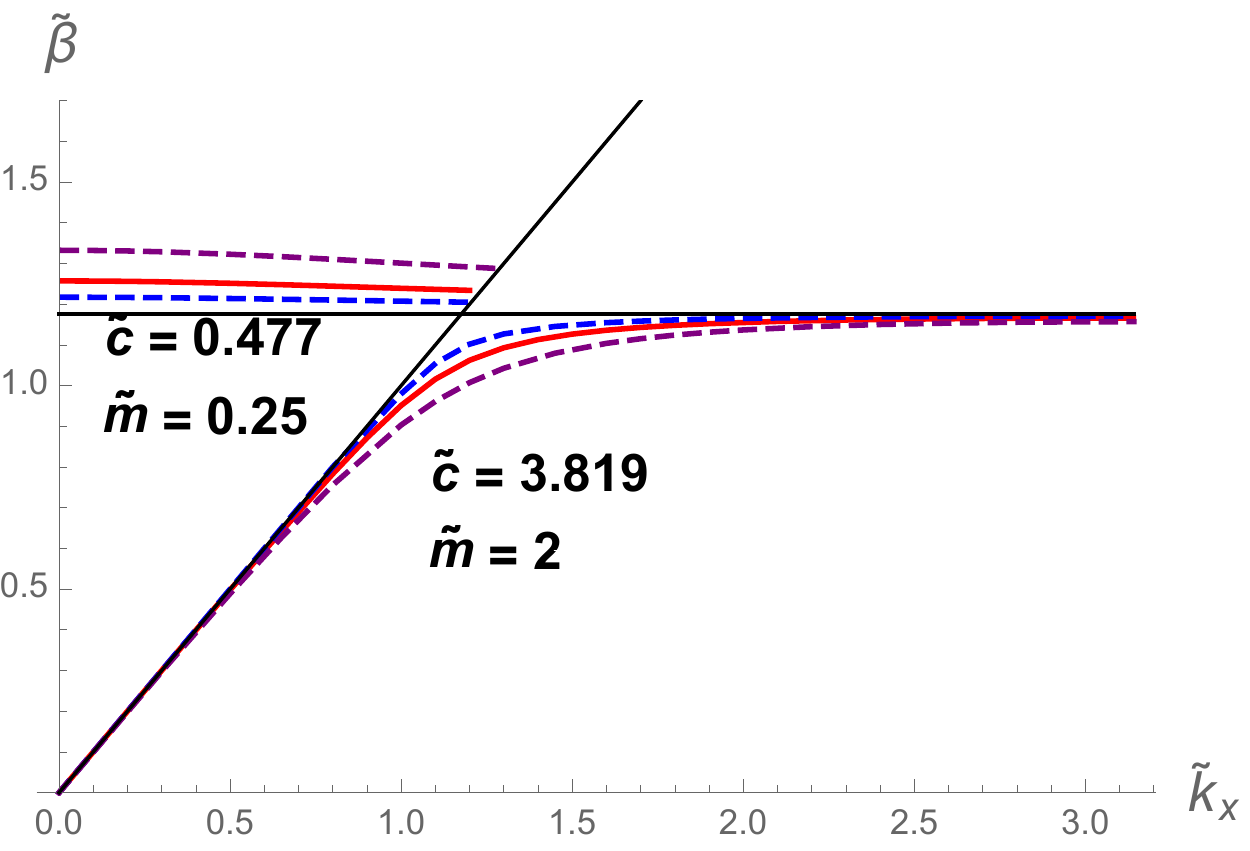}~
\includegraphics[width=10.5cm]{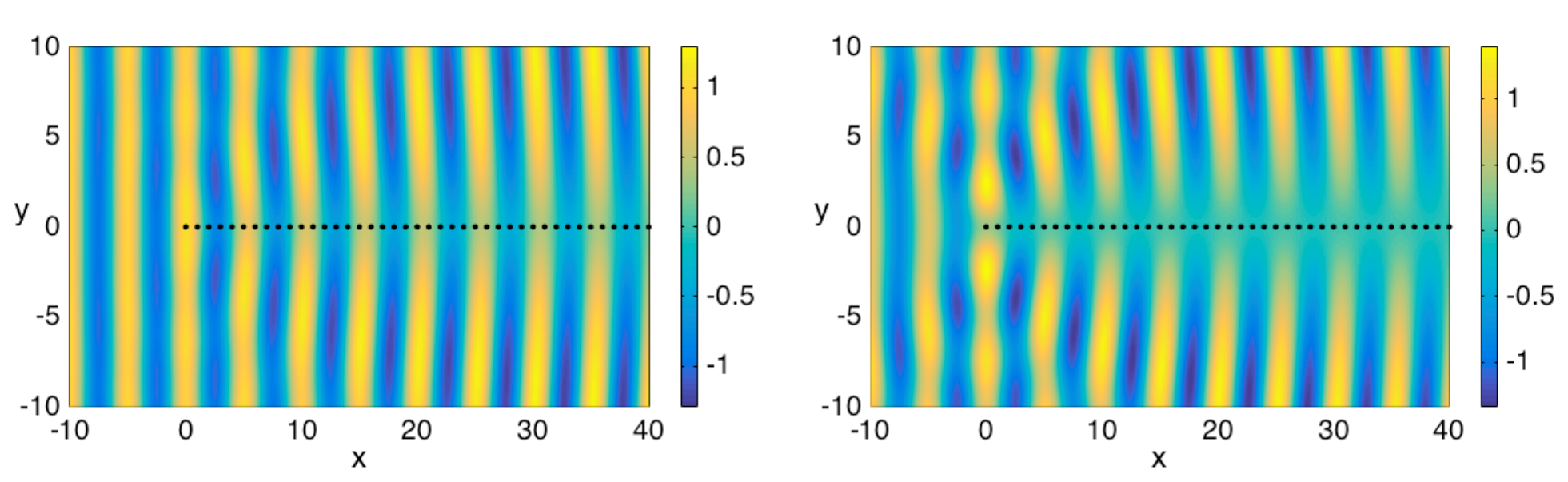}
\put(-140,37) {{\small(a)}}
\put(-85,37) {{\small(b)}}
\put(-30,37) {{\small(c)}}
\caption{\label{bg_width} Stop-band width control: 
(a) Dispersion curves for a line of point sprung masses for $\tilde{c}/\tilde{m} = 1.910$ with three pairs of dimensionless mass and stiffness: $\tilde{m}=1.0, \tilde{c} =1.910$ (solid curve);  $\tilde{m} = 0.25, \tilde{c} = 0.477$ (dashed blue curve); $\tilde{m} = 2.0, \tilde{c} = 3.819$ (dashed purple curve). Asymptotes denoted by solid straight lines. Real part of total displacement fields for $\tilde{\beta} = 1.25, \psi = 0$ for (b) $\tilde{m} = 0.25, \tilde{c} = 0.477$, (c) $\tilde{m} = 2.0, \tilde{c} = 3.819$.}
\end{center}
\end{figure}

The location of the stop band is determined by the zeros of $\tilde{m}\tilde{\beta}^4/\tilde{c} - 1$, with the branches sandwiching the resulting asymptote $\tilde{\beta} = \tilde{\beta}^*$. Thus, the ratio $\tilde{c}/\tilde{m}$ tells us where the stop band occurs, but because of the additional factor $\tilde{m}$ in equation~(\ref{case_wg1d_1}), the width of the band can be altered by varying $\tilde{m}$ and $\tilde{c}$ such that their ratio remains constant. This is illustrated in figure~\ref{bg_width}(a), where we plot the dispersion curves (dashed) for $\tilde{c} = 3.819$ ($c = 200$), $\tilde{m} = 2$ and $\tilde{c} = 0.477$ ($c = 25$), $\tilde{m} = 0.25$ along with the case $\tilde{c}/\tilde{m} = 1.910$ (solid) from figure~\ref{dc_springs}(a). For increased $\tilde{m}$ and $\tilde{c}$, the band gap is widened, with the opposite result for simultaneous reduction of $\tilde{m}$, $\tilde{c}$, whilst maintaining the constant $\tilde{c}/\tilde{m} = 1.910$. 
The facility to control the width of the band gap is useful in filtering applications for arrays of mass-spring resonators; for a specified $\tilde{c}/\tilde{m}$, we can design systems that filter $\tilde{\beta}$ values simply by redistributing the masses and stiffnesses of the resonators, as illustrated in figure~\ref{bg_width} (b, c). For $\tilde{c}/\tilde{m} = 1.910$ and $\tilde{\beta} = 1.25$, the system with $\tilde{m} = 2.0$, $\tilde{c}= 3.819$ blocks normally incident waves in figure~\ref{bg_width} (c), but allows them to pass for $\tilde{m} = 0.25$, $\tilde{c} = 0.477$ in figure~\ref{bg_width} (b). Similar observations about controlling the width of the stop band were made by \cite{Xiao} for the doubly periodic square array of mass-spring resonators.

\section{Special cases of transmission - negative refraction and interfacial localisation}
\label{results}
In this section, we provide a collection of illustrative examples for designing systems to harness notable transmissive effects, including negative refraction and interfacial localisation, for the rectangular lattice with $\xi = \sqrt{2}$. For the sake of computational efficiency, we present the total displacement fields for truncated semi-infinite systems, using the algebraic system of equations adopted by Foldy \cite{foldy}. We rewrite the general governing equation~(\ref{gov2d2}) in the truncated form: 
\begin{equation}
\Phi(\omega, m, c) \sum_{n=0}^{N-1} u_n G^q(\beta, |s - n|; k_y, d_x, d_y) = u_s - f^{(i)}_s  \,\,\,\,\,\,\,\, s = 1, 2,....,N, 
\end{equation} 
where we have replaced the incident field notation $f_s$ with $f^{(i)}_s$. 
Hence, in terms of matrices we obtain the equation
\begin{equation}
{\bf u}\left( \Phi  {\bf G}^q - {\bf I_N}\right) \, = -{\bf f}^{(i)}.
\label{fold_mat}
\end{equation}
Here ${\bf u}$ is a vector representing the total displacement field, ${\bf f}^{(i)}$ is the vector representing the corresponding incident waves and ${\bf G}^q$ is a matrix of quasi-periodic Green's functions. We solve the algebraic system of equations~(\ref{fold_mat}) to retrieve the displacements $u_s$. 
The displacement fields are then illustrated by plotting the real part of 
\begin{equation}
u({\bf r}) = f^{(i)}({\bf r}) + \Phi(\omega, m, c) \sum_{n=0}^{N} u_n G^q(\beta, |{\bf r} - (nd_x,0)|; k_y, d_x, d_y),
\end{equation} 
where each vertical grating labelled by $n$ is centred at $(nd_x,0)$ and associated with a corresponding quasi-periodic Green's function $G^q$.

\subsection{Negative refraction}

Referring to figure~\ref{figblock2}(c), the flat segments for $XM$, $MY$ for the first two dispersion curves (solid) for the double-sided plate, and the first dispersion curve for the mass-spring resonators with $N = 1$ (dashed), are of interest. 
These flat sections correspond to isofrequency contours with sharp corners that bring saddle points on the dispersion surfaces. It is well known from the photonic crystal literature, see, for example, Joannopoulos {\it et al.} \cite{jo}, that this anisotropy gives rise to a number of interesting wave effects. Following the method of Zengerle \cite{zeng}, one uses wave-vector diagrams, consisting of isofrequency contours for both the platonic crystal and the ambient medium (the unstructured part of the plate here), to investigate wave phenomena in planar waveguides. This technique is also outlined by \cite{jo} for photonic crystals in their chapter 10, and was employed recently by \cite{Has2015} for platonic crystals. The key point is that the predicted direction of propagation for the group velocity of refracted waves into the platonic crystal is perpendicular to the isofrequency contours, and in direction of increasing frequency.

\begin{figure}[h]
\begin{center}
\includegraphics[width= 6.2cm]{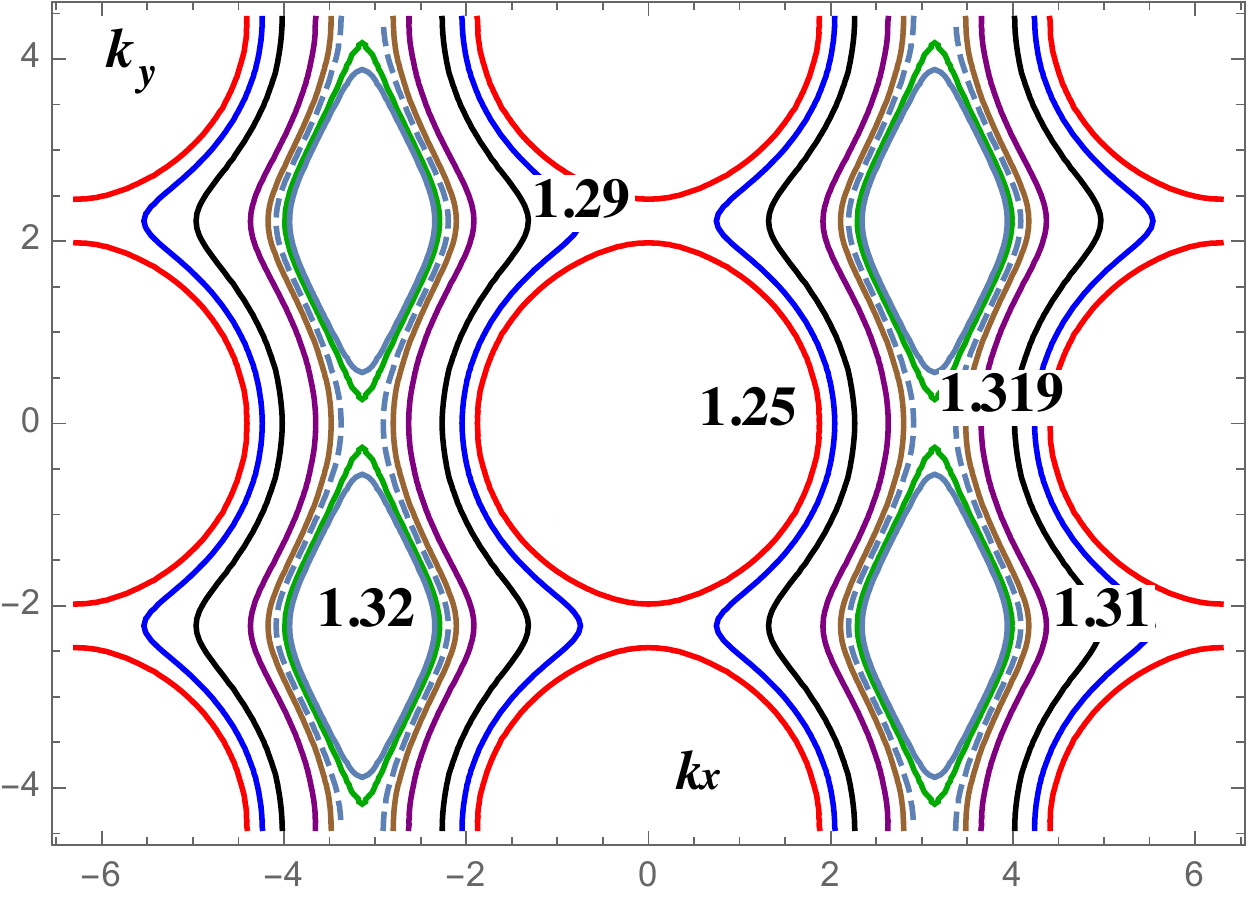}~
\includegraphics[height=4.65cm]{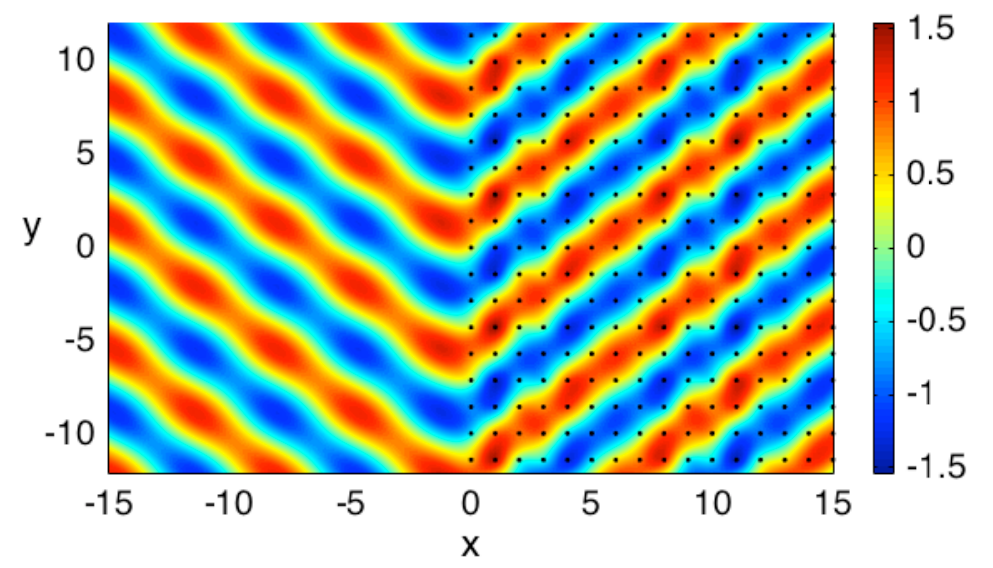}
\put(-118,50) {{\small(a)}}
\put(-45,50) {{\small(b)}}
\caption{\label{NRdsp} Semi-infinite rectangular array ($\xi = \sqrt{2}$) of mass-spring resonators on both faces of the plate with $\tilde{m}_{\scriptsize{\mbox{red}}} = 1.0, \tilde{c}_{\scriptsize{\mbox{red}}} = 1.910$. (a) Isofrequency contours for the corresponding infinite platonic crystal. (b) Real part of the total displacement field for $\psi = \pi/4$, $\tilde{\beta} = 1.319$.} 
\end{center}
\end{figure}

The sharp corners joining straight branches of constant $\beta$-contours, which are illustrated in figure~\ref{NRdsp}(a), are significant. Small changes in either the angle of incidence, or the frequency, of the incoming waves, thereby switching from one side of the sharp corner to the other on the isofrequency diagram, predict a strong modification of the direction of the refractive group velocity. 

We illustrate an example of negative refraction for the double-sided plate (DSP) in figure~\ref{NRdsp}, where we select the value of $\tilde{\beta} = 1.319$, labelled on the isofrequency contour diagram~\ref{NRdsp}(a). The contours illustrated in figure~\ref{NRdsp}(a) were selected with reference to the $\tilde{\beta}$ values for the relatively flat, although slightly increasing, branches of the second band in figure~\ref{neueffect}(c) (solid) i.e. $XM$ and $MY$. 
We choose the oblique angle of incidence $\psi = \pi/4$, which was used to demonstrate the neutrality condition in figure~\ref{neueffect}, for a semi-infinite array of mass-spring resonators attached to both faces of a plate (DSP). The mass and stiffness parameters are the same as those considered for that previous example, $\tilde{m}_{\scriptsize{\mbox{red}}} = 1.0, \tilde{c}_{\scriptsize{\mbox{red}}} = 1.910$, but the change in $\tilde{\beta}$ produces  significantly different propagation results.

We consider an array of 50 gratings positioned parallel to the $y$-axis, with period $d_y = \sqrt{2}$ and spacing $d_x = 1.0$. The real part of the total displacement field is shown in figure~\ref{NRdsp}(b), which demonstrates a negative refraction-like effect. This is consistent with the dynamic response of the periodic resonator structure at frequencies in the neighbourhood of saddle points on the corresponding dispersion surfaces, where preferential directions are identified for the  ``hyperbolic'' regime. This can also be interpreted as an example of ``mirror'' effects, which are often observed in semi-reflective optical media.

\subsection{Interfacial localisation and Dirac bridges}
\label{ifloc}
The platonic crystals featured in this paper display several Dirac-like points, 
some of which are illustrated in figures~\ref{figblock1}(d-f) for, respectively, square arrays of pins and point masses, and the rectangular array of Winkler-type masses. 
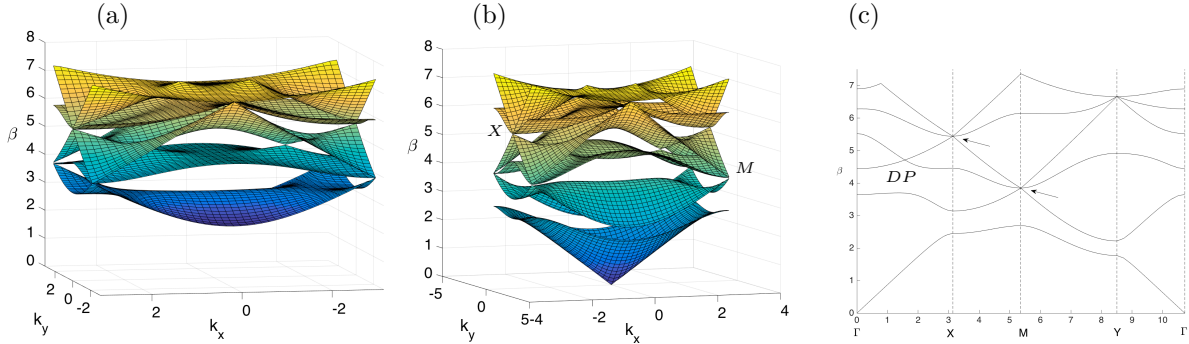
\begin{figure}[h]
\begin{center}
\includegraphics[width=5.2cm]{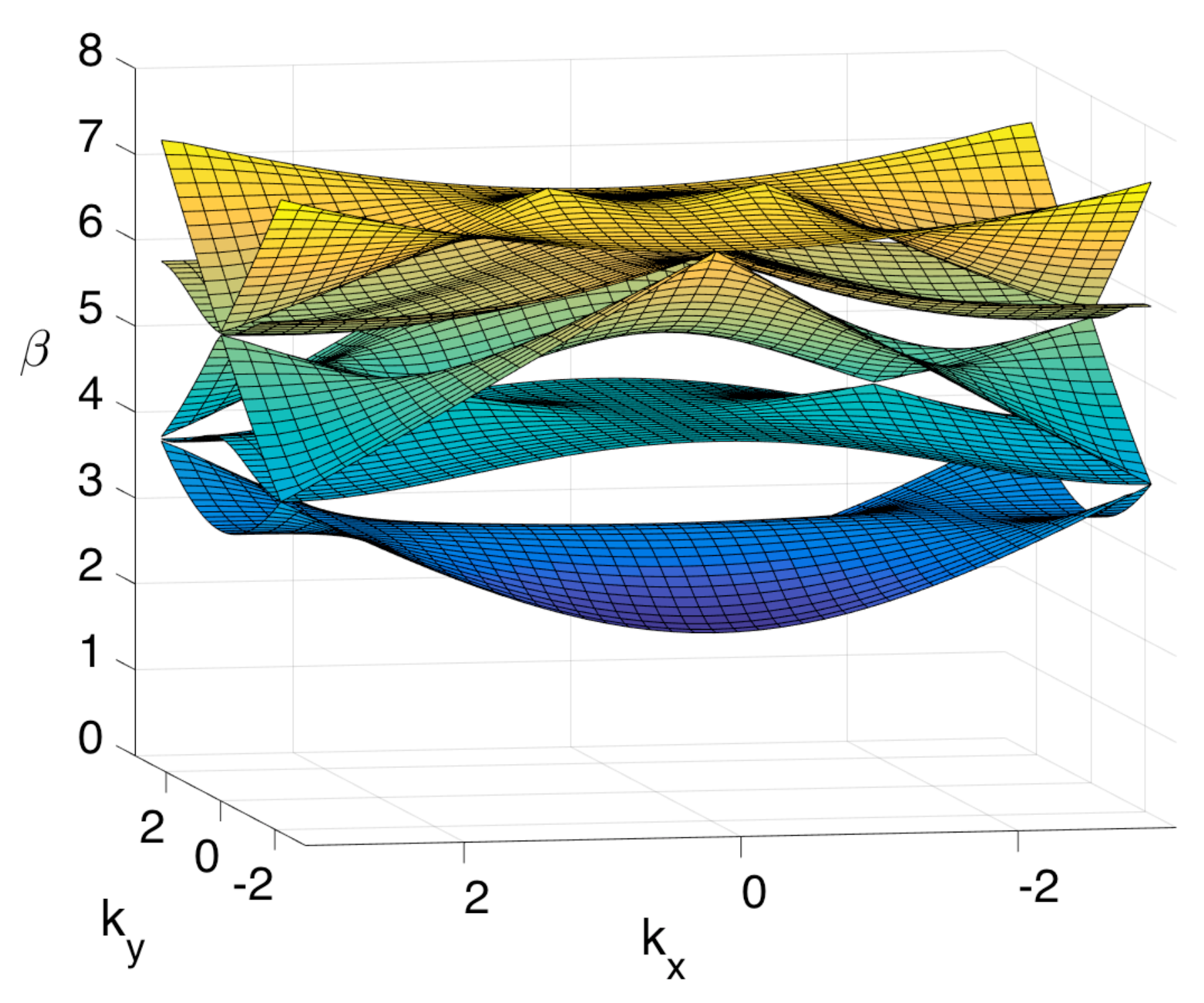}
\includegraphics[width=5.2cm]{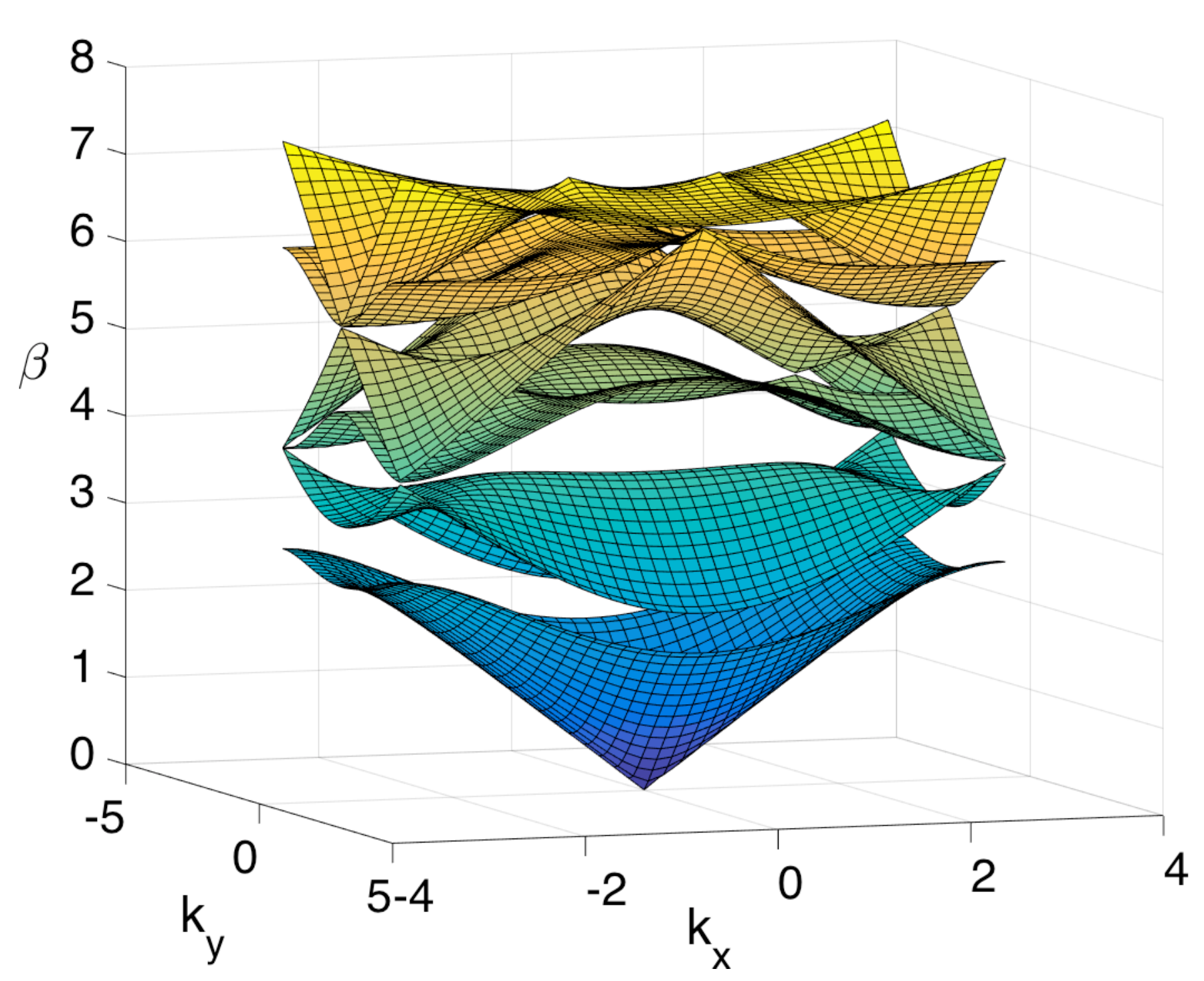}
\includegraphics[width=5.6cm]{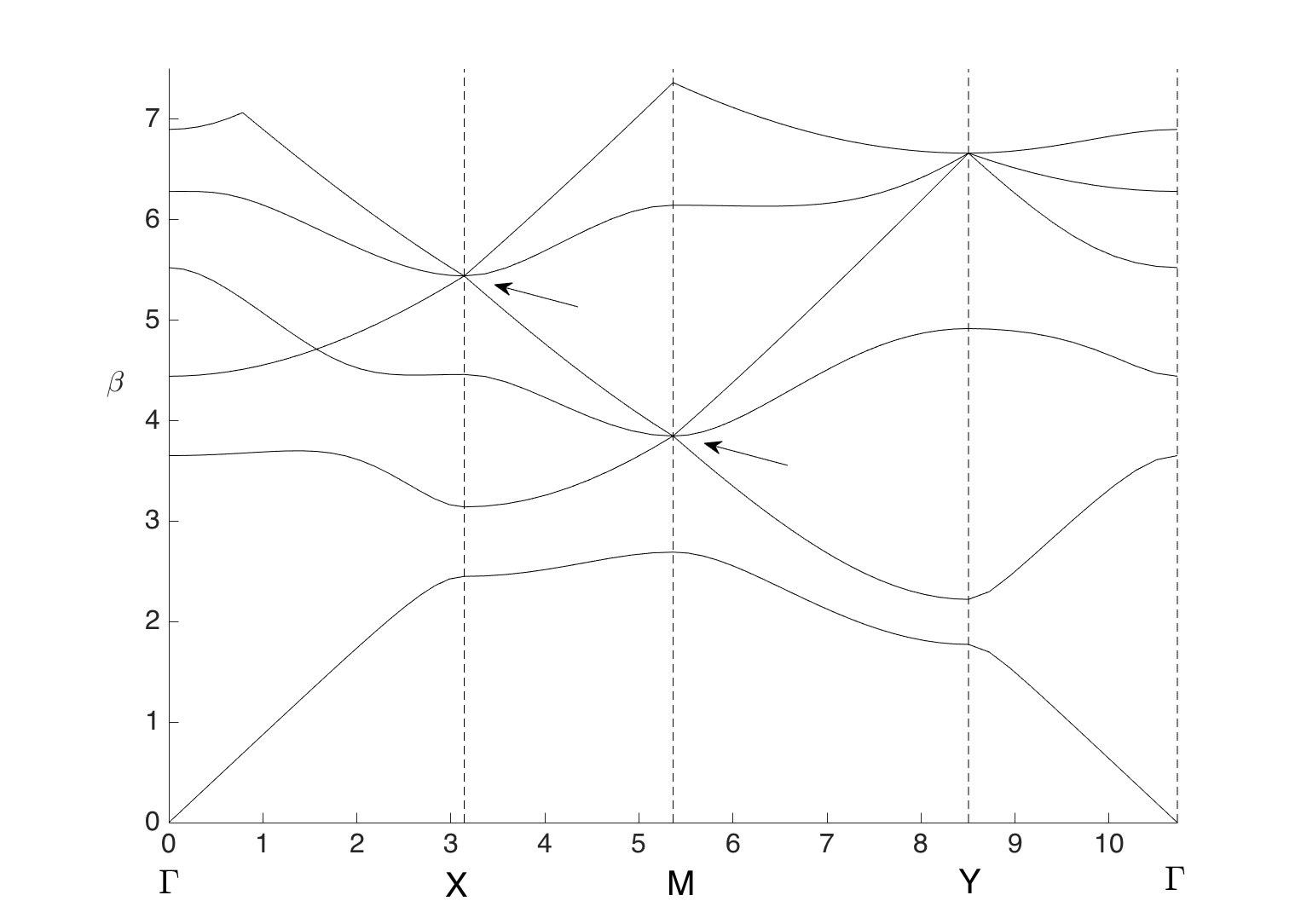}
\put(-150,43) {{\small(a)}}
\put(-100,43) {{\small(b)}}
\put(-50,43) {{\small(c)}}
\put(-65,23) {{\tiny $M$}}
\put(-98,28) {{\tiny $X$}}
\put(-45,21.8) {{\tiny $DP$}}
\caption{\label{surfsrect} 
Dispersion surfaces for rectangular array of point scatterers with $\xi = \sqrt{2}$ for (a) rigid pins (first five surfaces), (b) point masses, with $\tilde{m} = 1.0$ (first six surfaces), (c) Band diagram for point masses, with the Dirac-like points labelled at $X$ and $M$ in part (b) indicated by arrows. The Dirac point at $(\pi/2, 0)$, $\tilde{\beta} \approx 4.724$ is labelled by $DP$.}
\end{center}
\end{figure}
Here in figure~\ref{surfsrect} we consider the rectangular arrays ($\xi = \sqrt{2}$) of (a) pins and (b) point masses. Two Dirac-like points for the latter case are labelled at $X$ and $M$ of the irreducible Brillouin zone in figure~\ref{surfsrect} (b), and are also indicated by the arrows in the corresponding band diagram in figure~\ref{surfsrect}(c). The triple degeneracy, where the two Dirac-like cones are joined by another locally flat surface passing through, is clearly evident in figure~\ref{surfsrect}(c).

As discussed by Colquitt {\it et al.} (2016) for an elastic lattice, Dirac cones are often connected by relatively narrow flat regions on the
dispersion surfaces, which the authors term ``Dirac bridges". Dirac bridges possess resonances where the dispersion surfaces are locally parabolic, and give rise to highly localised unidirectional wave propagation. In this section, we consider one such regime for the point masses in the vicinity of a Dirac point at $(\pi/2, 0)$, $\tilde{\beta} \approx 4.724$, a feature present for both the masses and pins (labelled by $DP$ in figure~\ref{surfsrect}), and the two Dirac-like points highlighted in figures~\ref{surfsrect}(b,c).  
However, we observe additional steeply increasing sections of the third surface in figure~\ref{surfsrect}(b) for the case of point masses, which replace the flat parabolic profile parallel to $k_x = 0$ for the pins for the second surface in figure~\ref{surfsrect}(a).

\begin{figure}[h]
\begin{center}
\includegraphics[width= 7.2cm]{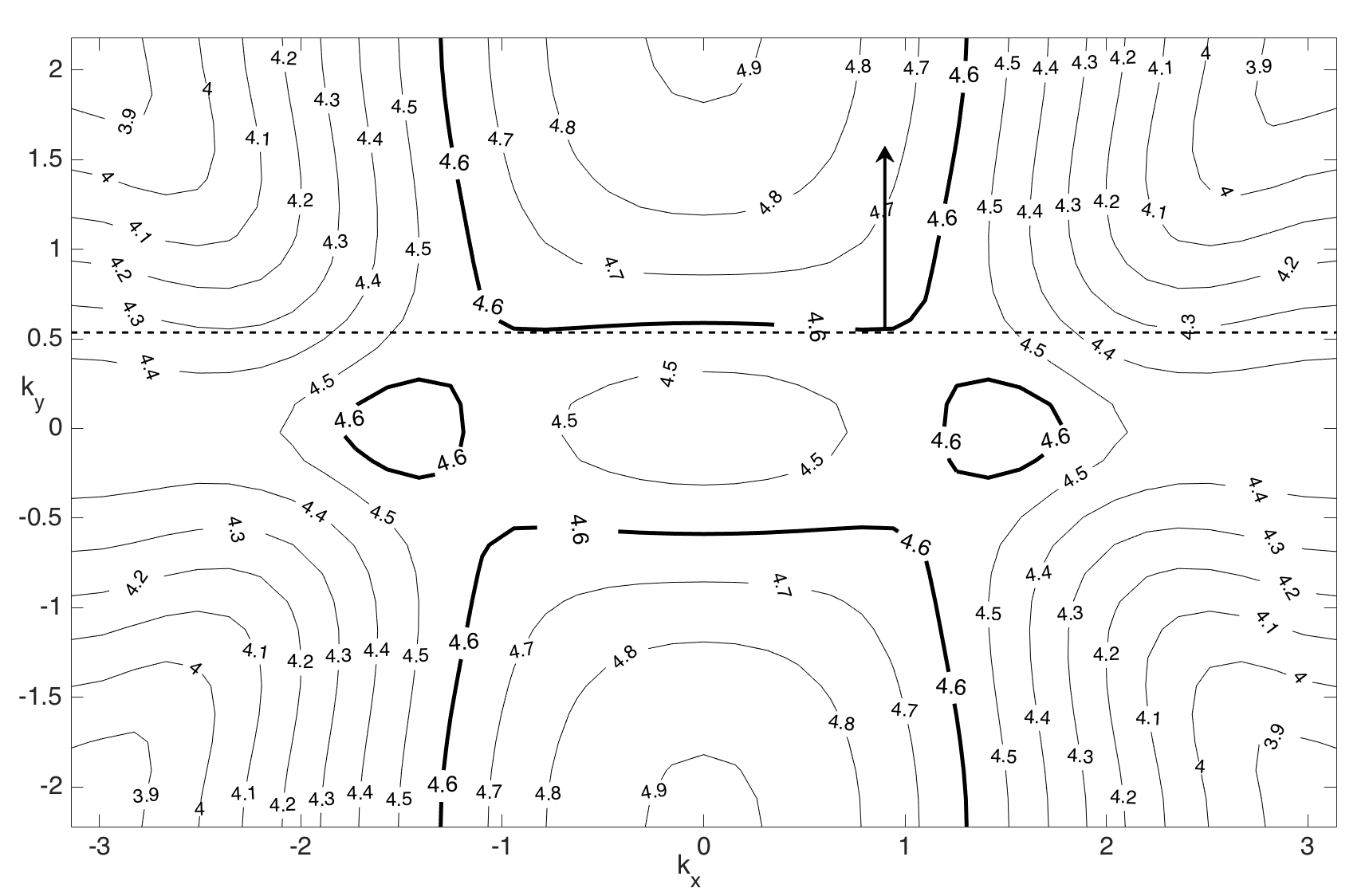}~~~~
\put(-0,52) {{\small(b)}}
\put(-66,52) {{\small(a)}}
\includegraphics[width= 6.9cm]{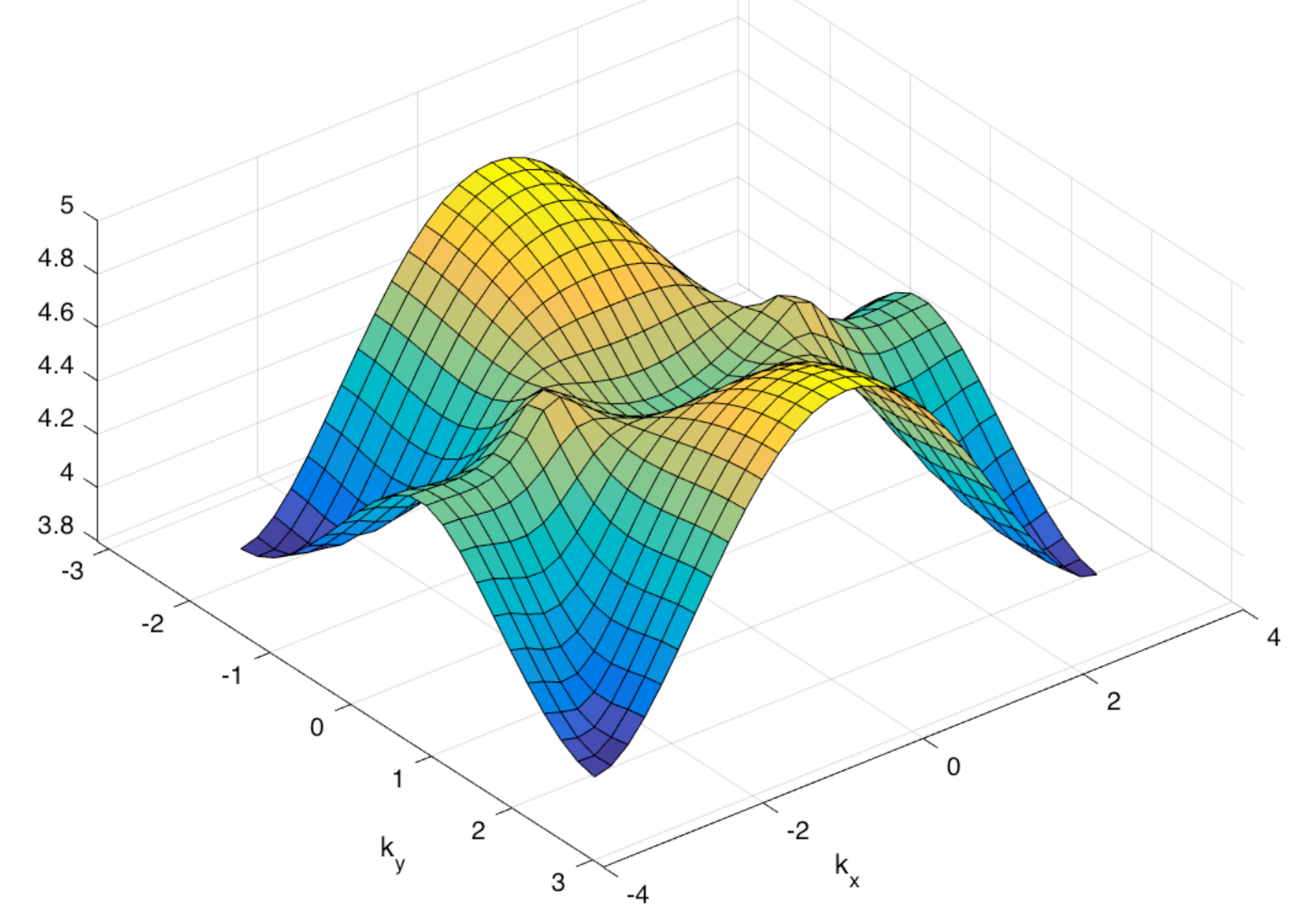}
\includegraphics[width= 7.8cm]{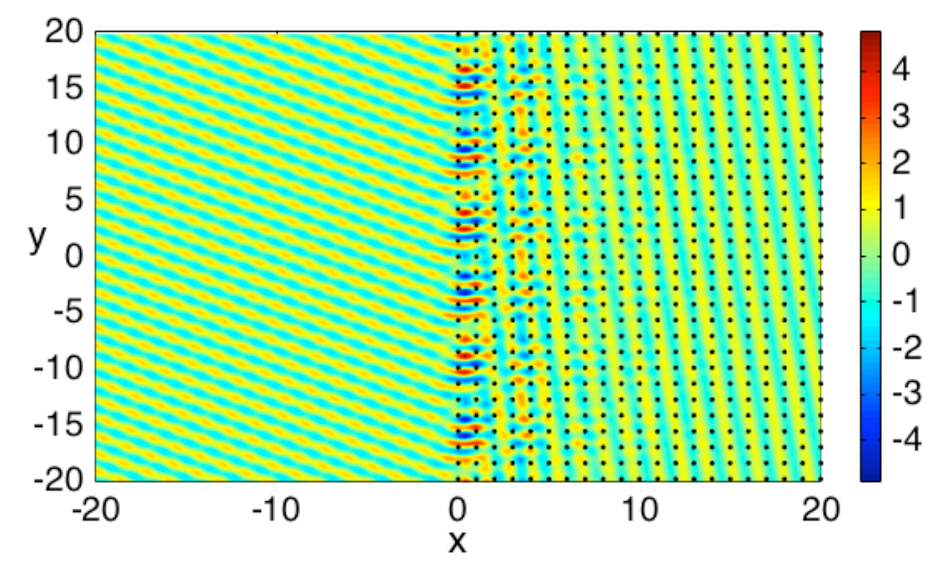}~~~
\includegraphics[width=7.2cm]{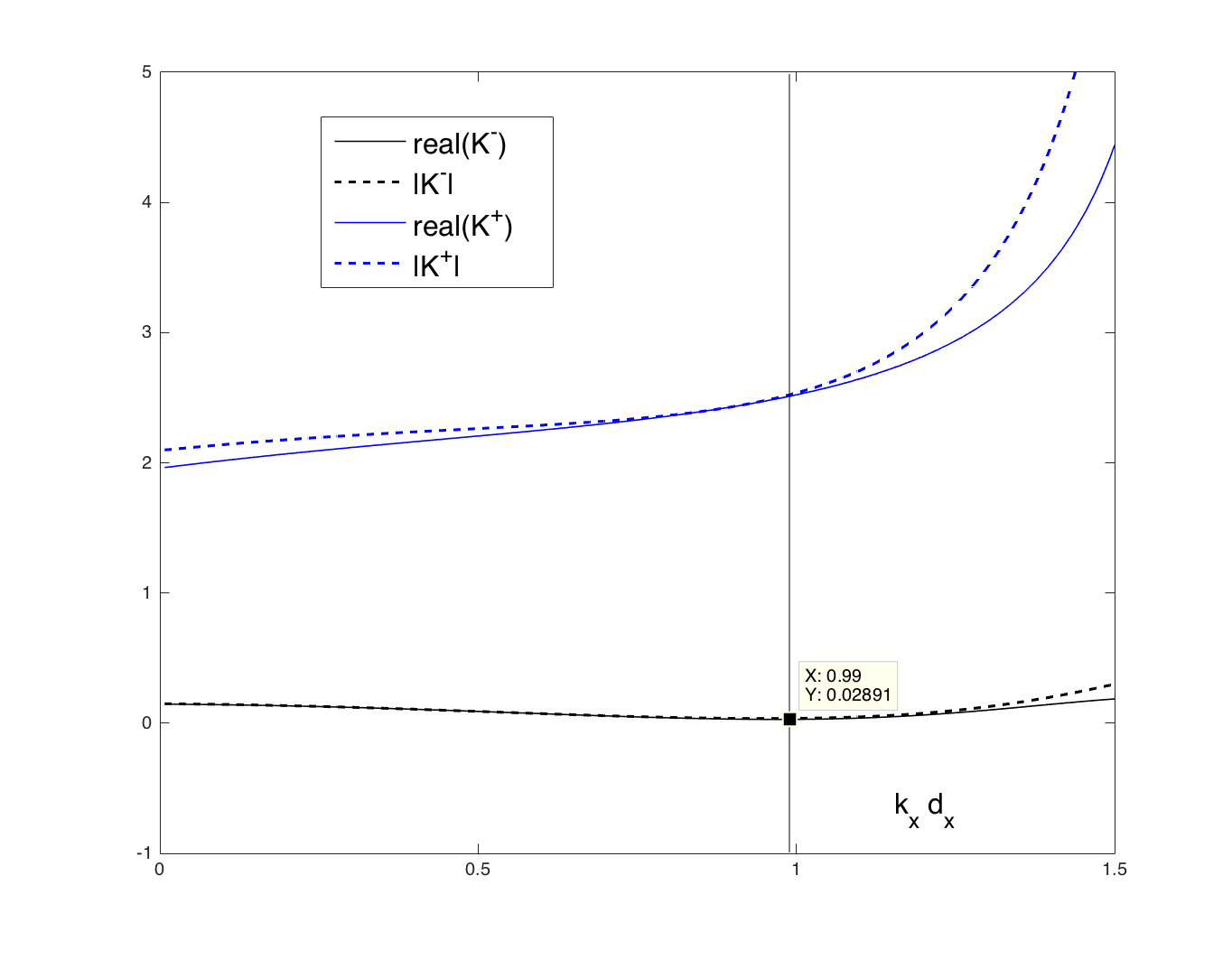}
\put(-140,50) {{\small(c)}}
\put(-75,50) {{\small(d)}}
\caption{\label{beta460_rect} Semi-infinite rectangular array of point masses with $\tilde{m} = 1.0$, $\xi = \sqrt{2}$. (a) Isofrequency contours for the third surface. The contour $\tilde{\beta} = 4.60$ is highlighted in bold. (b) Third dispersion surface. Real part of scattered field for $\tilde{\beta} = 4.60$ for (c) $\psi = 0.11$, $\tilde{k_y} = k_y d_x = 0.5050$. (d) ${\cal K}^-$ and ${\cal K}^+$ versus $k_x d_x \in [0, 1.5]$.}
\end{center}
\end{figure}

We investigate a semi-infinite rectangular array of 500 gratings of point masses with $\tilde{m} = 1.0$, $\xi = \sqrt{2}$ for $\tilde{\beta} = 4.60$. The third dispersion surface for the corresponding infinite system is shown in figures~\ref{beta460_rect}(a, b) by, respectively, isofrequency contours and the surface itself. With reference to the $\tilde{\beta} = 4.60$ contour of figure~\ref{beta460_rect}(a), the parameter setting of $\psi = 0.11$ (with associated Bloch parameter $k_y = \beta \sin{\psi} = 0.5050$ in the infinite $y$-direction) is selected to support a refracted wave directed parallel to the $k_y$-axis. Recall that since we are considering a finite array of 500 gratings, the information we obtain from the infinite doubly periodic system is only an approximate guide for the design choices of $\psi$ and $k_y$ for the corresponding finite system.

The designed system displays the interfacial localisation for the point masses in figure~\ref{beta460_rect}(c); the preferred direction for the group velocity of the resultant refracted wave is perpendicular to the isofrequency contour, and in the direction of increasing frequency, i.e. parallel to the $k_y$-axis (indicated by an arrow in part (a) of figure~\ref{beta460_rect}). In contrast, for a slight increase of $\psi$, and $k_y$ accordingly, the point of interest would move to the other side of the corner, parallel to the $k_y$-axis. The predicted direction would then be approximately parallel to the $k_x$-axis, and into the periodic part of the plate. We note that examples for interfacial localisation are easier to find for point masses rather than mass-spring resonators, which possess internal resonances at nearby frequencies ${\tilde{\beta}}^* = (\tilde{c}/\tilde{m})^{1/4}$.

\subsubsection{Eigenmodes and resonances for interfacial waves}
\label{eigen}

The observation of interfacial localisation, illustrated in figure~\ref{beta460_rect}, is linked to the analysis of the dispersion surfaces and stationary points of a certain type. Special attention was given to ``parabolic'' regimes, i.e. locally parabolic dispersion surfaces, which correspond to a unidrectional localisation of waveforms. Here, we offer an alternative viewpoint, based on the analysis of the homogeneous equation 
\begin{equation}
\hat{U}_- = \hat{U}_+ {\cal K},
\label{eig1}
\end{equation}
where the external forcing term is absent (compare with equation~(\ref{whgen}) for the general case). In the ring of analyticity, the kernel can be written as ${\cal K} = {\cal K}^+ {\cal K}^-$, and the factorised equation~(\ref{eig1}) takes the form:
\begin{equation}
\hat{U}_- /{\cal K}^-= \hat{U}_+ {\cal K}^+ =\mbox{\rm const} \ne 0.
\label{eig2}
\end{equation}
Here, $ \hat{U}_+$ and  $\hat{U}_-$ represent $z-$transforms of the displacements on the right and the left half-planes, respectively, as defined in~(\ref{whforms_hp}), and the kernel factors for a general $\Phi$ are analogous to those given for the mass-spring resonators in equation~(\ref{factkern}):
\begin{equation}
 \tilde{{\cal K}}^+ (\tilde{\beta}; \tilde{z}) =  \tilde{\Phi} (\tilde{\beta}, \tilde{m}, \tilde{c}) \,\, \tilde{\hat{{\cal G}}}_+(\tilde{\beta}; \tilde{z});  \,\,\,\,\,\,\,
\tilde{ {\cal K}}^-  (\tilde{\beta}; \tilde{z}) =
 \tilde{\hat{{\cal G}}}_-(\tilde{\beta}; \tilde{z}) - \frac{1}{\tilde{\Phi} (\tilde{\beta}, \tilde{m}, \tilde{c}) \, \tilde{\hat{{\cal G}}}_+(\tilde{\beta}; \tilde{z})} .
\label{kernfactsgen} 
\end{equation}

For the case of point masses illustrated in figures~\ref{beta460_rect}(a-c), $\tilde{\Phi} = \tilde{m} \tilde{\beta}^2$ in~(\ref{kernfactsgen}), and from~(\ref{eig2}) we seek a solution corresponding to localised interfacial waveforms such that ${\cal K}_-$ vanishes, whilst ${\cal K}_+$ remains finite. In turn, for this set of parameters the quantity $\hat{U}_-$ also vanishes, whereas a non-trivial solution $\hat{U}_+$ represents the interfacial waveform within the grating stack, as illustrated in figure~\ref{beta460_rect}(c).
In figure~\ref{beta460_rect}(d), we verify that the parameters $\tilde{\beta} = 4.6$ and $k_x \approx 0.99, k_y \approx 0.5050$ obtained from figure~\ref{beta460_rect}(a) satisfy these conditions. The real (solid) and absolute (dashed) parts of $\tilde{{\cal K}}_-$ and $\tilde{{\cal K}}_+$ are plotted, on the same figure~\ref{beta460_rect}(d), versus $k_x d_x \in [0, 1.5]$ for the vicinity of the estimate for $k_x d_x$ denoted by the position of the arrow in figure~\ref{beta460_rect}(a). For $k_x d_x = 0.99$, the former function does indeed have a local minimum $\approx 0$, whilst the latter function is finite and nonzero for the same $k_x d_x$.

\section{Concluding remarks}
\label{conc} 
The ability to control flexural wave propagation is important in numerous practical engineering structures such as bridges, aircraft wings and buildings, many of whose components may be modelled as structured elastic plates. In this article, we have modelled a collection of potential platonic crystals, where a Kirchhoff-Love plate is structured with a semi-infinite array of point scatterers, including concentrated point masses, mass-spring resonators positioned on either, or both, faces of the plate and Winkler-sprung masses. We have considered semi-infinite rectangular arrays, defined by periodicities $d_x, d_y$, but the methods are equally applicable for alternative geometries of the platonic crystal such as triangular or hexagonal lattices. 

The introduction of resonators, and their mass and spring stiffness parameters, significantly broadens the frequency range that supports interesting wave effects, compared with the simplified pinned plate model \cite{Has2015}. Here, we have shown examples of perfect transmission and negative refraction for various mass-spring resonator configurations at frequencies that would fall into the zero-frequency stop band imposed by the rigid pins.

A discrete Wiener-Hopf method was employed to determine the scattered and total displacement fields for a plane wave incident at a specified angle. The characteristic feature of each of the resulting functional equations is the kernel which, for all of the cases featured here, incorporates a doubly quasi-periodic Green's function:
\begin{equation}
{\cal K}(z) = \Phi(\omega, m, c) \hat{\cal G}(z) - 1,
\end{equation}
and a function $\Phi (\omega, m, c)$ of frequency, mass and stiffness determined by which of the four featured systems is being analysed. 
By identifying and deriving conditions for specific frequency regimes of the kernel function, we predict and demonstrate various scattering effects. In this article, we have illustrated examples of reflection, dynamic neutrality or perfect transmission, interfacial localisation and waveguide transmission. For certain regimes, we have also established a direct connection between alternative scatterers, including a condition for dynamic neutrality that occurs at the same frequency, shown in figure~\ref{neueffect}, for a plate with mass-spring resonators attached to both faces of the plate and Winkler-sprung masses.

The important observation that the semi-infinite system's kernel function is directly connected with the dispersion relation for the infinite doubly periodic platonic crystals, means that a thorough understanding of the Bloch-Floquet analysis provides great insight. Moreover, an understanding of the kernel function is sufficient to design the system for predicting and illustrating wave effects of interest, avoiding the necessity for lengthy computations for the evaluation of the explicit Wiener-Hopf solution. In section~\ref{eigen}, we introduce an alternative approach to predicting interfacial localisation frequency regimes, based on solving the homogeneous functional equation. This is an inherently interesting problem in itself, and we illustrate its viability with the example of figure~\ref{beta460_rect} obtained using wave-vector diagram analysis.

The numerous wave effects demonstrated here suggest that these semi-infinite platonic crystals have potential applications in the control and guiding of flexural waves in structures comprising thin plates. We have presented an overview of an assortment of practically interesting designs for semi-infinite platonic metamaterials. Any one of these models could be studied in its own right, with its parameters tuned to improve the resolution of the perfect transmission and interfacial localisation illustrated here. These effects are inherited by finite cluster subsets of the semi-infinite model, which could be used as a basis for the design and manufacture of semi-infinite platonic metamaterials.

\section*{Acknowledgements}
All of the authors thank the EPSRC (UK) for their support through the Programme Grant EP/L024926/1. SGH thanks Dr G. Carta and Dr D. J. Colquitt for valuable discussions about the use of finite element software packages.

\end{document}